# Prime Implicates and Prime Implicants:
# From Propositional to Modal Logic

**Meghyn Bienvenu**                                   MEGHYN@INFORMATIK.UNI-BREMEN.DE
*Department of Mathematics and Computer Science*
*University of Bremen, Germany*

## Abstract

Prime implicates and prime implicants have proven relevant to a number of areas of artificial intelligence, most notably abductive reasoning and knowledge compilation. The purpose of this paper is to examine how these notions might be appropriately extended from propositional logic to the modal logic $\mathcal{K}$. We begin the paper by considering a number of potential definitions of clauses and terms for $\mathcal{K}$. The different definitions are evaluated with respect to a set of syntactic, semantic, and complexity-theoretic properties characteristic of the propositional definition. We then compare the definitions with respect to the properties of the notions of prime implicates and prime implicants that they induce. While there is no definition that perfectly generalizes the propositional notions, we show that there does exist one definition which satisfies many of the desirable properties of the propositional case. In the second half of the paper, we consider the computational properties of the selected definition. To this end, we provide sound and complete algorithms for generating and recognizing prime implicates, and we show the prime implicate recognition task to be PSPACE-complete. We also prove upper and lower bounds on the size and number of prime implicates. While the paper focuses on the logic $\mathcal{K}$, all of our results hold equally well for multi-modal $\mathcal{K}$ and for concept expressions in the description logic $\mathcal{ALC}$.

## 1. Introduction

Prime implicates and prime implicants are important notions in artificial intelligence. They have given rise to a significant body of work in automated reasoning and have been applied to a number of different sub-areas in AI. Traditionally, these concepts have been studied in the context of propositional logic, but they have also been considered for many-valued (Ramesh & Murray, 1994) and first-order logic (Marquis, 1991a, 1991b). Not much is known, however, about prime implicates and prime implicants in other logics. In particular, no definition of prime implicate or prime implicant has ever been proposed for a modal or description logic, nor has it been shown that no reasonable definition can be provided. Given the increasing interest in modal and description logics as knowledge representation languages, one naturally wonders whether these notions can be suitably generalized to these more expressive logics.

We recall that in propositional logic the prime implicates of a formula are defined to be its logically strongest clausal consequences. The restriction to clauses is made in order to reduce redundant elements from a formula's set of consequences: there is no use in keeping around the consequence $a \wedge b$ when one already has the consequences $a$ and $b$. The decision to consider only the logically strongest clausal consequences is motivated by a desire to eliminate irrelevant weaker consequences: if we already have the consequence $a$, there is no point in retaining the consequences $a \vee b$ or $a \vee \neg b$. Prime implicates thus provide a complete yet compact representation of the set of logical consequences of a formula. What is particularly nice about this representation is that it makes many computational tasks simpler: satisfiability, tautology, entailment, and equivalence queries and the conditioning and forgetting transformations are all tractable for formulae represented by their prime implicates (Darwiche & Marquis, 2002). This is why prime implicates are considered an interesting target language for knowledge compilation (Cadoli & Donini, 1997; Darwiche & Marquis, 2002). Prime implicates have also proved relevant to other sub-areas of AI, like distributed reasoning (Adjiman, Chatalic, Goasdoué, Rousset, & Simon, 2006), belief revision (Bittencourt, 2007; Pagnucco, 2006), non-monotonic reasoning (cf. Przymusinski, 1989), and characterizations of relevance (Lakemeyer, 1995; Lang, Liberatore, & Marquis, 2003).

The dual notion to prime implicates is prime implicants, which are defined to be the logically weakest terms (= conjunctions of literals) which imply a given formula. The main application domain for prime implicants is in abduction and diagnosis. We recall that in abduction, one is given a background theory and an observation, and the objective is to find an explanation for the observation. In logical terms, an explanation is a formula which logically entails the observation when taken together with the background theory. As the set of explanations for an abduction problem can be very large, an important question is how to select a representative subset of explanations. One very common approach is to use prime implicants: the relevant explanations for an observation $o$ with respect to a background theory $t$ are taken to be the prime implicants of $t \to o$ (de Kleer, Mackworth, & Reiter, 1992; Eiter & Makino, 2002).

For many applications in AI, the expressive power of propositional logic proves insufficient. First-order logic provides a much greater level of expressivity, but at the price of undecidability. Modal and description logics offer an interesting trade-off between expressivity and complexity, as they are generally more expressive than propositional logic yet are better-behaved computationally than first-order logic. This explains the growing trend towards using such languages for knowledge representation.

A prototypical description logic is $\mathcal{ALC}$, which extends propositional logic with restricted forms of universal and existential quantification. An example expression in $\mathcal{ALC}$ is

$$Female \sqcap \exists hasChild.Female \sqcap \forall hasChild.(Doctor \sqcup Professor) \sqcap \exists hasPet.Dog$$

which describes the set of individuals who are female, have at least one daughter and one pet dog, and are such that all of their children are either doctors or professors. The above concept expression can be represented equally well in the modal logic $\mathcal{K}_2$ by the formula:

$$Female \wedge \Diamond_1 Female \wedge \Box_1(Doctor \vee Professor) \wedge \Diamond_2 Dog$$

Schild (1991) proved a general result which showed that the description logic $\mathcal{ALC}$ with $n$ binary relations is in fact a notational variant of the multi-modal logic $\mathcal{K}_n$. This means that results concerning $\mathcal{K}_n$ can be transferred to $\mathcal{ALC}$, and vice-versa.

In this paper, we investigate the notions of prime implicates and prime implicants for the modal logic $\mathcal{K} = \mathcal{K}_1$, but actually all of our results hold for formulae in $\mathcal{K}_n$, and hence also for concept expressions in $\mathcal{ALC}$. The decision to present our results in terms of $\mathcal{K}$ rather than in terms of $\mathcal{K}_n$ or $\mathcal{ALC}$ was motivated solely by a desire to simplify notation and increase the readability of the proofs.

The question of how the notions of prime implicates and prime implicants can be suitably defined for the logic $\mathcal{K}$ is clearly of interest from a theoretical point of view. We argue, however, that this question is also practically relevant. To support this claim, we briefly discuss two application areas in which the study of prime implicates and prime implicants in $\mathcal{K}$ might prove useful.

The first domain of application we will consider is abductive reasoning in $\mathcal{K}$. As noted above, one of the key foundational issues in abductive reasoning is the selection of an interesting subset of explanations. This issue is especially crucial for logics like $\mathcal{K}$ which allow for an infinite number of non-equivalent formulae, since this means that the number of non-equivalent explanations for an abduction problem is not just large but in fact infinite, making it simply impossible to enumerate the entire set of explanations. As prime implicants are a widely-accepted means of characterizing relevant explanations in propositional logic, a reasonable starting point for research into abductive reasoning in the logic $\mathcal{K}$ is the study of different possible definitions of prime implicant in $\mathcal{K}$ and their properties.

The investigation of prime implicates in $\mathcal{K}$ is also relevant to the development of knowledge compilation procedures for $\mathcal{K}$. We recall that knowledge compilation (cf. Darwiche & Marquis, 2002) is a general technique for coping with the intractability of reasoning which consists in an off-line phase in which a knowledge base is rewritten as an equivalent knowledge base which allows for tractable reasoning, followed by an online phase in which reasoning is performed on the compiled knowledge base. The idea is that the initial cost of compiling the knowledge base will be offset by computational savings on later queries. Currently, most work on knowledge compilation is restricted to propositional logic, even though this technique could prove highly relevant for modal and description logics, which generally suffer from an even higher computational complexity than propositional logic. As prime implicates are one of the better-known mechanisms for compiling formulae in propositional logic, it certainly makes sense to investigate whether this approach to knowledge compilation can be fruitfully extended to logics like $\mathcal{K}$.

Our paper is organized as follows. After some preliminaries, we consider how to appropriately generalize the notions of clauses and terms to $\mathcal{K}$. As there is no obvious definition, we enumerate a list of syntactic, semantic, and complexity-theoretic properties of the propositional definitions, which we then use to compare the different candidate definitions. We next consider the different definitions in light of the notions of prime implicate and prime implicant they induce. Once again, we list some basic properties from the propositional case that we would like to satisfy, and we see how the different definitions measure up. In the second half of the paper, we investigate the computational properties of the most satisfactory definition of prime implicates. We consider the problems of prime implicate generation and recognition, and we provide sound and complete algorithms for both tasks. We also study the complexity of the prime implicate recognition problem, showing it to be Pspace-complete and thus of the same complexity as satisfiability and deduction in $\mathcal{K}$. We conclude the paper with a discussion of the relevance of our results to the two application areas cited above and some directions for future research. In order to enhance the readability of the paper, proofs have been omitted from the body of the text. Full proofs can be found in the appendix.

## 2. Preliminary Definitions and Notation

We briefly recall the basics of the modal logic $\mathcal{K}$ (refer to Blackburn, de Rijke, & Venema, 2001; Blackburn, van Benthem, & Wolter, 2006, for good introductions to modal logic). Formulae in $\mathcal{K}$ are built up from a set of propositional variables $\mathcal{V}$, the standard logical connectives ($\neg$, $\wedge$, and $\vee$), and the modal operators $\Box$ and $\Diamond$. We will call a formula of the form $\Box\varphi$ (resp. $\Diamond\varphi$) a $\Box$-*formula* (resp. $\Diamond$-*formula*). Where convenient we will use $\varphi \rightarrow \psi$ as an abbreviation for $\neg\varphi \vee \psi$. We adopt the shorthand $\Box^k\varphi$ (resp. $\Diamond^k\varphi$) to refer to the formula consisting of $\varphi$ preceded by $k$ copies of $\Box$ (resp. $\Diamond$), with the convention that $\Box^0\varphi = \Diamond^0\varphi = \varphi$. We will use $var(\varphi)$ to refer to the set of propositional variables appearing in a formula $\varphi$. The modal *depth* of a formula $\varphi$, written $\delta(\varphi)$, is defined as the maximal number of nested modal operators appearing in $\varphi$, e.g. $\delta(\Diamond(a \wedge \Box a) \vee a) = 2$. We define the *length* of a formula $\varphi$, written $|\varphi|$, to be the number of occurrences of propositional variables, logical connectives, and modal operators in $\varphi$. For example, we would have $|(a \wedge \neg b)| = 4$ and $|\Diamond(a \vee b) \wedge \Box\neg a| = 8$.

*Negation normal form* (NNF) is defined just as in propositional logic: a formula is said to be in NNF if negation only appears directly before propositional variables. Every formula $\varphi$ in $\mathcal{K}$ can be transformed into an equivalent formula in NNF using the recursive procedure **Nnf** defined as follows:

$$
\begin{array}{ll}
\mathbf{Nnf}(l) = l \text{ (for propositional literals } l\text{)} & \mathbf{Nnf}(\Box\psi) = \Box\mathbf{Nnf}(\psi) \\
\mathbf{Nnf}(\psi_1 \wedge \psi_2) = \mathbf{Nnf}(\psi_1) \wedge \mathbf{Nnf}(\psi_2) & \mathbf{Nnf}(\neg\Box\psi) = \Diamond\mathbf{Nnf}(\neg\psi) \\
\mathbf{Nnf}(\neg(\psi_1 \wedge \psi_2)) = \mathbf{Nnf}(\neg\psi_1) \vee \mathbf{Nnf}(\neg\psi_2) & \mathbf{Nnf}(\Diamond\psi) = \Diamond\mathbf{Nnf}(\psi) \\
\mathbf{Nnf}(\psi_1 \vee \psi_2) = \mathbf{Nnf}(\psi_1) \vee \mathbf{Nnf}(\psi_2) & \mathbf{Nnf}(\neg\Diamond\psi) = \Box\mathbf{Nnf}(\neg\psi) \\
\mathbf{Nnf}(\neg(\psi_1 \vee \psi_2)) = \mathbf{Nnf}(\neg\psi_1) \wedge \mathbf{Nnf}(\neg\psi_2) & \mathbf{Nnf}(\neg\neg\psi) = \mathbf{Nnf}(\psi)
\end{array}
$$

For example, applying **Nnf** to the formula $\neg\Box(a \wedge \Diamond(\neg b \vee c))$ results in the formula $\Diamond(\neg a \vee \Box(b \wedge \neg c))$ which is in NNF. The transformation **Nnf** takes linear time, and yields a formula which is no more than double the size of the original formula and has the same modal depth and propositional variables as the original.

A *model* for $\mathcal{K}$ is a tuple $\mathfrak{M} = \langle \mathcal{W}, \mathcal{R}, v \rangle$, where $\mathcal{W}$ is a non-empty set of possible worlds, $\mathcal{R} \subseteq \mathcal{W} \times \mathcal{W}$ is a binary relation over worlds, and $v : \mathcal{W} \times \mathcal{V} \rightarrow \{true, false\}$ is a valuation of the propositional variables at each world. Models can be understood as labelled directed graphs, in which the vertices correspond to the elements of $\mathcal{W}$, the directed edges represent the binary relation $\mathcal{R}$, and the vertices are labeled by propositional valuations which specify the propositional variables which are true in the corresponding possible world.

Satisfaction of a formula $\varphi$ in a model $\mathfrak{M}$ at the world $w$ (written $\mathfrak{M}, w \models \varphi$) is defined inductively as follows:

- $\mathfrak{M}, w \models a$ if and only if $v(w, a) = true$

- $\mathfrak{M}, w \models \neg\varphi$ if and only if $\mathfrak{M}, w \not\models \varphi$

- $\mathfrak{M}, w \models \varphi \wedge \psi$ if and only if $\mathfrak{M}, w \models \varphi$ and $\mathfrak{M}, w \models \psi$

- $\mathfrak{M}, w \models \varphi \vee \psi$ if and only if $\mathfrak{M}, w \models \varphi$ or $\mathfrak{M}, w \models \psi$

- $\mathfrak{M}, w \models \Box\varphi$ if and only if $\mathfrak{M}, w' \models \varphi$ for all $w'$ such that $w\mathcal{R}w'$

- $\mathfrak{M}, w \models \Diamond\varphi$ if and only if $\mathfrak{M}, w' \models \varphi$ for some $w'$ such that $w\mathcal{R}w'$

If we think of models as labeled directed graphs, then determining the satisfaction of a formula $\Box\varphi$ at vertex $w$ consists in evaluating $\varphi$ at all of the vertices which can be reached from $w$ via an edge; $\Box\varphi$ is satisfied at $w$ just in the case that $\varphi$ holds in each of these successor vertices. Similarly, in order to decide whether a formula $\Diamond\varphi$ holds at a vertex $w$, we consider each of the successors of $w$ in the graph and check whether at least one of these vertices satisfies $\varphi$.

A formula $\varphi$ is said to be a *tautology*, written $\models \varphi$, if $\mathfrak{M}, w \models \varphi$ for every model $\mathfrak{M}$ and world $w$. A formula $\varphi$ is *satisfiable* if there is some model $\mathfrak{M}$ and some world $w$ such that $\mathfrak{M}, w \models \varphi$. If there is no $\mathfrak{M}$ and $w$ for which $\mathfrak{M}, w \models \varphi$, then $\varphi$ is called *unsatisfiable*, and we write $\varphi \models \bot$.

Ladner (1977) showed that satisfiability and unsatisfiability in $\mathcal{K}$ are PSPACE-complete. For PSPACE membership, Ladner exhibited a polynomial space tableaux-style algorithm for deciding satisfiability of $\mathcal{K}$ formulae. PSPACE-hardness was proven by means of a reduction from QBF validity (the canonical PSPACE-complete problem).

In modal logic, the notion of logical consequence (or entailment) can be defined in one of two ways:

- a formula $\psi$ is a *global consequence* of $\varphi$ if whenever $\mathfrak{M}, w \models \varphi$ for every world $w$ of a model $\mathfrak{M}$, then $\mathfrak{M}, w \models \psi$ for every world $w$ of $\mathfrak{M}$

- a formula $\psi$ is a *local consequence* of $\varphi$ if $\mathfrak{M}, w \models \varphi$ implies $\mathfrak{M}, w \models \psi$ for every model $\mathfrak{M}$ and world $w$

In this paper, we will only consider the notion of local consequence, and we will take $\varphi \models \psi$ to mean that $\psi$ is a local consequence of $\varphi$. When $\varphi \models \psi$, we will say that $\varphi$ *entails* $\psi$. Two formulae $\varphi$ and $\psi$ will be called *equivalent*, written $\varphi \equiv \psi$, if both $\varphi \models \psi$ and $\psi \models \varphi$. A formula $\varphi$ is said to be *logically stronger* than $\psi$ if $\varphi \models \psi$ and $\psi \not\models \varphi$.

We now highlight some basic properties of logical consequence and equivalence in $\mathcal{K}$ which will play an important role in the proofs of our results.

**Theorem 1.** *Let $\psi, \psi_1, ..., \psi_m, \chi, \chi_1, ..., \chi_n$ be formulae in $\mathcal{K}$, and let $\gamma$ be a propositional formula. Then*

1. $\psi \models \chi \Leftrightarrow \models \neg\psi \vee \chi \Leftrightarrow \psi \wedge \neg\chi \models \bot$

2. $\psi \models \chi \Leftrightarrow \Diamond\psi \models \Diamond\chi \Leftrightarrow \Box\psi \models \Box\chi$

3. $\gamma \wedge \Diamond\psi_1 \wedge ... \wedge \Diamond\psi_m \wedge \Box\chi_1 \wedge ... \wedge \Box\chi_n \models \bot \Leftrightarrow (\gamma \models \bot \text{ or } \psi_i \wedge \chi_1 \wedge ... \wedge \chi_n \models \bot \text{ for some } i)$

4. $\models \gamma \vee \Diamond\psi_1 \vee ... \vee \Diamond\psi_m \vee \Box\chi_1 \vee ... \vee \Box\chi_n \Leftrightarrow (\models \gamma \text{ or } \models \psi_1 \vee ... \vee \psi_m \vee \chi_i \text{ for some } i)$

5. $\Box\chi \models \Box\chi_1 \vee ... \vee \Box\chi_n \Leftrightarrow \chi \models \chi_i \text{ for some } i$

6. $\Diamond\psi_1 \vee ... \vee \Diamond\psi_m \vee \Box\chi_1 \vee ... \vee \Box\chi_n$
   $\equiv \Diamond\psi_1 \vee ... \vee \Diamond\psi_m \vee \Box(\chi_1 \vee \psi_1 \vee ... \vee \psi_m) \vee ... \vee \Box(\chi_n \vee \psi_1 \vee ... \vee \psi_m)$

Statement 1 of Theorem 1 shows how the three reasoning tasks of entailment, unsatisfiability, and tautology-testing can be rephrased in terms of one another. Statement 2 tells us how entailment between two $\Box$- or $\Diamond$-formulae can be reduced to entailment between those formulae with the first modality removed. Statements 3 and 4 define the conditions under which a conjunction (resp. disjunction) of propositional literals and $\Box$- and $\Diamond$-formulae is unsatisfiable (resp. a tautology). Statement 5 gives us the conditions under which a $\Box$-formula implies a disjunction of $\Box$-formulae. Statement 6 demonstrates the interaction between $\Box$- and $\Diamond$-formulae in a disjunction.

**Theorem 2.** *Let $\lambda$ be a disjunction of propositional literals and $\Box$- and $\Diamond$-formulae. Then each of the following statements holds:*

1. *If $\lambda \models \gamma$ for some non-tautological propositional clause $\gamma$, then every disjunct of $\lambda$ is either a propositional literal or an unsatisfiable $\Diamond$-formula*

2. *If $\lambda \models \Diamond\psi_1 \vee ... \vee \Diamond\psi_n$, then every disjunct of $\lambda$ is a $\Diamond$-formula*

3. *If $\lambda \models \Box\chi_1 \vee ... \vee \Box\chi_n$ and $\not\models \Box\chi_1 \vee ... \vee \Box\chi_n$, then every disjunct of $\lambda$ is either a $\Box$-formula or an unsatisfiable $\Diamond$-formula*

**Theorem 3.** *Let $\lambda = \gamma \vee \Diamond\psi_1 \vee ... \vee \Diamond\psi_m \vee \Box\chi_1 \vee ... \vee \Box\chi_n$ and $\lambda' = \gamma' \vee \Diamond\psi_1' \vee ... \vee \Diamond\psi_p' \vee \Box\chi_1' \vee ... \vee \Box\chi_q'$ be formulae in $\mathcal{K}$. If $\gamma$ and $\gamma'$ are both propositional and $\not\models \lambda'$, then*

$$\lambda \models \lambda' \Leftrightarrow \left\{ \begin{array}{l} \gamma \models \gamma' \text{ and} \\ \psi_1 \vee ... \vee \psi_m \models \psi_1' \vee ... \vee \psi_p' \text{ and} \\ \text{for every } \chi_i \text{ there is some } \chi_j' \text{ such that } \chi_i \models \psi_1' \vee ... \vee \psi_p' \vee \chi_j' \end{array} \right.$$

Theorems 2 and 3 concern entailment relations between formulae which are disjunctions of propositional literals and $\Box$- and $\Diamond$-formulae. Theorem 2 tells us what kinds of formulae of this type can entail a propositional clause, a disjunction of $\Diamond$-formulae, or a disjunction of $\Box$-formulae, while Theorem 3 outlines the conditions under which two formulae of this type can be related to each other by the entailment relation. We illustrate Theorem 3 on a small example.

**Example 4.** Consider the formula $\lambda = \neg b \vee \Diamond(a \wedge \Diamond c) \vee \Diamond(d \wedge \Box a) \vee \Box(c \vee d)$. Then according to Theorem 3, we have:

- $\lambda \models \neg b \vee \neg d \vee \Diamond(a \vee d) \vee \Box c$, since $\neg b \models \neg b \vee \neg d$ and $(a \wedge \Diamond c) \vee (d \wedge \Box a) \models a \vee d$ and $c \vee d \models c \vee (a \vee d)$

- $\lambda \not\models a \vee \Diamond c$, since $\neg b \not\models a$

- $\lambda \not\models a \vee \neg b \vee \Diamond(a \wedge c)$, since $(a \wedge \Diamond c) \vee (d \wedge \Box a) \not\models a \wedge c$

- $\lambda \not\models \neg b \vee \Diamond(a \vee \Box a) \vee \Box c$, since $c \vee d \not\models c \vee (a \vee \Box a)$

## 3. Literals, Clauses, and Terms in $\mathcal{K}$

As we have seen in the introduction, the notions of prime implicates and implicants are straightforwardly defined using the notions of clauses and terms. Thus, if we aim to provide suitable definitions of prime implicates and implicants for the logic $\mathcal{K}$, we first need to decide upon a suitable definition of clauses and terms in $\mathcal{K}$. Unfortunately, whereas clauses and terms are standard notions in both propositional and first-order logic[1], there is no generally accepted definition of clauses and terms in $\mathcal{K}$. Indeed, several quite different notions of clauses and terms have been proposed in the literature for different purposes.

Instead of blindly picking a definition and hoping that it is appropriate, we prefer to list a number of characteristics of literals, clauses, and terms in propositional logic, giving us a principled means of comparing different candidate definitions. Each of the properties below describes something of what it is to be a literal, clause, or term in propositional logic. Although our list cannot be considered exhaustive, we do believe that it covers the principal syntactic, semantic, and complexity-theoretic properties of the propositional definition.

**P1** Literals, clauses, and terms are in negation normal form.

**P2** Clauses do not contain $\wedge$, terms do not contain $\vee$, and literals contain neither $\wedge$ nor $\vee$.

**P3** Clauses (resp. terms) are disjunctions (resp. conjunctions) of literals.

**P4** The negation of a literal is equivalent to another literal. Negations of clauses (resp. terms) are equivalent to terms (resp. clauses).

**P5** Every formula is equivalent to a finite conjunction of clauses. Likewise, every formula is equivalent to a finite disjunction of terms.

**P6** The task of deciding whether a given formula is a literal, term, or clause can be accomplished in polynomial-time.

**P7** The task of deciding whether a clause (resp. term) entails another clause (resp. term) can be accomplished in polynomial-time.

One may wonder whether there exist definitions of literals, clauses, and terms for $\mathcal{K}$ satisfying all of these properties. Unfortunately, we can show this to be impossible.

**Theorem 5.** *Any definition of literals, clause, and terms for $\mathcal{K}$ that satisfies properties* **P1** *and* **P2** *cannot satisfy* **P5**.

The proof of Theorem 5 only makes use of the fact that $\wedge$ does not distribute over $\diamondsuit$ and $\vee$ does not distribute over $\square$, which means that our impossibility result holds equally well for most standard modal and description logics.

We will now consider a variety of possible definitions and evaluate them with respect to the above criteria. The first definition that we will consider is that proposed by Cialdea

---

1. One might wonder why we do not simply translate our formulae in $\mathcal{K}$ into first-order formulae and then put them into clausal form. The reason is simple: we are looking to define clauses and terms *within* the language of $\mathcal{K}$, and the clauses we obtain on passing by first-order logic are generally not expressible in $\mathcal{K}$. Moreover, if we were to define clauses in $\mathcal{K}$ as those first-order clauses which are representable in $\mathcal{K}$, we would obtain a set of clauses containing no $\diamondsuit$ modalities, thereby losing much of the expressivity of $\mathcal{K}$.

Mayer and Pirri (1995) in a paper on abductive reasoning in modal logic. They define terms to be the formulae which can be constructed from the propositional literals using only $\wedge$, $\square$, and $\diamond$. Modal clauses and literals are not used in the paper but can be defined analogously, yielding the following definition[2]:

$$\textbf{D1} \quad \begin{aligned} L &::= a \mid \neg a \mid \square L \mid \diamond L \\ C &::= a \mid \neg a \mid \square C \mid \diamond C \mid C \vee C \\ T &::= a \mid \neg a \mid \square T \mid \diamond T \mid T \wedge T \end{aligned}$$

It is easy to see by inspection that this definition satisfies properties **P1**-**P2**, **P4**, and **P6**. Property **P3** is not satisfied, however, since there are clauses that are not disjunctions of literals – take for instance $\square(a \vee b)$. From Theorem 5 and the fact that both **P1** and **P2** are satisfied, we can conclude that property **P5** cannot hold. At first glance, it may seem that entailment between clauses or terms could be accomplished in polynomial time, but this is not the case. In fact, we can show this problem to be NP-complete. The proof relies on the very strong resemblance between terms of **D1** and concept expressions in the description logic $\mathcal{ALE}$ (for which both unsatisfiability and deduction are known to be NP-complete).

By using a slightly different definition, we can gain **P3**:

$$\textbf{D2} \quad \begin{aligned} L &::= a \mid \neg a \mid \square L \mid \diamond L \\ C &::= L \mid C \vee C \\ T &::= L \mid T \wedge T \end{aligned}$$

It can be easily verified that definition **D2** satisfies properties **P1**-**P4** and **P6**. As definition **D1** does not satisfy **P5**, and definition **D2** is even less expressive, it follows that **D2** does not satisfy **P5** either. This reduced expressiveness does not however improve its computational complexity: property **P7** is still not satisfied as we can show that entailment between clauses or terms is NP-complete using the same reduction as was used for definition **D1**. The fact that even an extremely inexpressive definition like **D2** does not allow for polynomial entailment between clauses and terms suggests that property **P7** cannot be satisfied by any reasonable definition of clauses and terms for $\mathcal{K}$.

Let us now consider some more expressive options. We begin with the following definition of clauses that was proposed by Enjalbert and Fariñas del Cerro (1989) for the purpose of modal resolution:

$$\textbf{D3} \quad \begin{aligned} C &::= a \mid \neg a \mid \square C \mid \diamond ConjC \mid C \vee C \\ ConjC &::= C \mid ConjC \wedge ConjC \end{aligned}$$

This definition of clauses can be extended to a definition of terms and literals which satisfies **P3** or **P4**, but there is no extension which satisfies both properties. Let us first consider one of the possible extensions which satisfies **P4** and a maximal subset of **P1**-**P7**:

$$\textbf{D3a} \quad \begin{aligned} L &::= a \mid \neg a \mid \square L \mid \diamond L \\ C &::= a \mid \neg a \mid \square C \mid \diamond ConjC \mid C \vee C \\ ConjC &::= C \mid ConjC \wedge ConjC \\ T &::= a \mid \neg a \mid \square DisjT \mid \diamond T \mid T \wedge T \\ DisjT &::= T \mid DisjT \vee DisjT \end{aligned}$$

---

2. Note that here and in what follows, we let $a$ range over propositional variables and $L$, $C$, and $T$ range over the sets of literals, clauses, and terms, respectively.

This definition satisfies **P1** and **P4**-**P6** (satisfaction of **P5** was shown in Enjalbert & Fariñas del Cerro, 1989). It does not satisfy **P3** as there are clauses that are not disjunctions of literals – take for example $\Box(a \lor b)$. Given that definition **D3a** is strictly more expressive than definitions **D1** and **D2**, it follows that entailment between clauses or terms must be NP-hard, which means that **D3a** does not satisfy **P7**. In fact, we can show that entailment between clauses or terms of definition **D3a** is PSPACE-complete. To do so, we modify the polynomial translation of QBF into $\mathcal{K}$ used to prove PSPACE-hardness of $\mathcal{K}$ so that the translated formula is a conjunction of clauses with respect to **D3a**. We then notice that a formula $\varphi$ is unsatisfiable if and only if $\Diamond\varphi$ entails $\Diamond(a \land \neg a)$. We thus reduce QBF validity to entailment between clauses, making this task PSPACE-hard, and hence (being a subproblem of entailment in $\mathcal{K}$) PSPACE-complete. This same idea is used to show PSPACE-completeness for definitions **D3b** and **D5** below.

If instead we extend **D3** so as to enforce property **P3**, we obtain the following definition:

$$\begin{aligned}
&L ::= a \mid \neg a \mid \Box C \mid \Diamond ConjC\\
\textbf{D3b} \quad &C ::= a \mid \neg a \mid \Box C \mid \Diamond ConjC \mid C \lor C\\
&ConjC ::= C \mid ConjC \land ConjC\\
&T ::= L \mid T \land T
\end{aligned}$$

This definition satisfies all of the properties except **P2**, **P4**, and **P7**. Property **P4** fails to hold because the negation of the literal $\Diamond(a \lor b)$ is not equivalent to any literal. The proof that **P5** holds is constructive: we use standard logical equivalences to rewrite formulae as equivalent conjunctions of clauses and disjunctions of terms (this is also what we do for definitions **D4** and **D5** below).

We now consider two rather simple definitions that satisfy properties **P3**, **P4**, and **P5**. The first definition, which is inspired by the notion of modal atom proposed by Giunchiglia and Sebastiani (1996), defines literals as the set of formulae in NNF that cannot be decomposed propositionally.

$$\begin{aligned}
&L ::= a \mid \neg a \mid \Box F \mid \Diamond F\\
\textbf{D4} \quad &C ::= L \mid C \lor C\\
&T ::= L \mid T \land T\\
&F ::= a \mid \neg a \mid F \land F \mid F \lor F \mid \Box F \mid \Diamond F
\end{aligned}$$

**D4** satisfies all of the properties except **P2** and **P7**. For **P7**, we note that an arbitrary formula $\varphi$ in NNF is unsatisfiable (a PSPACE-complete problem) if and only if $\Diamond\varphi \models \Diamond(a \land \neg a)$.

Definition **D4** is very liberal, imposing no structure on the formulae behind modal operators. If we define literals to be the formulae in NNF that cannot be decomposed *modally* (instead of propositionally), we obtain a much stricter definition which satisfies exactly the same properties as **D4**.

$$\begin{aligned}
&L ::= a \mid \neg a \mid \Box C \mid \Diamond T\\
\textbf{D5} \quad &C ::= L \mid C \lor C\\
&T ::= L \mid T \land T
\end{aligned}$$

A summary of our analysis of the different definitions with respect to properties **P1**-**P7** is provided in the following table.

**Theorem 6.** *The results in Figure 1 hold.*

|        | D1              | D2  | D3a | D3b | D4  | D5  |
|--------|-----------------|-----|-----|-----|-----|-----|
| **P1** | yes             | yes | yes | yes | yes | yes |
| **P2** | yes             | yes | no  | no  | no  | no  |
| **P3** | no              | yes | no  | yes | yes | yes |
| **P4** | yes             | yes | yes | no  | yes | yes |
| **P5** | no              | no  | yes | yes | yes | yes |
| **P6** | yes             | yes | yes | yes | yes | yes |
| **P7** | no (unless P=NP) |     | no (unless P=Pspace) |     |     |     |

Figure 1: Properties of the different definitions of literals, clauses, and terms.

Clearly deciding between different candidate definitions is more complicated than counting up the number of properties that the definitions satisfy, the simple reason being that some properties are more important than others. Take for instance property **P5** which requires clauses and terms to be expressive enough to represent all of the formulae in $\mathcal{K}$. If we just use the standard propositional definition of clauses and terms (thereby disregarding the modal operators), then we find that it satisfies every property except **P5**, and hence more properties than any of the definitions considered in this section, and yet we would be hard-pressed to find someone who considers the propositional definition an appropriate definition for $\mathcal{K}$. This demonstrates that expressiveness is a particularly important property, so important in fact that we should be willing to sacrifice properties **P2** and **P7** to keep it. Among the definitions that satisfy **P5**, we prefer definitions **D4** and **D5** to definitions **D3a** and **D3b**, as the latter definitions have less in common with the propositional definition and present no advantages over **D4** and **D5**.

Of course, when it comes down to it, the choice of a definition must depend on the particular application in mind. There may very well be circumstances in which a less expressive or less elegant definition may prove to be the most suitable. In this paper we are using clauses and terms to define prime implicates and prime implicants, so for us the most important criteria for choosing a definition will be the quality of the notions of prime implicates and prime implicants that the definition induces.

## 4. Prime Implicates/Implicants in $\mathcal{K}$

Once a definition of clauses and terms has been fixed, we can define prime implicates and prime implicants in exactly the same manner as in propositional logic:

**Definition 7.** A clause $\lambda$ is an implicate of a formula $\varphi$ if and only if $\varphi \models \lambda$. $\lambda$ is a *prime implicate* of $\varphi$ if and only if:

1. $\lambda$ is an implicate of $\varphi$

2. If $\lambda'$ is an implicate of $\varphi$ such that $\lambda' \models \lambda$, then $\lambda \models \lambda'$

**Definition 8.** A term $\kappa$ is an implicant of the formula $\varphi$ if and only if $\kappa \models \varphi$. $\kappa$ is a *prime implicant* of $\varphi$ if and only if:

1. $\kappa$ is an implicant of $\varphi$

2. If $\kappa'$ is an implicant of $\varphi$ such that $\kappa \models \kappa'$, then $\kappa' \models \kappa$

Of course, the notion of prime implicate (resp. implicant) that we get will be determined by the definition of clause (resp. term) that we have chosen. We will compare different definitions using the following well-known properties of prime implicates/implicants in propositional logic:

**Finiteness** The number of prime implicates (resp. prime implicants) of a formula is finite modulo logical equivalence.

**Covering** Every implicate of a formula is entailed by some prime implicate of the formula. Conversely, every implicant of a formula entails some prime implicant of the formula.

**Equivalence** A model $\mathfrak{M}$ is a model of $\varphi$ if and only if $\mathfrak{M}$ is a model of all the prime implicates of $\varphi$ if and only if $\mathfrak{M}$ is a model of some prime implicant of $\varphi$[3].

**Implicant-Implicate Duality** Every prime implicant of a formula is equivalent to the negation of some prime implicate of the negated formula. Conversely, every prime implicate of a formula is equivalent to the negation of a prime implicant of the negated formula.

**Distribution** If $\lambda$ is a prime implicate of $\varphi_1 \vee ... \vee \varphi_n$, then there exist prime implicates $\lambda_1, ..., \lambda_n$ of $\varphi_1, ..., \varphi_n$ such that $\lambda \equiv \lambda_1 \vee ... \vee \lambda_n$. Likewise, if $\kappa$ is a prime implicant of $\varphi_1 \wedge ... \wedge \varphi_n$, then there exist prime implicants $\kappa_1, ..., \kappa_n$ of $\varphi_1, ..., \varphi_n$ such that $\kappa \equiv \kappa_1 \wedge ... \wedge \kappa_n$

**Finiteness** ensures that the prime implicates/implicants of a formula can be finitely represented, while **Covering** means the prime implicates provide a complete representation of the formula's implicates. **Equivalence** guarantees that no information is lost in replacing a formula by its prime implicates/implicants, whereas **Implicant-Implicate Duality** allows us to transfer results and algorithms for prime implicates to prime implicants, and vice-versa. Finally, **Distribution** relates the prime implicates/implicants of a formula to the prime implicates/implicants of its sub-formulae. This property will play a key role in the prime implicate generation algorithm presented in the next section.

We can show that definition **D4** satisfies all five properties. For **Finiteness** and **Covering**, we first demonstrate that every implicate $\lambda$ of a formula $\varphi$ is entailed by some implicate $\lambda'$ of $\varphi$ with $var(\lambda') \subseteq var(\varphi)$ and having depth at most $\delta(\varphi) + 1$ (and similarly for implicants). As there are only finitely many non-equivalent formulae on a finite language and with bounded depth, it follows that there are only finitely many prime implicates/implicants of a given formula, and that there can be no infinite chains of increasingly stronger implicates (or increasingly weaker implicants). **Equivalence** follows directly from **Covering** and the property **P5** of the previous section: we use **P5** to rewrite $\varphi$ as a conjunction of clauses, each of which is implied by some prime implicate of $\varphi$ because of **Covering**. The property **Implicant-Implicate Duality** is an immediate consequence of the duality between clauses and terms (**P4**). **Distribution** can be shown using **Covering** plus the fact that a disjunction of clauses is a clause and a conjunction of terms is a term (**P3**).

---

3. The property **Equivalence** is more commonly taken to mean that a formula is equivalent to the conjunction of its prime implicates and the disjunction of its prime implicants. We have chosen a model-theoretic formulation in order to allow for the possibility that the set of prime implicates/implicants is infinite.

**Theorem 9.** *The notions of prime implicates and prime implicants induced by definition* **D4** *satisfy* **Finiteness**, **Covering**, **Equivalence**, **Implicant-Implicate Duality**, *and* **Distribution**.

We remark by way of contrast that in first-order logic the notion of prime implicate induced by the standard definition of clauses has been shown to falsify **Finiteness**, **Covering**, and **Equivalence** (Marquis, 1991a, 1991b).

We now show that definition **D4** is the only one of our definitions to satisfy all five properties. For definitions **D1** and **D2**, we show that **Equivalence** does not hold. This is a fairly straightforward consequence of the fact that these definitions do not satisfy property **P5**.

**Theorem 10.** *The notions of prime implicates and prime implicants induced by definitions* **D1** *and* **D2** *do not satisfy* **Equivalence**.

For the notions of prime implicates induced by definitions **D3a**, **D3b**, and **D5**, we show in the appendix that the clause $\Box \Diamond^k a \lor \Diamond(a \land b \land \Box^k \neg a)$ is a prime implicate of $\Box(a \land b)$ for every $k \geq 1^4$. We thereby demonstrate not only that these definitions admit formulae with infinitely many distinct prime implicates but also that they allow seemingly irrelevant clauses to be counted as prime implicates. This gives us strong grounds for dismissing these definitions as much of the utility of prime implicates in applications comes from their ability to eliminate such irrelevant consequences.

**Theorem 11.** *The notions of prime implicates and prime implicants induced by* **D3a**, **D3b**, *and* **D5** *falsify* **Finiteness**.

While the comparison in the last section suggested that **D5** was at least as suitable as **D4** as a definition of clauses and terms, the results of this section rule out **D5** as a suitable definition for prime implicates and prime implicants. In the remainder of the paper, we will concentrate our attention on the notions of prime implicates and prime implicants induced by definition **D4**, as these have been shown to be the most satisfactory generalizations of the propositional case.

## 5. Prime Implicate Generation and Recognition

In this section, we investigate the computational aspects of modal prime implicates. As we will be primarily focusing on the notion of prime implicate induced by definition **D4**, for the remainder of the paper we will use the words "clause", "term", and "prime implicate" to mean clause, term, and prime implicate with respect to definition **D4**, except where explicitly stated otherwise.

We remark that, without loss of generality, we can restrict our attention to prime implicates since by **Implicant-Implicate Duality** (Theorem 9) any algorithm for generating or recognizing prime implicates can be easily adapted into an algorithm for generating or recognizing prime implicants.

---

4. For **D4**, the only prime implicate of $\Box(a \land b)$ is itself.

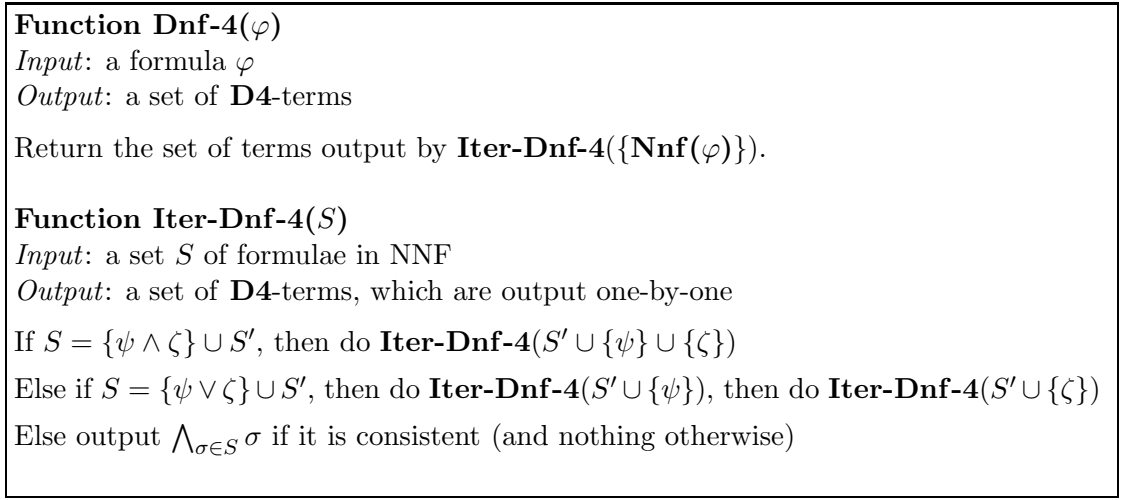

**Function Dnf-4($\varphi$)**
*Input*: a formula $\varphi$
*Output*: a set of **D4**-terms

Return the set of terms output by **Iter-Dnf-4**($\{\mathbf{Nnf}(\varphi)\}$).

**Function Iter-Dnf-4($S$)**
*Input*: a set $S$ of formulae in NNF
*Output*: a set of **D4**-terms, which are output one-by-one

If $S = \{\psi \wedge \zeta\} \cup S'$, then do **Iter-Dnf-4**($S' \cup \{\psi\} \cup \{\zeta\}$)

Else if $S = \{\psi \vee \zeta\} \cup S'$, then do **Iter-Dnf-4**($S' \cup \{\psi\}$), then do **Iter-Dnf-4**($S' \cup \{\zeta\}$)

Else output $\bigwedge_{\sigma \in S} \sigma$ if it is consistent (and nothing otherwise)

Figure 2: Helper functions **Dnf-4** and **Iter-Dnf-4**.

## 5.1 Generating Prime Implicates

We start by considering the problem of generating the set of prime implicates of a given formula. This task is important if we want to produce abductive explanations, or if we want to compile a formula into its set of prime implicates.

For our generation algorithm, we will require a means of transforming the input formula into an equivalent disjunction of "simpler" formulae. To this end, we introduce in Figure 2 the helper function **Dnf-4**($\varphi$) which returns a set of satisfiable terms with respect to **D4** whose disjunction is equivalent to $\varphi$. The function **Dnf-4** is defined in terms of another function **Iter-Dnf-4** which takes an input $S$ of formulae in NNF and returns in an iterative fashion a set of satisfiable terms whose disjunction is equivalent to $S$. The following lemmas highlight some important properties of these functions.

**Lemma 12. Iter-Dnf-4** *terminates and requires only polynomial space in the size of its input.*

**Lemma 13.** *The output of* **Dnf-4** *on input $\varphi$ is a set of satisfiable terms with respect to* **D4** *whose disjunction is equivalent to $\varphi$.*

**Lemma 14.** *There are at most $2^{|\varphi|}$ terms in* **Dnf-4**($\varphi$). *Each of the terms has length at most $2|\varphi|$, depth at most $\delta(\varphi)$, and contains only those propositional variables appearing in $var(\varphi)$.*

We present in Figure 3 the algorithm **GenPI** which computes the set of prime implicates of a given formula. Our algorithm works as follows: in Step 1, we check whether $\varphi$ is unsatisfiable, outputting a contradictory clause if this is the case. For satisfiable $\varphi$, we set $\mathcal{T}$ equal to a set of satisfiable terms whose disjunction is equivalent to $\varphi$. Because of **Distribution**, we know that every prime implicate of $\varphi$ is equivalent to some disjunction of prime implicates of the terms in $\mathcal{T}$. In Step 2, we set $\Delta(T)$ equal to the propositional literals in $T$ ($L_T$) plus the strongest $\Box$-literal implied by $T$ ($\Box\beta_T$) plus the strongest $\Diamond$-literals implied by $T$ ($\{\Diamond(\zeta \wedge \beta_T) \,|\, \zeta \in D_T\}$). It is not too hard to see that every prime implicate of $T$ must be equivalent to one of the elements in $\Delta(T)$. This means that in Step 3 we are guaranteed that every prime implicate of the input formula is equivalent to some candidate prime implicate in CANDIDATES. During the comparison phase in Step 4, non-prime candidates are eliminated, and exactly one prime implicate of each equivalence class will be retained.

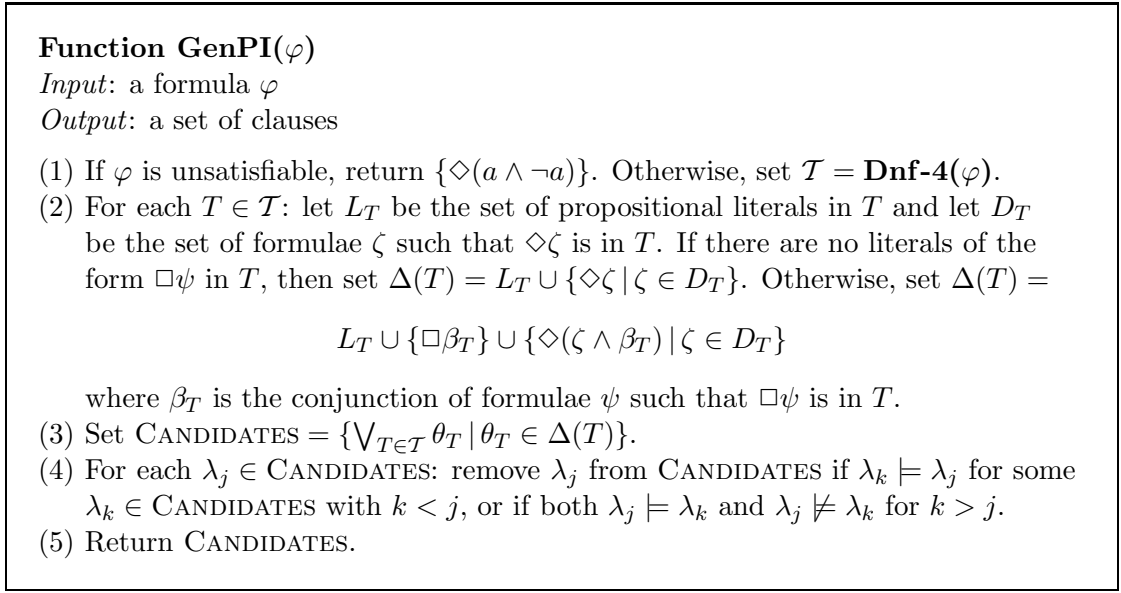

Figure 3: Algorithm for prime implicate generation.

We illustrate the behavior of **GenPI** on an example:

**Example 15.** We run **GenPI** on input $\varphi = a \wedge ((\Diamond(b \wedge c) \wedge \Diamond b) \vee (\Diamond b \wedge \Diamond(c \vee d) \wedge \Box e \wedge \Box f))$.

*Step 1:* As $\varphi$ is satisfiable, we call the function **Dnf-4** on $\varphi$, and it returns the two terms $T_1 = a \wedge \Diamond(b \wedge c) \wedge \Diamond b$ and $T_2 = a \wedge \Diamond b \wedge \Diamond(c \vee d) \wedge \Box e \wedge \Box f$.

*Step 2:* We have $L_{T_1} = \{a\}$, $D_{T_1} = \{b \wedge c, b\}$, and there are no $\Box$-literals in $T_1$, so we get $\Delta(T_1) = \{a, \Diamond(b \wedge c), \Diamond b\}$. For $T_2$, we have $L_{T_2} = \{a\}$, $D_{T_2} = \{b, c \vee d\}$, and $\beta_{T_2} = e \wedge f$, giving us $\Delta(T_2) = \{a, \Box(e \wedge f), \Diamond(b \wedge e \wedge f), \Diamond((c \vee d) \wedge e \wedge f)\}$.

*Step 3:* The set CANDIDATES will contain all the different possible disjunctions of elements in $\Delta(T_1)$ with elements in $\Delta(T_2)$, of which there are 12: $a \vee a$, $a \vee \Box(e \wedge f)$, $a \vee \Diamond(b \wedge e \wedge f)$, $a \vee \Diamond((c \vee d) \wedge e \wedge f)$, $\Diamond(b \wedge c) \vee a$, $\Diamond(b \wedge c) \vee \Box(e \wedge f)$, $\Diamond(b \wedge c) \vee \Diamond(b \wedge e \wedge f)$, $\Diamond(b \wedge c) \vee \Diamond((c \vee d) \wedge e \wedge f)$, $\Diamond b \vee a$, $\Diamond b \vee \Box(e \wedge f)$, $\Diamond b \vee \Diamond(b \wedge e \wedge f)$, and $\Diamond b \vee \Diamond((c \vee d) \wedge e \wedge f)$.

*Step 4:* We will remove from CANDIDATES the clauses $a \vee \Box(e \wedge f)$, $a \vee \Diamond(b \wedge e \wedge f)$, $a \vee \Diamond((c \vee d) \wedge e \wedge f)$, $\Diamond(b \wedge c) \vee a$, and $\Diamond b \vee a$ since they are strictly weaker than $a \vee a$. We will also eliminate the clauses $\Diamond b \vee \Box(e \wedge f)$, $\Diamond b \vee \Diamond(b \wedge e \wedge f)$, and $\diamond b \lor \diamond((c \lor d) \land e \land f)$ since they are weaker than the clauses $\diamond(b \land c) \lor \square(e \land f)$, $\diamond(b \land c) \lor \diamond(b \land e \land f)$, $\diamond(b \land c) \lor \diamond((c \lor d) \land e \land f)$.

*Step 5:* **GenPI** will return the four remaining clauses in Candidates, which are $a \lor a$, $\diamond(b \land c) \lor \square(e \land f)$, $\diamond(b \land c) \lor \diamond(b \land e \land f)$, and $\diamond(b \land c) \lor \diamond((c \lor d) \land e \land f)$.

Our algorithm can be shown to be a sound and complete procedure for generating prime implicates.

**Theorem 16.** *The algorithm* **GenPI** *always terminates and outputs exactly the set of prime implicates of the input formula.*

By examining the prime implicates produced by the algorithm, we can place an upper bound on the length of a formula's prime implicates.

**Theorem 17.** *The length of the smallest clausal representation of a prime implicate of a formula is at most single exponential in the length of the formula.*

This upper bound is optimal as we can find formulae with exponentially large prime implicates. This situation contrasts with propositional logic, where the length of prime implicates is linearly bounded by the number of propositional variables in the formula.

**Theorem 18.** *The length of the smallest clausal representation of a prime implicate of a formula can be exponential in the length of the formula.*

It is interesting to note that the formula used in the proof of Theorem 18 has a depth of 1, which means that we cannot avoid this worst-case spatial complexity by restricting our attention to formulae of shallow depth. Nor can we escape this exponential worst-case spatial complexity by dropping down to one of the less expressive notions of prime implicates examined in the previous section, as the following theorem attests.

**Theorem 19.** *If prime implicates are defined using either* **D1** *or* **D2***, then the length of the smallest clausal representation of a prime implicate of a formula can be exponential in the length of the formula.*

An examination of the set of candidate prime implicates constructed by our algorithm allows us to place a bound on the maximal number of non-equivalent prime implicates a formula can possess.

**Theorem 20.** *The number of non-equivalent prime implicates of a formula is at most double exponential in the length of the formula.*

This bound can also be shown to be optimal. This situation contrasts with propositional logic, where there can be at most single exponentially many non-equivalent prime implicates of a given formula.

**Theorem 21.** *The number of non-equivalent prime implicates of a formula may be double exponential in the length of the formula.*

Again, this worst-case result is robust in that it can be improved neither by restricting the depth of formulae, nor by using less expressive notions of prime implicate, as the following theorem demonstrates.

**Theorem 22.** *If prime implicates are defined using either **D1** or **D2**, then the number of non-equivalent prime implicates of a formula may be double exponential in the length of the formula.*

Theorems 19 and 22 together suggest that definitions **D1** or **D2** do not yield especially interesting approximate notions of prime implicate, as they induce a significant loss of expressivity without any improvement in the size or number of prime implicates in the worst-case.

Our generation algorithm **GenPI** corresponds to the simplest possible implementation of the distribution property, and it is quite clear that it does not represent a practicable way for producing prime implicates. One major source of inefficiency is the large number of clauses that are generated, so if we want to design a more efficient algorithm, we need to find ways to generate fewer candidate clauses. There are a couple of different techniques that could be used. One very simple method which could yield a smaller number of clauses is to eliminate from $\Delta(T)$ those elements which are not prime implicates of $T$, thereby decreasing the cardinalities of the $\Delta(T)$ and hence of CANDIDATES. To do this, we simply test whether $\beta_T$ is a tautology (and remove it if it is) and then compare the $\diamond$-literals in $\Delta(T)$, discarding any weaker elements. If we apply this technique to Example 15, we would remove $\diamond b$ from $\Delta(T_1)$, thereby reducing the cardinality of CANDIDATES from 12 to 8.

More substantial savings could be achieved by using a technique developed in the framework of propositional logic (cf. Marquis, 2000) which consists in calculating the prime implicates of $T_1$, then the prime implicates of $T_1 \vee T_2$, then those of $T_1 \vee T_2 \vee T_3$, and so on until we get the prime implicates of the full disjunction of terms. By interleaving comparison and construction, we can eliminate early on a partial clause that cannot give rise to prime implicates instead of producing all of the extensions of the partial clause and then deleting them one by one during the comparison phase. In our example, there were only two terms, but imagine that there was a third term $T_3$. Then by applying this technique, we would first produce the 4 prime implicates of $T_1 \vee T_2$ and then we would compare the $4|\Delta(T_3)|$ candidate clauses of $T_1 \vee T_2 \vee T_3$. Compare this with the current algorithm which generates and then compares $12|\Delta(T_3)|$ candidate clauses.

Given that the number of elements in CANDIDATES can be double exponential in the length of the input, cutting down on the size of the input to **GenPI** could yield significant savings. One obvious idea would be to break conjunctions of formulae into their conjuncts, and then calculate the prime implicates of each of the conjuncts. Unfortunately, however, we cannot apply this method to every formula as the prime implicates of the conjuncts are not necessarily prime implicates of the full conjunction. One solution which was proposed in the context of approximation of description logic concepts (cf. Brandt & Turhan, 2002) is to identify simple syntactic conditions that guarantee that we will get the same result if we break the formula into its conjuncts. For instance, one possible condition is that the conjuncts do not share any propositional variables. The formula $\varphi$ in our example satisfies this condition since the variables in $a$ and $((\diamond(b \wedge c) \wedge \diamond b) \vee (\diamond b \wedge \diamond(c \vee d) \wedge \Box e \wedge \Box f))$ are disjoint. By generating the prime implicates of the conjuncts separately, we can directly identify the prime implicate $a$, and we only have 6 candidate clauses of $((\diamond(b \wedge c) \wedge \diamond b) \vee (\diamond b \wedge \diamond(c \vee d) \wedge \Box e \wedge \Box f))$ to compare. If we also remove weaker elements from the $\Delta(T_i)$ as suggested above, we get only 3 candidate clauses for $((\Diamond(b \wedge c) \wedge \Diamond b) \vee (\Diamond b \wedge \Diamond(c \vee d) \wedge \Box e \wedge \Box f))$, all of which are prime implicates of $\varphi$.

Another important source of inefficiency in our algorithm is the comparison phase in which we compare all candidate clauses one-by-one in order to identify the strongest ones. The problem with this is of course that in the worst-case there can be a double exponential number of candidate clauses, simply because there may be double exponentially many distinct prime implicates, and each prime implicate must be equivalent to some candidate clause. Keeping all of these double exponentially many clauses in memory will generally not be feasible. Fortunately, however, it is not necessary to keep all of the candidate clauses in memory at once since we can generate them on demand from the sets $\Delta(T)$. Indeed, as we demonstrate in the appendix, by implementing our algorithm in a more clever fashion, we obtain an algorithm which outputs the prime implicates iteratively while requiring only single-exponential space (the output of the algorithm could of course be double exponentially large because of Theorem 21).

**Theorem 23.** *There exists an algorithm which runs in single-exponential space in the size of the input and incrementally outputs, without duplicates, the set of prime implicates of the input formula.*

Although our modified algorithm has a much better spatial complexity than the original, it still does not yield a practicable means for generating prime implicates. The reason is that we still need to compare each of the candidate clauses against all the other candidate clauses in order to decide whether a candidate is a prime implicate or not. Given that the set of candidate clauses may be double exponential in number, this means that our algorithm may need to perform double exponentially many entailment tests before producing even a single prime implicate. A much more promising approach would be to test directly whether or not a candidate clause is a prime implicate *without considering all of the other candidate clauses*. In order to implement such an approach, we must of course come up with a procedure for determining whether or not a given clause is a prime implicate. This will be our objective in the following section.

## 5.2 Recognizing Prime Implicates

The focus of this section is the problem of recognizing prime implicates, that is, the problem of deciding whether a clause $\lambda$ is a prime implicate of a formula $\varphi$. As has been discussed in the previous subsection, this problem is of central importance, as any algorithm for generating prime implicates must contain (implicitly or explicitly) some mechanism for ensuring that the generated clauses are indeed prime implicates.

In propositional logic, prime implicate recognition is $BH_2$-complete (Marquis, 2000), being as hard as both satisfiability and deduction. In $\mathcal{K}$, satisfiability and unsatisfiability are both PSPACE-complete, so we cannot hope to find a prime implicate recognition algorithm with a complexity of less than PSPACE.

**Theorem 24.** *Prime implicate recognition is* PSPACE-*hard.*

In order to obtain a first upper bound, we can exploit Theorem 17 which tells us that there exists a polynomial function $f$ such that every prime implicate of a formula $\varphi$ is equivalent to some clauses of length at most $2^{f(|\varphi|)}$. This leads to a simple procedure for determining if a clause $\lambda$ is a prime implicate of a formula $\varphi$. We simply check for every clause $\lambda'$ of length at most $2^{f(|\varphi|)}$ whether $\lambda'$ is an implicate of $\varphi$ which implies $\lambda$ but is not implied by $\lambda$. If this is the case, then $\lambda$ is not a prime implicate (we have found a logically stronger implicate of $\varphi$), otherwise, there exists no stronger implicate, so $\lambda$ is a prime implicate. It is not too hard to see that this algorithm can be carried out in exponential space, which gives us an Expspace upper bound.

Of course, the problem with this naive approach is that it does not at all take into account the structure of $\lambda$, so we end up comparing a huge amount of irrelevant clauses, which is exactly what we were hoping to avoid. The algorithm that we propose later in this section avoids this problem by exploiting the information in the input formula and clause in order to cut down on the number of clauses to test. The key to our algorithm is the following theorem which shows how the general problem of prime implicate recognition can be reduced to the more specialized tasks of prime implicate recognition for propositional formulae, $\Box$-formulae, and $\Diamond$-formulae. To simplify the presentation of the theorem, we let $\Pi(\varphi)$ refer to the set of prime implicates of $\varphi$, and we use the notation $\lambda \setminus \{l_1, ..., l_n\}$ to refer to the clause obtained by removing each of the literals $l_i$ from $\lambda$. For example $(a \lor b \lor \Diamond c) \setminus \{a, \Diamond c\}$ refers to the clause $b$.

**Theorem 25.** *Let $\varphi$ be a formula of $\mathcal{K}$, and let $\lambda = \gamma_1 \lor ... \lor \gamma_k \lor \Diamond \psi_1 \lor ... \lor \Diamond \psi_n \lor \Box \chi_1 \lor ... \lor \Box \chi_m$ ($\gamma_j$ propositional literals) be a non-tautologous clause such that (a) $\chi_i \equiv \chi_i \lor \psi_1 ... \lor \psi_n$ for all $i$, and (b) there is no literal $l$ in $\lambda$ such that $\lambda \equiv \lambda \setminus \{l\}$. Then $\lambda \in \Pi(\varphi)$ if and only if the following conditions hold:*

1. $\gamma_1 \lor ... \lor \gamma_k \in \Pi(\varphi \land \neg(\lambda \setminus \{\gamma_1, ..., \gamma_k\}))$

2. $\Box(\chi_i \land \neg \psi_1 \land ... \land \neg \psi_n) \in \Pi(\varphi \land \neg(\lambda \setminus \{\Box \chi_i\}))$ *for every $i$*

3. $\Diamond(\psi_1 \lor ... \lor \psi_n) \in \Pi(\varphi \land \neg(\lambda \setminus \{\Diamond \psi_1, ..., \Diamond \psi_n\}))$

We remark that the restriction to clauses for which $\chi_i \equiv \chi_i \lor \psi_1 \lor ... \lor \psi_m$ for all $i$ and for which $\lambda \not\equiv \lambda \setminus \{l\}$ for all $l$ is required. If we drop the first requirement, then there are some non-prime implicates that satisfy all three conditions, and if we drop the second, there are prime implicates which fail to satisfy one of the conditions[5]. These restrictions are without loss of generality however since every clause can be transformed into an equivalent clause satisfying them. For the first condition, we replace each $\Box \chi_i$ by $\Box(\chi_i \lor \psi_1 \lor ... \lor \psi_m)$, thereby transforming a clause $\gamma_1 \lor ... \lor \gamma_k \lor \Diamond \psi_1 \lor ... \Diamond \psi_m \lor \Box \chi_1 \lor ... \lor \Box \chi_n$ into the equivalent $\gamma_1 \lor ... \lor \gamma_k \lor \Diamond \psi_1 \lor ... \Diamond \psi_m \lor \Box(\chi_1 \lor \psi_1 \lor ... \lor \psi_m) \lor ... \lor \Box(\chi_n \lor \psi_1 \lor ... \lor \psi_m)$. Then to make the clause satisfy the second condition, we simply remove from $\lambda$ those literals for which $\lambda \equiv \lambda \setminus \{l\}$ until no such literal remains.

Theorem 25 shows how prime implicate recognition can be split into three more specialized sub-tasks, but it does not tell us how to carry out these tasks. Thus, in order to turn this theorem into an algorithm for prime implicate recognition, we need to figure out how to test whether a propositional clause, a $\Box$-formula, or a $\Diamond$-formula is a prime implicate of a formula.

---

5. For the first restriction, consider the formula $\varphi = \Diamond(a \land b \land c) \lor \Box a$ and the clause $\lambda = \Diamond(a \land b) \lor \Box(a \land \neg b)$. It can be easily shown that $\lambda$ is an implicate of $\varphi$, but $\lambda$ is not a prime implicate of $\varphi$ since there exist stronger implicates (e.g. $\varphi$ itself). Nonetheless, it can be verified that both $\Box(a \land \neg b \land \neg(a \land b)) \in \Pi(\varphi \land \neg(\lambda \setminus \{\Box(a \land \neg b)\}))$ and $\Diamond(a \land b) \in \Pi(\varphi \land \neg(\lambda \setminus \{\Diamond(a \land b)\}))$. For the second restriction, consider the formula $\Box a$ and the clause $\Box a \lor \Box(a \land b)$. We have $\Box(a \land b) \notin \Pi(\Box a \land \neg(\Box a))$ even though $\Box a \lor \Box(a \land b) \equiv \Box a$ is a prime implicate of $\Box a$.

Determining whether a propositional clause is a prime implicate of a formula in $\mathcal{K}$ is conceptually no more difficult than determining whether a propositional clause is a prime implicate of a propositional formula. We first ensure that the clause is an implicate of the formula and then make sure that all literals appearing in the clause are necessary.

**Theorem 26.** *Let $\varphi$ be a formula of $\mathcal{K}$, and let $\gamma$ be a non-tautologous propositional clause such that $\varphi \models \gamma$ and such that there is no literal $l$ in $\gamma$ such that $\gamma \equiv \gamma \setminus \{l\}$. Then $\gamma \in \Pi(\varphi)$ if and only if $\varphi \not\models \gamma \setminus \{l\}$ for all $l$ in $\gamma$.*

We now move on to the problem of deciding whether a clause of the form $\Box\chi$ is a prime implicate of a formula $\varphi$. We remark that if $\Box\chi$ is implied by $\varphi$, then it must also be implied by each of the terms $T_i \in \mathbf{Dnf\text{-}4}(\varphi)$. But if $T_i \models \Box\chi$, then by Theorem 1, it must be the case that the conjunction of the $\Box$-literals in $T_i$ implies $\Box\chi$. This means that the formula $\Box\beta_1 \lor ... \lor \Box\beta_n$ (where $\beta_i$ is the conjunction of the formulae $\zeta$ such that $\Box\zeta$ is in $T_i$) is an implicate of $\varphi$ which implies $\Box\chi$, and moreover it is the strongest such implicate. It follows then that $\Box\chi$ is a prime implicate of $\varphi$ just in the case that $\Box\chi \models \Box\beta_1 \lor ... \lor \Box\beta_n$, which is true if and only if $\chi \models \beta_i$ for some $i$ (by Theorem 1). Thus, by comparing the formula $\chi$ with the formulae $\beta_i$ associated with the terms of $\varphi$, we can decide whether or not $\Box\chi$ is a prime implicate of $\varphi$.

**Theorem 27.** *Let $\varphi$ be a formula of $\mathcal{K}$, and let $\lambda = \Box\chi$ be a non-tautologous clause such that $\varphi \models \lambda$. Then $\lambda \in \Pi(\varphi)$ if and only if there exists some term $T \in \mathbf{Dnf\text{-}4}(\varphi)$ such that $\chi \models \beta_T$, where $\beta_T$ is the conjunction of formulae $\psi$ such that $\Box\psi$ is in $T$.*

Finally let us turn to the problem of deciding whether a clause $\Diamond\psi$ is a prime implicate of a formula $\varphi$. Now we know by **Covering** that if $\Diamond\psi$ is an implicate of $\varphi$, then there must be some prime implicate $\pi$ of $\varphi$ which implies $\Diamond\psi$. It follows from Theorem 2 that $\pi$ must be a disjunction of $\Diamond$-literals, and from Theorem 16 that $\pi$ is equivalent to a disjunction $\bigvee_{T \in \mathbf{Dnf\text{-}4}(\varphi)} \Diamond d_T$ where $\Diamond d_T$ is an element of $\Delta(T)$ for every $T$ (refer back to Figure 3 for the definition of $\Delta(T)$). According to Definition 7, $\Diamond\psi$ is a prime implicate of $\varphi$ just in the case that $\Diamond\psi \models \bigvee_{T \in \mathbf{Dnf\text{-}4}(\varphi)} \Diamond d_T$, or equivalently $\psi \models \bigvee_{T \in \mathbf{Dnf\text{-}4}(\varphi)} d_T$. Thus, $\Diamond\psi$ is not a prime implicate of $\varphi$ just in the case that there is a choice of $\Diamond d_T \in \Delta(T)$ for each $T \in \mathbf{Dnf\text{-}4}(\varphi)$ such that $\bigvee_{T \in \mathbf{Dnf\text{-}4}(\varphi)} d_T \models \psi$ and $\psi \not\models \bigvee_{T \in \mathbf{Dnf\text{-}4}(\varphi)} d_T$.

Testing directly whether $\psi$ entails some formula $\bigvee_{T \in \mathbf{Dnf\text{-}4}(\varphi)} d_T$ could take exponential space in the worst case since there may be exponentially many terms in $\mathbf{Dnf\text{-}4}(\varphi)$. Luckily, however, we can get around this problem by exploiting the structure of the formula $\bigvee_{T \in \mathbf{Dnf\text{-}4}(\varphi)} d_T$. We remark that because of the way $\Delta(T)$ is defined the formula $d_T$ must be a conjunction of formulae $\zeta$ such that $\Box\zeta$ or $\Diamond\zeta$ appears in $\mathbf{Nnf}(\varphi)$ outside the scope of modal operators – we will use $\mathcal{X}$ to denote the set of formulae $\zeta$ satisfying this condition. We show in the appendix that $\psi \not\models \bigvee_{T \in \mathbf{Dnf\text{-}4}(\varphi)} d_T$ implies the existence of a subset $S \subseteq \mathcal{X}$ such that (a) $\psi \not\models \bigvee_{\lambda \in S} \lambda$ and (b) every $d_T$ has at least one conjunct from the set $S$. Conversely, the existence of such a subset of $\mathcal{X}$ implies $\psi \not\models \bigvee_{T \in \mathbf{Dnf\text{-}4}(\varphi)} d_T$. This observation is the basis for the algorithm **Test$\Diamond$PI** given in Figure 4. The basic idea behind the algorithm is to try out each of the different subsets of $\mathcal{X}$ in order to see whether some subset satisfies the aforementioned conditions. If we find a suitable subset, this proves that $\diamond\psi$ is not a prime implicate, and if no such subset exists, then we can be sure there is no stronger implicate than $\diamond\psi$. The algorithm can be shown to run in polynomial space since there can be at most $|\varphi|$ elements in $\mathcal{X}$, and we can consider the terms in **Dnf-4**$(\varphi)$ one at a time.

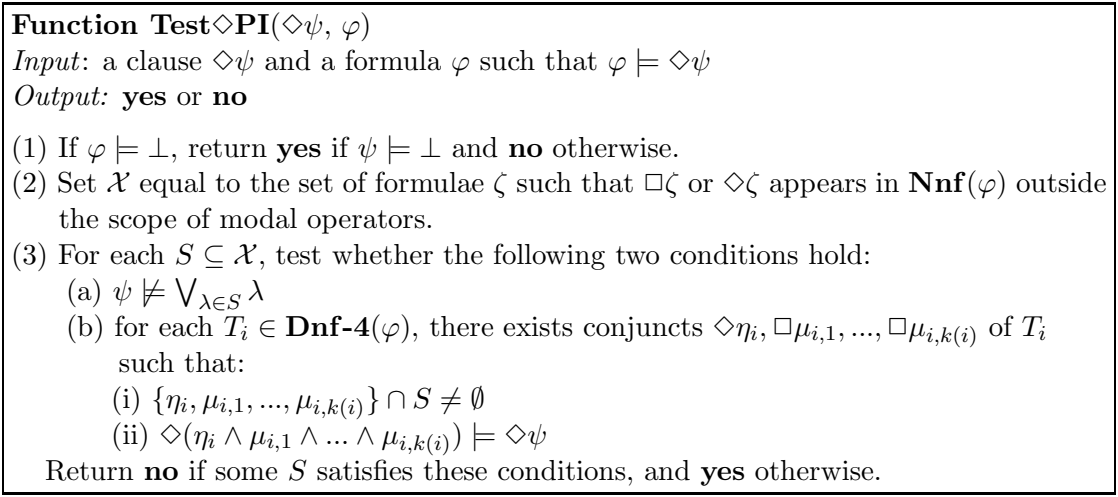

Figure 4: Algorithm for identifying prime implicates of the form $\diamond\psi$.

**Theorem 28.** *Let $\varphi$ be a formula, and let $\diamond\psi$ be an implicate of $\varphi$. Then the algorithm* **Test$\diamond$PI** *returns* **yes** *on input $(\diamond\psi,\ \varphi)$ if and only if $\diamond\psi$ is a prime implicate of $\varphi$.*

**Theorem 29.** *The algorithm* **Test$\diamond$PI** *runs in polynomial space.*

We now illustrate the algorithm **Test$\diamond$PI** with two examples.

**Example 30.** We use **Test$\diamond$PI** to test whether the clause $\lambda = \diamond(a \wedge b)$ is a prime implicate of $\varphi = a \wedge (\Box(b \wedge c) \vee \Box(e \vee f)) \wedge \diamond(a \wedge b)$.

*Step 1:* As $\varphi$ is satisfiable, we pass directly to Step 2.

*Step 2:* We set $\mathcal{X}$ equal to the set of formulae $\zeta$ such that $\Box\zeta$ or $\diamond\zeta$ appears in **Nnf**$(\varphi)$ outside the scope of modal operators. In our case, we set $\mathcal{X} = \{b \wedge c, e \vee f, a \wedge b\}$ since $\varphi =$**Nnf**$(\varphi)$ and $b \wedge c$, $e \vee f$, and $a \wedge b$ are the only formulae satisfying the requirements.

*Step 3:* We examine each of the different subsets of $\mathcal{X}$ to determine whether they satisfy conditions (a) and (b). In particular, we consider the subset $S = \{b \wedge c, e \vee f\}$. We remark that this subset satisfies condition (a) since $a \wedge b \not\models (b \wedge c) \vee (e \vee f)$. In order to check condition (b), we first call the function **Dnf-4** on $\varphi$ which returns the two terms $T_1 = a \wedge \Box(b \wedge c) \wedge \diamond(a \wedge b)$ and $T_2 = a \wedge \Box(e \vee f) \wedge \diamond(a \wedge b)$. We notice that the conjuncts $\diamond(a \wedge b)$ and $\Box(b \wedge c)$ of $T_1$ satisfy conditions (i) and (ii) since $b \wedge c \in S$ and $\diamond(a \wedge b \wedge (b \wedge c)) \models \lambda$. We then notice that the conjuncts $\diamond(a \wedge b)$ and $\Box(e \vee f)$ of $T_2$ also satisfy conditions (i) and (ii) since $e \vee f \in S$ and $\diamond(a \wedge b \wedge (e \vee f)) \models \lambda$. That

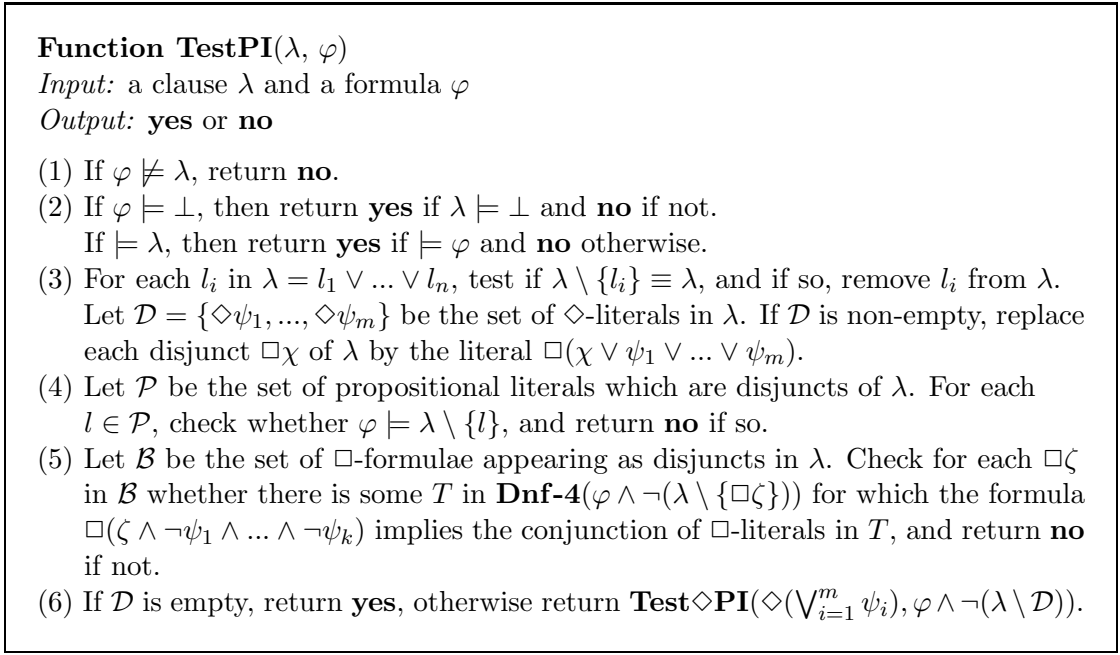

**Function TestPI**($\lambda, \varphi$)

*Input:* a clause $\lambda$ and a formula $\varphi$

*Output:* **yes** or **no**

(1) If $\varphi \not\models \lambda$, return **no**.

(2) If $\varphi \models \bot$, then return **yes** if $\lambda \models \bot$ and **no** if not.
If $\models \lambda$, then return **yes** if $\models \varphi$ and **no** otherwise.

(3) For each $l_i$ in $\lambda = l_1 \vee \dots \vee l_n$, test if $\lambda \setminus \{l_i\} \equiv \lambda$, and if so, remove $l_i$ from $\lambda$.
Let $\mathcal{D} = \{\Diamond\psi_1, \dots, \Diamond\psi_m\}$ be the set of $\Diamond$-literals in $\lambda$. If $\mathcal{D}$ is non-empty, replace each disjunct $\Box\chi$ of $\lambda$ by the literal $\Box(\chi \vee \psi_1 \vee \dots \vee \psi_m)$.

(4) Let $\mathcal{P}$ be the set of propositional literals which are disjuncts of $\lambda$. For each $l \in \mathcal{P}$, check whether $\varphi \models \lambda \setminus \{l\}$, and return **no** if so.

(5) Let $\mathcal{B}$ be the set of $\Box$-formulae appearing as disjuncts in $\lambda$. Check for each $\Box\zeta$ in $\mathcal{B}$ whether there is some $T$ in **Dnf-4**($\varphi \wedge \neg(\lambda \setminus \{\Box\zeta\})$) for which the formula $\Box(\zeta \wedge \neg\psi_1 \wedge \dots \wedge \neg\psi_k)$ implies the conjunction of $\Box$-literals in $T$, and return **no** if not.

(6) If $\mathcal{D}$ is empty, return **yes**, otherwise return **Test$\Diamond$PI**($\Diamond(\bigvee_{i=1}^{m} \psi_i), \varphi \wedge \neg(\lambda \setminus \mathcal{D})$).

Figure 5: Algorithm for identifying prime implicates.

means that we have found a subset $S$ of $\mathcal{X}$ which satisfies conditions (a) and (b), so the algorithm returns **no**. This is the correct output since $\Diamond(a \wedge b \wedge ((b \wedge c) \vee (e \vee f)))$ is an implicate of $\varphi$ which is strictly stronger than $\lambda$.

**Example 31.** We use **Test$\Diamond$PI** to test whether the clause $\lambda = \Diamond(a \wedge b \wedge c)$ is a prime implicate of $\varphi = a \wedge (\Box(b \wedge c) \vee \Box(e \vee f)) \wedge \Diamond(a \wedge b) \wedge \neg\Box(e \vee f \vee (a \wedge b \wedge c))$.

*Step 1:* We proceed directly to Step 2 since $\varphi$ is satisfiable.

*Step 2:* We set $\mathcal{X} = \{b \wedge c, e \vee f, a \wedge b, \neg e \wedge \neg f \wedge (\neg a \vee \neg b \vee \neg c)\}$ since **Nnf**($\varphi$)=$a \wedge (\Box(b \wedge c) \vee \Box(e \vee f)) \wedge \Diamond(a \wedge b) \wedge \Diamond(\neg e \wedge \neg f \wedge (\neg a \vee \neg b \vee \neg c))$.

*Step 3:* We check whether there is some subset of $\mathcal{X}$ satisfying conditions (a) and (b). We claim that there is no such subset. To see why, notice that $a \wedge \Box(b \wedge c) \wedge \Diamond(a \wedge b) \wedge \Diamond(\neg e \wedge \neg f \wedge (\neg a \vee \neg b \vee \neg c))$ is the only term in **Dnf-4**($\varphi$). Moreover, there is only one set of conjuncts of this term which implies $\Diamond(a \wedge b \wedge c)$, namely $\{\Diamond(a \wedge b), \Box(b \wedge c)\}$. But that means that $S$ must contain either $a \wedge b$ or $b \wedge c$ in order to satisfy condition (b)(i). As $a \wedge b \wedge c$ implies both $a \wedge b$ and $b \wedge c$, we are guaranteed that $a \wedge b \wedge c$ will imply the disjunction of elements in $S$, thereby falsifying condition (a). It follows that there is no subset of $\mathcal{X}$ satisfying the necessary conditions, so **Test$\Diamond$PI** returns **yes**, which is the desired result.

In Figure 5, we present our algorithm for testing whether a clause $\lambda$ is a prime implicate of a formula $\varphi$. The first two steps of the algorithm treat the limit cases where $\lambda$ is not an implicate or where one or both of $\varphi$ and $\lambda$ is a tautology or contradiction. In Step 3, we apply equivalence-preserving transformations to $\lambda$ to make it satisfy the requirements of Theorem 25. Then in Steps 4, 5, and 6 we use the procedures from Theorems 26, 27, and 28 to test whether the three conditions in Theorem 25 are verified. If the three tests succeed, then by Theorem 25, the clause is a prime implicate, so we return **yes**. If some test fails, we return **no** as the clause has been shown not to be a prime implicate.

**Theorem 32.** *The algorithm* **TestPI** *always terminates, and it returns* **yes** *on input* $(\lambda, \varphi)$ *if and only if* $\lambda$ *is a prime implicate of* $\varphi$.

We demonstrate the use of **TestPI** on an example.

**Example 33.** We use **TestPI** to test if the clauses $\lambda_1 = b$, $\lambda_2 = \Box b \vee \Box(e \vee f)$, $\lambda_3 = a \vee \Diamond c$, $\lambda_4 = \Diamond(a \wedge b)$, and $\lambda_5 = \Diamond(a \wedge b \wedge c) \vee \Diamond(a \wedge b \wedge c \wedge f) \vee \Box(e \vee f)$ are prime implicates of $\varphi = a \wedge (\Box(b \wedge c) \vee \Box(e \vee f)) \wedge \Diamond(a \wedge b)$.

$\lambda_1$: We output **no** in Step 1 since $\varphi \not\models \lambda_1$.

$\lambda_2$: We skip Steps 1 and 2 since $\lambda \models \lambda_2$ and neither $\varphi \models \bot$ nor $\models \lambda_2$. In Step 3, we make no changes to $\lambda_2$ since it contains no redundant literals nor any $\Diamond$-literals. We skip Step 4 since $\lambda_2$ has no propositional disjuncts. In Step 5, we return **no** since **Dnf-4**$(\varphi \wedge \neg(\lambda_2 \setminus \{\Box b\})) = \{a \wedge \Box(b \wedge c) \wedge \Diamond(a \wedge b) \wedge \Diamond(\neg e \wedge \neg f)\}$ and $\Box b \not\models \Box(b \wedge c)$.

$\lambda_3$: We proceed directly to Step 3 since $\lambda \models \lambda_3$, $\varphi \not\models \bot$, and $\not\models \lambda_3$. No modifications are made to $\lambda_3$ in Step 3 as it does not contain any redundant literals or $\Box$-literals. In Step 4, we test whether or not $\varphi \models \lambda_3 \setminus \{a\}$. As $\varphi \not\models \Diamond c$, we proceed on to Step 5, and then directly on to Step 6 since $\lambda_3$ contains no $\Box$-literals. In Step 6, we call **Test$\Diamond$PI**$(\Diamond c, \varphi \wedge \neg(\lambda_3 \setminus \{\Diamond c\}))$, which outputs **no** since $\varphi \wedge \neg(\lambda_3 \setminus \{\Diamond c\}) \models \bot$ and $c \not\models \bot$.

$\lambda_4$: Steps 1-5 are all inapplicable, so we skip directly to Step 6. In this step, we call **Test$\Diamond$PI** with as input the clause $\Diamond(a \wedge b)$ and the formula $\varphi \wedge \neg(\lambda_4 \setminus \{\Diamond(a \wedge b)\}) = \varphi$. We have already seen in Example 30 above that **Test$\Diamond$PI** returns **no** on this input, which means that **TestPI** also returns **no**.

$\lambda_5$: We proceed directly to Step 3, where we delete the redundant literal $\Diamond(a \wedge b \wedge c \wedge f)$ and then modify the literal $\Box(e \vee f)$. At the end of this step, we have $\lambda_5 = \Diamond(a \wedge b \wedge c) \vee \Box((e \vee f) \vee (a \wedge b \wedge c))$. Step 4 is not applicable since there are no propositional disjuncts in $\lambda_5$. In Step 5, we continue since **Dnf-4**$(\varphi \wedge \neg(\lambda_5 \setminus \{\Box((e \vee f \vee (a \wedge b \wedge c))\})) = \{a \wedge \Box(e \vee f) \wedge \Diamond(a \wedge b) \wedge \Box(\neg a \vee \neg b \vee \neg c)\}$, and $\Box(((e \vee f \vee (a \wedge b \wedge c)) \wedge (\neg a \vee \neg b \vee \neg c)) \models \Box(e \vee f) \wedge \Box(\neg a \vee \neg b \vee \neg c)$. In Step 6, we return **yes** since we call **Test$\Diamond$PI** on input $(\Diamond(a \wedge b \wedge c), \varphi \wedge \neg(\lambda_5 \setminus \{\Diamond(a \wedge b \wedge c)\}))$, and we have previously shown in Example 31 that **Test$\Diamond$PI** returns **yes** on this input.

We show in the appendix that the algorithm **TestPI** runs in polynomial space. As we have already shown that **TestPI** decides prime implicate recognition, it follows that this problem is in Pspace:

**Theorem 34.** *Prime implicate recognition is in* Pspace.

By putting together Theorems 24 and 34, we obtain a tight complexity bound for the prime implicate recognition task.

**Corollary 35.** *Prime implicate recognition is* Pspace-*complete*.

## 6. Conclusion and Future Work

The first contribution of this work is a detailed comparison of several different possible definitions of clauses, terms, prime implicates, and prime implicants for the modal logic $\mathcal{K}$. The results of this investigation were largely positive: although we have shown that no perfect definition exists, we did exhibit a very simple definition (**D4**) which satisfies most of the desirable properties of the propositional case. The second contribution of our work is a thorough investigation of the computational aspects of the selected definition **D4**. To this end, we presented a sound and complete algorithm for generating prime implicates, as well as a number of optimizations to improve the efficiency of the algorithm. An examination of the structure of the prime implicates generated by our algorithm allowed us to place upper bounds on the length of prime implicates and on the number of prime implicates a formula can possess. We showed these bounds to be optimal by exhibiting matching lower bounds, and we further proved that the lower bounds hold even for some much less expressive notions of prime implicates. Finally, we constructed a polynomial-space algorithm for deciding prime implicate recognition, thereby showing this problem to be Pspace-complete, which is the lowest complexity that could reasonably be expected. Although the focus of the paper was on the logic $\mathcal{K}$, all of our results can be easily lifted to multi-modal $\mathcal{K}$ and to concept expressions in the well-known description logic $\mathcal{ALC}$.

As was mentioned in the introduction, one of the main applications of prime implicants in propositional logic is to the area of abductive reasoning, where prime implicants play the role of abductive explanations. The results of our paper can be directly applied to the problem of abduction in $\mathcal{K}$: our notion of prime implicants can be used as a definition of abductive explanations in $\mathcal{K}$, and our prime implicate generation algorithm provides a means of producing all of the abductive explanations to a given abduction problem. Moreover, because the notion of term underlying our definition of abductive explanations is more expressive than that used by Cialdea Mayer and Pirri (1995), we are able to find explanations which are overlooked by their method. For instance, if we look for an explanation of the observation $c$ given the background information $\Box(a \lor b) \to c$, we obtain $\Box(a \lor b)$, whereas their framework yields $\Box a$ and $\Box b$. This is an argument in favor of our approach since generally in abduction one is looking to find the *weakest* conditions guaranteeing the truth of the observation given the background information.

Also of interest are our results on the size and number of prime implicates, as these yield corresponding lower bounds on the size and number of abductive explanations. In particular, our results imply that the abductive explanations of Cialdea Mayer and Pirri (1995) can have exponential size and be double exponentially many in number in the worst case, and thus behave no better in these respects than the notion of abductive explanation induced by our preferred definition **D4**. Moreover, the fact that these lower bounds hold even in the case of the extremely inexpressive notion of abductive explanations induced by definition **D2** suggests that these high worst-case complexity results really cannot be avoided. In light of these intractability results, an interesting question for future research would be to study the problem of generating a single prime implicate, since in some applications it may prove sufficient to produce a single minimal explanation for an observation. Another interesting subject for future work which is relevant from the point of view of abduction is the investigation of the notion of prime implicate over a fixed vocabulary. The development of generation algorithms for this more refined notion of prime implicate would allow one to generate only those abductive explanations which are built up from a given set of propositional variables.

The second domain of application which was mentioned in the introduction was the area of knowledge compilation. In propositional logic, one well-known target language for knowledge compilation is prime implicate normal form, in which a formula is represented as the conjunction of its prime implicates. A natural idea would be to use our selected definition of prime implicate to define in an analogous manner a notion of prime implicate normal form for $\mathcal{K}$ formulae. Unfortunately, the normal form we obtain satisfies few of the nice properties of the propositional case. For instance, we find that entailment between two formulae in prime implicate normal form is no easier than between arbitrary $\mathcal{K}$ formulae. To see why, consider any pair of formulae $\varphi$ and $\psi$ in negation normal form. The formulae $\Diamond\varphi$ and $\Diamond\psi$ are their own prime implicates and hence are in prime implicate normal form according to the naive definition. As $\varphi \models \psi$ just in the case that $\Diamond\varphi \models \Diamond\psi$, we can reduce entailment between arbitrary $\mathcal{K}$ formulae in NNF to entailment between formulae in prime implicate normal form. As the former problem is known to be Pspace-complete, it follows that the latter is Pspace-complete as well.

At first sight, this appears to be quite a disappointing result as one would hope that the computational difficulty of representing a formula by its prime implicates would be offset by some good computational properties of the resulting formula. As it turns out, however, the problem lies not in our definition of prime implicates but rather in the naive way of defining prime implicate normal form. Indeed, in a continuation of the present work (Bienvenu, 2008), we proposed a more sophisticated definition of prime implicate normal form, in which we specify which of the many different clausal representations of a prime implicate should be used. This normal form was shown to enjoy a number of desirable properties which make it interesting from the viewpoint of knowledge compilation. Most notably, it was proven that entailment between formulae in $\mathcal{K}$ in our prime implicate normal form can be carried out in polynomial time using a simple structural comparison algorithm which is reminiscent of the structural subsumption algorithms used in subpropositional description logics. It should be noted that the proof of this and other results by Bienvenu (2008) make ample use of the material presented in the current paper.

In this work, we studied prime implicates with respect to the local consequence relation, so a natural direction for future work would be the investigation of prime implicates with respect to the global consequence relation. This question is particularly interesting given that global consequence is the type of consequence used in description logic ontologies. Unfortunately, our preliminary investigations suggest that defining and generating prime implicates with respect to the global consequence relation will likely prove more difficult than for the local consequence relation. For one thing, if we use a definition of clause which is reasonably expressive, then the notion of prime implicate we obtain does not satisfy **Covering** since we can construct infinite sequences of stronger and stronger implicates. Take for instance the formula $(\neg a \lor b) \land (\neg b \lor \Diamond b)$ which implies (using the global consequence relation) each of the increasingly stronger clauses in the infinite sequence $\neg a \lor \Diamond b$, $\neg a \lor \Diamond(b \land \Diamond b)$, $\neg a \lor \Diamond(b \land \Diamond(b \land \Diamond b))$, ... This is a familiar situation for description logic practitioners since these infinite sequences are responsible for the non-existence of most specific concepts in many common DLs (cf. Küsters & Molitor, 2002) and the lack of uniform interpolation for $\mathcal{ALC}$ TBoxes (Ghilardi, Lutz, & Wolter, 2006). A standard solution to this problem is to simply place a bound on the depth of formulae to be considered, effectively blocking these problematic infinite sequences. This will not allow us to regain **Covering**, but it will give us a weaker version of this property, which should be sufficient for most applications. The development of generation algorithms for the global consequence relation may also prove challenging, since it is unclear at this point whether we will be able to draw inspiration from pre-existing methods. Despite these potential difficulties, we feel that this subject is worth exploring since it could contribute to the development of more flexible ways of accessing and structuring information in description logic ontologies.

Finally, another natural direction for future research would be to extend our investigation of prime implicates and prime implicants to other popular modal and description logics. Particularly of interest are modal logics of knowledge and belief and expressive description logics used for the semantic web. We are confident that the experience gained from our investigation of prime implicates and prime implicants in $\mathcal{K}$ will prove a valuable asset in the exploration of other modal and description logics.

## Acknowledgments

This paper corrects and significantly extends an earlier conference publication (Bienvenu, 2007). This paper was written while the author was a PhD student working at IRIT, Université Paul Sabatier, France. The author would like to thank her thesis supervisors Andreas Herzig, Jérôme Lang, and Jérôme Mengin, as well as the anonymous reviewers for very helpful feedback.

## Appendix A. Proofs

**Theorem 1** *Let $\psi$, $\psi_1$, ..., $\psi_m$, $\chi$, $\chi_1$, ..., $\chi_n$ be formulae in $\mathcal{K}$, and let $\gamma$ be a propositional formula. Then*

1. *$\psi \models \chi \Leftrightarrow \models \neg\psi \lor \chi \Leftrightarrow \psi \land \neg\chi \models \bot$*

2. *$\psi \models \chi \Leftrightarrow \Diamond\psi \models \Diamond\chi \Leftrightarrow \Box\psi \models \Box\chi$*

3. *$\gamma \land \Diamond\psi_1 \land ... \land \Diamond\psi_m \land \Box\chi_1 \land ... \land \Box\chi_n \models \bot \Leftrightarrow (\gamma \models \bot \text{ or } \psi_i \land \chi_1 \land ... \land \chi_n \models \bot \text{ for some } i)$*

4. *$\models \gamma \lor \Diamond\psi_1 \lor ... \lor \Diamond\psi_m \lor \Box\chi_1 \lor ... \lor \Box\chi_n \Leftrightarrow (\models \gamma \text{ or } \models \psi_1 \lor ... \lor \psi_m \lor \chi_i \text{ for some } i)$*

5. *$\Box\chi \models \Box\chi_1 \lor ... \lor \Box\chi_n \Leftrightarrow \chi \models \chi_i \text{ for some } i$*

6. $\Diamond\psi_1 \vee ... \vee \Diamond\psi_m \vee \Box\chi_1 \vee ... \vee \Box\chi_n$
   $\equiv \Diamond\psi_1 \vee ... \vee \Diamond\psi_m \vee \Box(\chi_1 \vee \psi_1 \vee ... \vee \psi_m) \vee ... \vee \Box(\chi_n \vee \psi_1 \vee ... \vee \psi_m)$

*Proof.* The first statement is a well-known property of local consequence, but we prove it here for completeness:

$$
\begin{aligned}
\psi \models \chi \quad &\Leftrightarrow \quad \mathfrak{M}, w \models \psi \text{ implies } \mathfrak{M}, w \models \chi \text{ for all } \mathfrak{M}, w \\
&\Leftrightarrow \quad \mathfrak{M}, w \not\models \psi \text{ or } \mathfrak{M}, w \models \chi \text{ for all } \mathfrak{M}, w \\
&\Leftrightarrow \quad \mathfrak{M}, w \models \neg\psi \text{ or } \mathfrak{M}, w \models \chi \text{ for all } \mathfrak{M}, w \\
&\Leftrightarrow \quad \models \neg\psi \vee \chi \\
&\Leftrightarrow \quad \mathfrak{M}, w \not\models \psi \wedge \neg\chi \text{ for all } \mathfrak{M}, w \\
&\Leftrightarrow \quad \psi \wedge \neg\chi \models \bot
\end{aligned}
$$

For the second statement, if $\psi \not\models \chi$, then there is some $\mathfrak{M}, w$ such that $\mathfrak{M}, w \models \psi \wedge \neg\chi$. Create a new model $\mathfrak{M}'$ from $\mathfrak{M}$ by adding a new world $w'$ and placing a single arc from $w'$ to $w$. Then $\mathfrak{M}', w' \models \Diamond\psi \wedge \Box\neg\chi$, which means that $\Diamond\psi \wedge \Box\neg\chi$ is satisfiable and hence $\Diamond\psi \not\models \Diamond\chi$ (since $\neg\Box\neg\chi \equiv \Diamond\chi$). For the other direction, suppose $\Diamond\psi \not\models \Diamond\chi$. Then there exists $\mathfrak{M}, w$ such that $\mathfrak{M}, w \models \Diamond\psi \wedge \neg\Diamond\chi \equiv \Diamond\psi \wedge \Box\neg\chi$. But this means that there is some $w'$ for which $\psi \wedge \neg\chi$, hence $\psi \not\models \chi$. To complete the proof, we use the following chain of equivalences: $\Box\psi \models \Box\chi \Leftrightarrow \neg\Box\chi \models \neg\Box\psi \Leftrightarrow \Diamond\neg\chi \models \Diamond\neg\psi \Leftrightarrow \neg\psi \models \neg\chi \Leftrightarrow \psi \models \chi$.

For 3, suppose that $\gamma \wedge \Diamond\psi_1 \wedge ... \wedge \Diamond\psi_m \wedge \Box\chi_1 \wedge ... \wedge \Box\chi_n \not\models \bot$. Then there exist $\mathfrak{M}, w$ such that $\mathfrak{M}, w \models \gamma \wedge \Diamond\psi_1 \wedge ...\Diamond\psi_m \wedge \Box\chi_1 \wedge ... \wedge \Box\chi_n$. As $\mathfrak{M}, w \models \gamma$, we cannot have $\gamma \models \bot$, nor can we have $\psi_i \wedge \chi_1 \wedge ... \wedge \chi_n \models \bot$ since for each $i$ there is some $w'$ such that $\mathfrak{M}, w' \models \psi_i \wedge \chi_1 \wedge ... \wedge \chi_n$. Now for the other direction suppose that $\gamma$ and all of the $\psi_i \wedge \chi_1 \wedge ... \wedge \chi_n$ are satisfiable. Then there is some propositional model $w$ of $\gamma$, and for each $i$, we can find $\mathfrak{M}_i, w_i$ such that $\mathfrak{M}_i, w_i \models \psi_i \wedge \chi_1 \wedge ... \wedge \chi_n$. Now we construct a new Kripke structure which contains the models $\mathfrak{M}_i$ and the world $w$ and in which there are arcs going from $w$ to each of the $w_i$. It is not hard to see that in this new model $\mathfrak{M}_{new}$ we have $\mathfrak{M}_{new}, w \models \gamma \wedge \Diamond\psi_1 \wedge ...\Diamond\psi_m \wedge \Box\chi_1 \wedge ...\Box\chi_n$, so $\gamma \wedge \Diamond\psi_1 \wedge ...\Diamond\psi_m \wedge \Box\chi_1 \wedge ... \wedge \Box\chi_n \not\models \bot$.

Statement 4 follows easily from the third statement. We simply notice that $\gamma \vee \Diamond\psi_1 \vee ... \vee \Diamond\psi_m \vee \Box\chi_1 \vee ... \vee \Box\chi_n$ is a tautology just in the case that its negation $\neg\gamma \wedge \Diamond\neg\chi_1 \wedge ... \wedge \Diamond\neg\chi_n \wedge \Box\neg\psi_1 \wedge ... \wedge \Box\neg\psi_m$ is unsatisfiable.

For 5, we use statements 1 and 4 to get the following chain of equivalences:

$$
\begin{aligned}
\Box\chi &\models \Box\chi_1 \vee ... \vee \Box\chi_n \\
\Leftrightarrow \quad &\models \Diamond\neg\chi \vee \Box\chi_1 \vee ... \vee \Box\chi_n \\
\Leftrightarrow \quad &\models \neg\chi \vee \chi_i \text{ for some } i \\
\Leftrightarrow \quad &\chi \models \chi_i \text{ for some } i
\end{aligned}
$$

The first implication of the equivalence in 6 is immediate since $\Diamond\psi_1 \vee ... \vee \Diamond\psi_m \models \Diamond\psi_1 \vee ... \vee \Diamond\psi_m$ and $\Box\chi_i \models \Box(\chi_i \vee \psi_1 \vee ... \vee \psi_m)$ for all $i$. For the other direction, we remark that by using statements 1 and 3, we get the following equivalences:

$$
\begin{aligned}
&\Box(\chi_i \vee \psi_1 \vee ... \vee \psi_m) \models \Box\chi_i \vee \Diamond\psi_1 \vee ... \vee \Diamond\psi_m \\
\Leftrightarrow \quad &\Box(\chi_i \vee \psi_1 \vee ... \vee \psi_m) \wedge \neg(\Box\chi_i \vee \Diamond\psi_1 \vee ... \vee \Diamond\psi_m) \models \bot \\
\Leftrightarrow \quad &\Box(\chi_i \vee \psi_1 \vee ... \vee \psi_m) \wedge \Diamond\neg\chi_i \wedge \Box\neg\psi_1 \wedge ... \wedge \Box\neg\psi_m \models \bot \\
\Leftrightarrow \quad &(\chi_i \vee \psi_1 \vee ... \vee \psi_m) \wedge \neg\chi_i \wedge \neg\psi_1 \wedge ... \wedge \neg\psi_m \models \bot
\end{aligned}
$$

As $(\chi_i \vee \psi_1 \vee ... \vee \psi_m) \wedge \neg\chi_i \wedge \neg\psi_1 \wedge ... \wedge \neg\psi_m$ is clearly unsatisfiable, it follows that $\Box(\chi_i \vee \psi_1 \vee ... \vee \psi_m) \models \Box\chi_i \vee \Diamond\psi_1 \vee ... \vee \Diamond\psi_m$ for every $i$ and hence that $\Diamond\psi_1 \vee ... \vee \Diamond\psi_m \vee \Box(\chi_1 \vee \psi_1 \vee ... \vee \psi_m) \vee ... \vee \Box(\chi_n \vee \psi_1 \vee ... \vee \psi_m) \models \Diamond\psi_1 \vee ... \vee \Diamond\psi_m \vee \Box\chi_1 \vee ... \vee \Box\chi_n$, completing the proof. $\qquad\blacksquare$

**Theorem 2** *Let $\lambda$ be a disjunction of propositional literals and $\Box$- and $\Diamond$-formulae. Then each of the following statements holds:*

1. *If $\lambda \models \gamma$ for some non-tautological propositional clause $\gamma$, then every disjunct of $\lambda$ is either a propositional literal or an unsatisfiable $\Diamond$-formula*

2. *If $\lambda \models \Diamond\psi_1 \vee ... \vee \Diamond\psi_n$, then every disjunct of $\lambda$ is a $\Diamond$-formula*

3. *If $\lambda \models \Box\chi_1 \vee ... \vee \Box\chi_n$ and $\not\models \Box\chi_1 \vee ... \vee \Box\chi_n$, then every disjunct of $\lambda$ is either a $\Box$-formula or an unsatisfiable $\Diamond$-formula*

*Proof.* For (1), let $\gamma$ be a non-tautologous propositional clause such that $\lambda \models \gamma$, and suppose for a contradiction that $\lambda$ contains a disjunct $\Box\chi$ or a disjunct $\Diamond\psi$ where $\psi \not\models \bot$. In the first case, we have $\Box\chi \models \gamma$, and hence $\models \Diamond\neg\chi \vee \gamma$. It follows from Theorem 1 that $\models \gamma$, contradicting our assumption that $\gamma$ is not a tautology. In the second case, we have $\Diamond\psi \models \gamma$, and thus $\models \Box\neg\psi \vee \gamma$. By Theorem 1, either $\models \neg\psi$ or $\models \gamma$. In both cases, we reach a contradiction since we have assumed that $\psi \not\models \bot$ and $\not\models \gamma$. It follows then that $\lambda$ cannot have any $\Box$-formulae or satisfiable $\Diamond$-formulae as disjuncts.

The proofs of (2) and (3) proceed similarly. $\qquad\blacksquare$

**Theorem 3** *Let $\lambda = \gamma \vee \Diamond\psi_1 \vee ... \vee \Diamond\psi_m \vee \Box\chi_1 \vee ... \vee \Box\chi_n$ and $\lambda' = \gamma' \vee \Diamond\psi'_1 \vee ... \vee \Diamond\psi'_p \vee \Box\chi'_1 \vee ... \vee \Box\chi'_q$ be formulae in $\mathcal{K}$. If $\gamma$ are $\gamma'$ are both propositional and $\not\models \lambda'$, then*

$$\lambda \models \lambda' \Leftrightarrow \begin{cases} \gamma \models \gamma' \text{ and} \\ \psi_1 \vee ... \vee \psi_m \models \psi'_1 \vee ... \vee \psi'_p \text{ and} \\ \text{for every } \chi_i \text{ there is some } \chi'_j \text{ such that } \chi_i \models \psi'_1 \vee ... \vee \psi'_p \vee \chi'_j \end{cases}$$

*Proof.* Since we have $\not\models \lambda'$, we know that $\not\models \gamma'$ and $\not\models \psi'_1 \vee ... \vee \psi'_p \vee \chi'_i$ for all $i$. Using this information together with Theorem 1, we get the following equivalences:

$$\begin{aligned}
\gamma \models \lambda' \quad &\Leftrightarrow \quad \models \neg\gamma \vee \gamma' \vee \Diamond\psi'_1 \vee ... \vee \Diamond\psi'_p \vee \Box\chi'_1 \vee ... \vee \Box\chi'_m \\
&\Leftrightarrow \quad \models \neg\gamma \vee \gamma' \\
&\Leftrightarrow \quad \gamma \models \gamma' \\
\Diamond\psi_1 \vee ... \vee \Diamond\psi_m \models \lambda' \quad &\Leftrightarrow \quad \Diamond(\psi_1 \vee ... \vee \psi_m) \models \lambda' \\
&\Leftrightarrow \quad \models \gamma' \vee \Diamond\psi'_1 \vee ... \vee \Diamond\psi'_p \vee \Box\neg(\psi_1 \vee ... \vee \psi_m) \vee \Box\chi'_1 \vee ... \vee \Box\chi'_q \\
&\Leftrightarrow \quad \models \psi'_1 \vee ... \vee \psi'_p \vee \neg(\psi_1 \vee ... \vee \psi_m) \\
&\Leftrightarrow \quad \psi_1 \vee ... \vee \psi_m \models \psi'_1 \vee ... \vee \psi'_p \\
\Box\chi_i \models \lambda' \quad &\Leftrightarrow \quad \models \gamma' \vee \Diamond(\psi'_1 \vee ... \vee \psi'_p \vee \neg\chi_i) \vee \Box\chi'_1 \vee ... \vee \Box\chi'_q \\
&\Leftrightarrow \quad \text{there is some } j \text{ such that } \models \psi'_1 \vee ... \vee \psi'_p \vee \neg\chi_i \vee \chi'_j \\
&\Leftrightarrow \quad \text{there is some } j \text{ such that } \chi_i \models \psi'_1 \vee ... \vee \psi'_p \vee \chi'_j
\end{aligned}$$

To complete the proof, we use the fact $\lambda \models \lambda'$ if and only if $\gamma \models \lambda'$, $\Diamond\psi_1 \vee ... \vee \Diamond\psi_m \models \lambda'$, and $\Box\chi_i \models \lambda'$ for every $i$. $\qquad\blacksquare$

**Theorem 5** *Any definition of literals, clause, and terms for $\mathcal{K}$ that satisfies properties $P1$ and $P2$ cannot satisfy $P5$.*

*Proof.* We remark that the set of clauses (resp. terms) with respect to definition $D1$ is precisely the set of formulae in NNF which do not contain $\wedge$ (resp. $\vee$), i.e. $D1$ is the most expressive definition satisfying both $P1$ and $P2$. Thus, to show the result, it suffices to show that $D1$ does not satisfy $P5$.

Suppose for a contradiction that $D1$ does satisfy $P5$. Then there must exist clauses $\lambda_1, ..., \lambda_n$ such that $\Diamond(a \wedge b) \equiv \lambda_1 \wedge ... \wedge \lambda_n$. Each of the clauses $\lambda_i$ is a disjunction $l_{i,1} \vee .... \vee l_{i,p_i}$. By distributing $\wedge$ over $\vee$, we obtain the following:

$$\Diamond(a \wedge b) \equiv \bigvee_{(j_1,...,j_n) \in \{1,...,p_1\} \times ... \times \{1,...,p_n\}} \bigwedge_{i=1}^{n} l_{i,j_i}$$

from which we can infer that for each $(j_1, ..., j_n) \in \{1, ..., p_1\} \times ... \times \{1, ..., p_n\}$ we have

$$\bigwedge_{i=1}^{n} l_{i,j_i} \models \Diamond(a \wedge b)$$

Consider some $(j_1, ..., j_n)$ such that $\bigwedge_{i=1}^{n} l_{i,j_i}$ is consistent (there must be at least one such tuple, otherwise we would have $\Diamond(a \wedge b) \equiv \bot$). The formulae $l_{i,j_i}$ are either propositional literals or formulae of the form $\Box\kappa$ or $\Diamond\kappa$ for some clause $\kappa$. It follows that $\bigwedge_{i=1}^{n} l_{i,j_i}$ must have the following form:

$$\gamma_1 \wedge ... \wedge \gamma_k \wedge \Diamond\psi_1 \wedge ... \wedge \Diamond\psi_m \wedge \Box\chi_1 \wedge ... \wedge \Box\chi_n$$

where $\gamma_1, ..., \gamma_k$ are propositional literals and $\psi_1, ..., \psi_m, \chi_1, ..., \chi_n$ are clauses with respect to $D1$. As we know that $\bigwedge_{i=1}^{n} l_{i,j_i} \models \Diamond(a \wedge b)$ and $\bigwedge_{i=1}^{n} l_{i,j_i} \not\models \bot$, by Theorem 1, there must be some $\Diamond\psi_q$ such that

$$\Diamond\psi_q \wedge \Box\chi_1 \wedge ... \wedge \Box\chi_n \models \Diamond(a \wedge b)$$

We now show that $\Diamond\psi_q \not\models \Diamond(a \wedge b)$ (and hence that $\not\models \chi_1 \wedge ... \wedge \chi_n$). Suppose for a contradiction that this is not the case. Then we must have $\psi_q \models a$ and $\psi_q \models b$. But by Theorem 1, every disjunct of $\psi_q$ (which we recall is a $D1$-clause) must either be unsatisfiable or equal to both $a$ and $b$. As the latter is impossible, it follows that $\psi_q \models \bot$, which is a contradiction since we assumed that $\bigwedge_{i=1}^{n} l_{i,j_i}$ is satisfiable. It follows then that in order to get $\Diamond\psi_q \wedge \Box\chi_1 \wedge ... \wedge \Box\chi_n \models \Diamond(a \wedge b)$, there must be some $\chi_r$ which is not a tautology.

Now let us consider the formula

$$\tau = \bigvee_{\{(j_1,...,j_n)| \bigwedge_{i=1}^{n} l_{i,j_i} \not\equiv \bot\}} \Box\chi_{j_1,...,j_n}$$

where $\Box\chi_{j_1,...,j_n}$ is a non-tautological $\Box$-formula appearing in $\bigwedge_{i=1}^{n} l_{i,j_i}$ (we have just shown that such a formula must exist). Clearly it must be the case that

$$\bigvee_{(j_1,...,j_n) \in \{1,...,p_1\} \times ... \times \{1,...,p_n\}} \bigwedge_{i=1}^{n} l_{i,j_i} \models \tau$$

from which we get:

$$\Diamond(a \wedge b) \models \tau$$

But according to Theorem 2, a satisfiable $\Diamond$-formula cannot imply a disjunction of $\Box$-formulae unless that disjunction is a tautology, so we must have $\models \tau$. However, this is impossible since it would imply (Theorem 1) that there is some $\chi_{j_1,...,j_n}$ which is a tautology, contradicting our earlier assumption to the contrary. We can thus conclude that there is no set of clauses $\lambda_1, ..., \lambda_n$ with respect to **D1** such that $\Diamond(a \wedge b) \equiv \lambda_1 \wedge ... \wedge \lambda_n$, and hence that any definition which satisfies **P1** and **P2** cannot satisfy **P5**. □

In order to prove Theorem 6, we will make use of the following lemmas:

**Lemma 6.1** *Definition **D5** satisfies **P5**.*

*Proof.* We demonstrate that any formula in $\mathcal{K}$ in NNF is equivalent to a conjunction of clauses with respect to definition **D5**. The restriction to formulae in NNF is without loss of generality as every formula is equivalent to a formula in NNF. The proof proceeds by induction on the structural complexity of formulae. The base case is propositional literals, which are already conjunctions of clauses since every propositional literal is a clause with respect to **D5**. We now suppose that the statement holds for formulae $\psi_1$ and $\psi_2$ and show that it holds for more complex formulae.

We first consider $\varphi = \psi_1 \wedge \psi_2$. By assumption, we can find clauses $\rho_i$ and $\zeta_j$ such that $\psi_1 \equiv \rho_1 \wedge ... \wedge \rho_n$ and $\psi_2 \equiv \zeta_1 \wedge ... \wedge \zeta_m$. Thus, $\varphi$ is equivalent to the formula $\rho_1 \wedge ... \wedge \rho_n \wedge \zeta_1 \wedge ... \wedge \zeta_m$, which is a conjunction of clauses with respect to **D5**.

Next we consider $\varphi = \psi_1 \vee \psi_2$. By the induction hypothesis, we have $\psi_1 \equiv \rho_1 \wedge ... \wedge \rho_n$ and $\psi_2 \equiv \zeta_1 \wedge ... \wedge \zeta_m$ for some clauses $\rho_i$ and $\zeta_j$. Thus, $\varphi \equiv (\rho_1 \wedge ... \wedge \rho_n) \vee (\zeta_1 \wedge ... \wedge \zeta_m)$, which can be written equivalently as $\varphi \equiv \wedge_{(i,j) \in \{1,...,n\} \times \{1,...,m\}} (\rho_i \vee \zeta_j)$. Since the union of two clauses produces another clause, all of the $\rho_i \vee \zeta_j$ are clauses, completing the proof.

We now consider the case where $\varphi = \Box \psi_1$. By assumption, $\psi_1 \equiv \rho_1 \wedge ... \wedge \rho_n$, where each $\rho_i$ is a clause. So $\varphi \equiv \Box(\rho_1 \wedge ... \wedge \rho_n)$. But we also know that $\Box(\rho_1 \wedge ... \wedge \rho_n) \equiv \Box\rho_1 \wedge ... \wedge \Box\rho_n$. It follows that $\varphi$ is equivalent to $\Box\rho_1 \wedge ... \wedge \Box\rho_n$, which is a conjunction of clauses since the $\Box\rho_i$ are all clauses.

Finally, we consider $\varphi = \Diamond \psi_1$. Using the induction hypothesis, we have $\varphi \equiv \Diamond(\rho_1 \wedge ... \wedge \rho_n)$ for clauses $\rho_i$. But since the $\rho_i$ are clauses, each $\rho_i$ is a disjunction of literals $l_{i,1} \vee ... \vee l_{i,p_i}$. After distributing $\wedge$ over $\vee$ and $\vee$ over $\Diamond$, we find that $\varphi$ is equivalent to the formula

$$\bigvee_{(j_1,...,j_n) \in \{1,...,p_1\} \times ... \times \{1,...,p_n\}} \Diamond(l_{1,j_1} \wedge l_{2,j_2} \wedge ... \wedge l_{n,j_n})$$

which is a clause with respect to **D5**.

The proof that every formula is equivalent to a disjunction of terms with respect to **D5** proceeds analogously. □

**Lemma 6.2** *Every clause (resp. term) with respect to **D5** is a clause (resp. term) with respect to definitions **D3a**, **D3b**, and **D4**.*

*Proof.* We will show by induction on the structural complexity of formulae that:

1. every clause $C$ with respect to **D5** is a clause with respect to definitions **D3a**, **D3b**, and **D4** *and* a disjunction of terms with respect to **D3a**

2. every term $T$ with respect to **D5** is a term with respect to definitions **D3a**, **D3b**, and **D4** *and* a conjunction of clauses with respect to **D3a** and **D3b**

We require this stronger formulation of the statement to prove some of the sub-cases.

The base case for our induction is propositional literals, which are both clauses and terms with respect to **D5**. It is easy to see that (1) and (2) are verified since propositional literals are both clauses and terms with respect to definitions **D3a**, **D3b**, and **D4** (and hence they are also disjunctions of terms with respect to **D3a** and conjunctions of clauses with respect to **D3a** and **D3b**).

For the induction step, we will show that the above statements hold for arbitrary clauses or terms with respect to **D5** under the assumption that the statments hold for all of their proper sub-clauses and sub-terms.

We begin with clauses. Let $C$ be a **D5**-clause such that all proper sub-clauses and sub-terms of $C$ satisfy (1) and (2). Now since $C$ is a clause with respect to **D5**, it can either be a propositional literal or a formula of the form $C_1 \vee C_2$ for clauses $C_1$ and $C_2$, $\Box C_1$ for some clause $C_1$, or $\Diamond T_1$ for some term $T_1$. The case where $C$ is a propositional literal has already been treated in the base case. Let us thus consider the case where $C = C_1 \vee C_2$. The first part of (1) holds since by the induction hypothesis both $C_1$ and $C_2$ are clauses with respect to definitions **D3a**, **D3b**, and **D4**, and for all three definitions the disjunction of two clauses is a clause. The second half of (1) is also verified since both $C_1$ and $C_2$ are disjunctions of terms with respect to **D3a**, and thus so is their disjunction $C_1 \vee C_2$. We next consider the case where $C = \Box C_1$ for some clause $C_1$ with respect to **D5**. The first part of (1) follows easily as we know that $C_1$ must also be a clause with respect to **D3a**, **D3b**, and **D4**, and for all of these definitions putting a $\Box$ before a clause yields another clause. The second part of (1) holds as well since $C_1$ is a disjunction of terms with respect to **D3a** and thus $\Box C_1$ is a term with respect to this same definition. We now suppose that $C = \Diamond T_1$ for some term $T_1$ with respect to **D5**. For definitions **D3a** and **D3b**, we know from the induction hypothesis that $T_1$ is a conjunction of clauses with respect to **D3a** and **D3b** and hence that $\Diamond T_1$ is a clause with respect to these definitions. For **D4**, the result obviously holds since we are allowed to put any formula in NNF behind $\Diamond$. The second part of (1) holds since by the induction hypothesis $T_1$ is a term with respect to **D3a** and hence $\Diamond T_1$ is also a term with respect to this definition.

We next consider terms. Let $T$ be a **D5**-term such that all proper sub-clauses and sub-terms of $T$ satisfy (1) and (2). Then $T$ must be either a propositional literal or a formula of the form $T_1 \wedge T_2$ for terms $T_1$ and $T_2$, $\Box C_1$ for some clause $C_1$, or $\Diamond T_1$ for some term $T_1$. If $T = T_1 \wedge T_2$, the first half of (2) holds since we know $T_1$ and $T_2$ to be terms with respect to **D3a**, **D3b**, and **D4**, and conjunctions of terms are also terms for all three definitions. The second half is also verified since both $T_1$ and $T_2$ are assumed to be conjunctions of clauses with respect to **D3a** and **D3b**, which means that $T$ is also a conjunction of clauses with respect to these definitions. Next suppose that $T = \Box C_1$. For definitions **D3b** and **D4**, it is easy to see that $T$ is a literal and hence a term. For **D3a**, the induction hypothesis tells us that $C_1$ is a disjunction of terms, from which we can deduce that $\Box C_1$ is a term. Moreover, since $C_1$ is known to be a clause with respect to **D3a** and **D3b**, then $\Box C_1$ must also be a clause with respect to these definitions, so $T$ is a conjunction of clauses with respect to both **D3a** and **D3b**. Finally, we treat the case where $T = \Diamond T_1$. For **D3a**, we use the fact that $T_1$ is a term with respect to **D3a**, which means that $\Diamond T_1$ must also be a term. For **D3b**, we use the supposition that $T_1$ is a conjunction of clauses with respect to **D3b**, from which we get that $\Diamond T_1$ is a literal and hence a term. The first part of (2) clearly also holds for **D4** since any formula behind $\Diamond$ yields a literal and thus a term. The second half of (2) follows from the fact that by the induction hypothesis $T_1$ is a conjunction of clauses with respect to **D3a** and **D3b**, so $\Diamond T_1$ is a clause (and hence a conjunction of clauses) with respect to these definitions. $\qquad\square$

$$\varphi_{\mathcal{U},\mathcal{S}} = \varphi_{1,1} \wedge ... \wedge \varphi_{1,m} \wedge \psi$$

where the $\varphi_{i,j}$ are defined inductively as follows

$$\varphi_{i,j} = \begin{cases} \Diamond \varphi_{i+1,j}, \text{if either } i \leq n, u_i \in S_j, \text{ or } i > n \text{ and } u_{i-n} \in S_j \\ \Box \varphi_{i+1,j}, \text{if either } i \leq n, u_i \notin S_j, \text{ or } i > n \text{ and } u_{i-n} \notin S_j \end{cases}$$

for $i \in \{1, ..., 2n\}$ and $\varphi_{2n+1,j} = \top$, and $\psi = \underbrace{\Box...\Box}_{2n} \bot$.

Figure 6: The formula $\varphi_{\mathcal{U},\mathcal{S}}$ which codes an instance $\mathcal{U} = \{u_1, ..., u_n\}$, $\mathcal{S} = \{S_1, ..., S_m\}$ of the exact cover problem.

**Lemma 6.3** *Entailment between terms or clauses is* NP-*complete for both definitions* **D1** *and* **D2**.

*Proof.* In the proofs of both NP-membership and NP-hardness, we will exploit the relationship between terms with respect to definitions **D1** and **D2** and concept expressions in the description logic $\mathcal{ALE}$ (cf. Baader, McGuiness, Nardi, & Patel-Schneider, 2003). We recall that concept expressions in this logic are constructed as follows (we use a modal logic syntax and assume a single modal operator in order to facilitate comparison between the formalisms):

$$\varphi ::= \top \mid \bot \mid a \mid \neg a \mid \varphi \wedge \varphi \mid \Box \varphi \mid \Diamond \varphi$$

The semantics of the symbols $\top$ and $\bot$ is as one would expect: $\mathfrak{M}, w \models \top$ and $\mathfrak{M}, w \not\models \bot$ for every model $\mathfrak{M}$ and world $w$. The semantics of atomic literals, conjunctions, and universal and existential modalities is exactly the same as for $\mathcal{K}$.

It is not hard to see that every term with respect to **D1** or **D2** is a concept expression in $\mathcal{ALE}$. As entailment between $\mathcal{ALE}$ expressions is decidable in nondeterministic polynomial time (cf. Donini, Lenzerini, Nardi, Hollunder, Nutt, & Marchetti Spaccamela, 1992), it follows that deciding entailment between terms with respect to either **D1** or **D2** can also be accomplished in nondeterministic polynomial time, i.e. these problems belong to NP.

It remains to be shown that these problems are NP-hard. To prove this, we show how the polynomial-time reduction of Donini (2003) (adapted from the original NP-hardness proof by Donini et al., 1992) of the NP-complete exact cover (XC) problem (Garey &

Johnson, 1979) to unsatisfiability in $\mathcal{ALE}$ can be modified so as to give a polynomial-time reduction from XC to entailment between terms with respect to **D1** or **D2**.

The exact cover problem is the following: given a set $\mathcal{U} = \{u_1, ..., u_n\}$ and a set $\mathcal{S} = \{S_1, ..., S_m\}$ of subsets of $\mathcal{U}$, determine whether there exists an exact cover, that is, a subset $\{S_{i_1}, ..., S_{i_q}\}$ of $\mathcal{S}$ such that $S_{i_h} \cap S_{i_k} = \emptyset$ for $h \neq k$ and $\bigcup_{k=1}^{q} S_{i_k} = \mathcal{U}$. Donini has proven (2003) that $\mathcal{U}, \mathcal{S}$ has an exact cover if and only if the formula $\varphi_{\mathcal{U}, \mathcal{S}}$ pictured in Figure 6 is unsatisfiable. Notice that $\varphi_{\mathcal{U}, \mathcal{S}}$ is not a term with respect to either **D1** and **D2** as it uses the symbols $\top$ and $\bot$. We would like to find a similar formula which is a term with respect to our definitions and which is satisfiable if and only if $\varphi_{\mathcal{U}, \mathcal{S}}$ is. Consider the formula

$$\varphi'_{\mathcal{U}, \mathcal{S}} = \varphi'_{1,1} \wedge ... \wedge \varphi'_{1,m} \wedge \psi'$$

where $\varphi'_{i,j}$ and $\psi'$ are defined exactly like $\varphi_{i,j}$ and $\psi$ except that we replace $\top$ by $a$ and $\bot$ by $\neg a$. It is easy to verify that $\varphi'_{\mathcal{U}, \mathcal{S}}$ is indeed a term with respect to both **D1** and **D2**. Moreover, it is not too hard to see that $\varphi_{1,1} \wedge ... \wedge \varphi_{1,m} \models \diamond^{2n}\top$ if and only if $\varphi'_{1,1} \wedge ... \wedge \varphi'_{1,m} \models \diamond^{2n}a$ and hence that $\varphi_{\mathcal{U}, \mathcal{S}}$ and $\varphi'_{\mathcal{U}, \mathcal{S}}$ are equisatisfiable. As $\mathcal{U}, \mathcal{S}$ has an exact cover if and only if $\varphi_{\mathcal{U}, \mathcal{S}}$ is unsatisfiable, and $\varphi_{\mathcal{U}, \mathcal{S}}$ is unsatisfiable just in the case that $\varphi'_{\mathcal{U}, \mathcal{S}}$ is, it follows that $\mathcal{U}, \mathcal{S}$ has an exact cover if and only if $\varphi'_{\mathcal{U}, \mathcal{S}}$ is unsatisfiable. Moreover, $\varphi'_{\mathcal{U}, \mathcal{S}}$ can be produced in linear time from $\varphi_{\mathcal{U}, \mathcal{S}}$, so we have a polynomial-time reduction from XC to unsatisfiability of terms in **D1** or **D2**. But a formula is unsatisfiable just in the case that it entails the term $a \wedge \neg a$. So, XC can be polynomially-reduced to entailment between terms with respect to either **D1** or **D2**, making these problems NP-hard and hence NP-complete.

In order to show the NP-completeness of clausal entailment, we remark that for both definitions **D1** and **D2**, the function **Nnf** transforms negations of clauses into terms and negations of terms into clauses. This means that we can test whether a clause $\lambda$ entails a clause $\lambda'$ by testing whether the term $\mathbf{Nnf}(\neg\lambda')$ entails the term $\mathbf{Nnf}(\neg\lambda)$. Likewise, we can test whether a term $\kappa$ entails another term $\kappa'$ by testing whether the clause $\mathbf{Nnf}(\neg\kappa')$ entails the clause $\mathbf{Nnf}(\neg\kappa)$. As the NNF transformation is polynomial, it follows that entailment between clauses is exactly as difficult as entailment between terms, so clausal entailment is NP-complete. □

**Lemma 6.4** *For definition* **D5**, *entailment between clauses or terms is* Pspace-*complete.*

*Proof.* Membership in Pspace is immediate since entailment between arbitrary formulae in $\mathcal{K}$ can be decided in polynomial space. To prove Pspace-hardness, we adapt an existing proof of Pspace-hardness of $\mathcal{K}$.

Figure 7 presents an encoding of a QBF $\beta = Q_1 p_1 ... Q_m p_m \theta$ in a $\mathcal{K}$-formula $f(\beta)$ that is used in section 6.7 of (Blackburn et al., 2001) to demonstrate the Pspace-hardness of $\mathcal{K}$. The formula $f(\beta)$ has the property that it is satisfiable just in the case that $\beta$ is a QBF-validity. As the formula $f(\beta)$ can be generated in polynomial-time from $\beta$, and the QBF-validity problem is known to be Pspace-hard, it follows that satisfiability of formulae in $\mathcal{K}$ is Pspace-hard as well.

In Figure 8, we show a modified encoding. We claim the following:

(1) $f(\beta)$ and $f'(\beta)$ are logically equivalent

$\boxed{\begin{array}{l}
\text{(i) } q_0 \\[4pt]
\text{(ii) } \bigwedge_{i=0}^{m}((q_i \rightarrow \wedge_{j \neq i} \neg q_j) \wedge \square(q_i \rightarrow \wedge_{j \neq i} \neg q_j) \wedge ... \wedge \square^m(q_i \rightarrow \wedge_{j \neq i} \neg q_j)) \\[4pt]
\text{(iiia) } \bigwedge_{i=0}^{m}((q_i \rightarrow \lozenge q_{i+1}) \wedge \square(q_i \rightarrow \lozenge q_{i+1}) \wedge ... \wedge \square^m(q_i \rightarrow \lozenge q_{i+1})) \\[4pt]
\text{(iiib) } \bigwedge_{\{i | Q_i = \forall\}} \square^i(q_i \rightarrow (\lozenge(q_{i+1} \wedge p_{i+1}) \wedge \lozenge(q_{i+1} \wedge \neg p_{i+1}))) \\[4pt]
\text{(iv) } \bigwedge_{i=1}^{m-1}(\bigwedge_{j=i}^{m-1} \square^j((p_i \rightarrow \square p_i) \wedge (\neg p_i \rightarrow \square \neg p_i))) \\[4pt]
\text{(v) } \square^m(q_m \rightarrow \theta)
\end{array}}$

Figure 7: The formula $f(\beta)$ is the conjunction of the above formulae.

$\boxed{\begin{array}{l}
\text{(i') } q_0 \\[4pt]
\text{(ii') } \bigwedge_{i=0}^{m}(\bigwedge_{j \neq i}((\neg q_i \vee \neg q_j) \wedge \square(\neg q_i \vee \neg q_j) \wedge ... \wedge \square^m(\neg q_i \vee \neg q_j))) \\[4pt]
\text{(iiia') } \bigwedge_{i=0}^{m}((\neg q_i \vee \lozenge q_{i+1}) \wedge \square(\neg q_i \vee \lozenge q_{i+1}) \wedge ... \wedge \square^m(\neg q_i \vee \lozenge q_{i+1})) \\[4pt]
\text{(iiib') } \bigwedge_{\{i | Q_i = \forall\}} \square^i(\neg q_i \vee \lozenge(q_{i+1} \wedge p_{i+1})) \wedge \square^i(\neg q_i \vee \lozenge(q_{i+1} \wedge \neg p_{i+1})) \\[4pt]
\text{(iv') } \bigwedge_{i=1}^{m-1}(\bigwedge_{j=i}^{m-1}(\square^j(\neg p_i \vee \square p_i) \wedge \square^j(p_i \vee \square \neg p_i))) \\[4pt]
\text{(v') } \square^m(\neg q_m \vee \theta_1) \wedge .... \wedge \square^m(\neg q_m \vee \theta_l)
\end{array}}$

Figure 8: The formula $f'(\beta)$ is the conjunction of the above formulae, where the formulae $\theta_i$ in (v') are propositional clauses such that $\theta \equiv \theta_1 \wedge ... \wedge \theta_l$.

(2) if $\theta$ is in CNF, then $f'(\beta)$ is a conjunction of clauses with respect to **D5**

(3) if $\theta$ is in CNF, then $f'(\beta)$ can be generated in polynomial time from $f(\beta)$

To show (1), it suffices to show that (i)≡(i'), (ii)≡(ii'), (iiia)≡(iiia'), (iiib)≡(iiib'), (iv)≡(iv'), and (v)≡(v'). The first equivalence is immediate since (i) and (i') are identical. (ii)≡(ii') follows from the fact that $\square^k(q_i \rightarrow \wedge_{j \neq i} \neg q_j) \equiv \wedge_{j \neq i} \square^k(\neg q_i \vee \neg q_j)$. (iiia)≡(iiia') holds since (iiia') is just (iiia) with $q_i \rightarrow \lozenge q_{i+1}$ replaced with $\neg q_i \vee \lozenge q_{i+1}$. We have (iiib)≡(iiib') since $\square^i(q_i \rightarrow (\lozenge(q_{i+1} \wedge p_{i+1}) \wedge \lozenge(q_{i+1} \wedge \neg p_{i+1}))) \equiv \square^i(\neg q_i \vee \lozenge(q_{i+1} \wedge p_{i+1})) \wedge \square^i(\neg q_i \vee \lozenge(q_{i+1} \wedge \neg p_{i+1}))$. The equivalence (iv)≡(iv') holds as $\square^j((p_i \rightarrow \square p_i) \wedge (\neg p_i \rightarrow \square \neg p_i)) \equiv \square^j(\neg p_i \vee \square p_i) \wedge \square^j(p_i \vee \square \neg p_i)$. Finally, we have (v)≡(v') since $\theta \equiv \theta_1 \wedge ... \wedge \theta_l$. Thus, $f(\beta)$ and $f'(\beta)$ are logically equivalent.

To prove (2), we show that each of the component formulae in $f'(\beta)$ is a conjunction of clauses with respect to **D5**, provided that $\theta$ is in CNF. Clearly this is the case for (i') as (i') is a propositional literal. The formula (ii') is also a conjunction of clauses with respect to **D5** since it is a conjunction formulae of the form $\square^k(\neg q_i \vee \neg q_j)$. Similarly, (iiia'), (iiib'), and (iv') are all conjunctions of clauses since the formulae $\square^k(\neg q_i \vee \lozenge q_{i+1})$, $\square^i(\neg q_i \vee \lozenge(q_{i+1} \wedge p_{i+1}))$, $\square^i(\neg q_i \vee \lozenge(q_{i+1} \wedge \neg p_{i+1}))$, $\square^k(\neg p_i \vee \square p_i)$, and $\square^k(p_i \vee \square \neg p_i)$ are all clauses with respect to **D5**. The formula (v') must also be a conjunction of clauses since the $\theta_i$ are assumed to be propositional clauses, making each $\square^m(\neg q_m \vee \theta_i)$ a clause with respect to **D5**, and (v') a conjunction of clauses with respect to **D5**.

For (3), it is clear that we can transform (i), (iiia), (iiib), and (iv) into (i'), (iiia'), (iiib'), and (iv') in polynomial time as the transformations involve only simple syntactic operations and the resulting formulae are at most twice as large. The transformation from (ii) to (ii') is very slightly more involved, but it is not too hard to see the resulting formula is at most $m$ times as large as the original (and $m$ can be no greater than the length of $f(\beta)$). The only step which could potentially result in an exponential blow-up is the transformation from (v) to (v'), as we put $\theta$ into CNF. But under the assumption that $\theta$ is already in CNF, the transformation can be executed in polynomial time and space, as all we have to do is separate $\theta$ into its conjuncts and rewrite the $(q_m \rightarrow \theta_i)$ as $(\neg q_m \vee \theta_i)$.

Now let $\beta = Q_1 p_1 ... Q_m p_m \theta$ be a QBF such that $\theta = \theta_1 \wedge ... \wedge \theta_l$ for some propositional clauses $\theta_i$. Let $f'(\beta)$ be the formula as defined in Figure 8. By (2) above, we know that $f'(\beta) = \lambda_1 \wedge ... \wedge \lambda_p$ for some clauses $\lambda_i$ with respect to **D5**. Now consider the following formula

$$\zeta = \Diamond(\Box \lambda_1 \wedge ... \wedge \Box \lambda_p \wedge \Diamond \Box(a \vee \neg a))$$

We can show that $f'(\beta)$ is satisfiable if and only if $\zeta$ is satisfiable as follows:

$$\zeta \text{ is unsatisfiable}$$
$$\Leftrightarrow \quad \Box \lambda_1 \wedge ... \wedge \Box \lambda_p \wedge \Diamond \Box(a \vee \neg a) \text{ is unsatisfiable}$$
$$\Leftrightarrow \quad \lambda_1 \wedge ... \wedge \lambda_p \wedge \Box(a \vee \neg a) \text{ is unsatisfiable}$$
$$\Leftrightarrow \quad \lambda_1 \wedge ... \wedge \lambda_p \text{ is unsatisfiable}$$
$$\Leftrightarrow \quad f'(\beta) \text{ is unsatisfiable}$$

But we also know from (1) above that $f'(\beta) \equiv f(\beta)$, and from (Blackburn et al., 2001) that $f(\beta)$ is satisfiable just in the case that $\beta$ is a QBF validity. It is also easy to see that $\zeta$ is satisfiable if and only if $\zeta$ does not entail the contradiction $\Diamond(a \wedge \neg a)$. Putting this altogether, we find that $\beta$ is valid just in the case that $\zeta$ does not entail $\Diamond(a \wedge \neg a)$. As $\zeta$ and $\Diamond(a \wedge \neg a)$ are both clauses and terms with respect to **D5**, we have shown that the QBF-validity problem for QBF with propositional formulae in CNF can be reduced to the problems of entailment of clauses or terms with respect to **D5**. Moreover, this is a polynomial time reduction since it follows from (3) that the transformation from $\beta$ to $\zeta$ can be accomplished in polynomial time. This suffices to show PSPACE-hardness, since it is well-known that QBF-validity remains PSPACE-hard even when we restrict the propositional part $\theta$ to be a formula in CNF (cf. Papadimitriou, 1994). ∎

**Theorem 6** *The results in Figure 1 hold.*

*Proof.* The satisfaction or dissatisfaction of properties **P1** and **P2** can be immediately determined by inspection of the definitions, as can the satisfaction of **P3** by definitions **D2**, **D3b**, **D4**, and **D5**. Counterexamples to **P3** for definitions **D1** and **D3a** were provided in body of the paper: the formula $\Box(a \vee b)$ is a clause but not a disjunction of literals with respect to both definitions.

In order to show that definition **D3b** does not satisfy **P4**, we remark that the negation of the literal $\Diamond(a \vee b)$ is equivalent to $\Box(\neg a \wedge \neg b)$ which cannot be expressed as a literal in **D3b**. For each of the other definitions, it can be shown (by a straightforward inductive proof) that **Nnf**$(\neg L)$ is a literal whenever $L$ is a literal, that **Nnf**$(\neg C)$ is a term whenever $C$ is a clause, and that $\mathbf{Nnf}(\neg T)$ is a clause whenever $T$ is a term. This is enough to prove that these definitions satisfy $\mathbf{P4}$ since $\mathbf{Nnf}(\varphi)$ is equivalent to $\varphi$.

Since we know that definitions $\mathbf{D1}$ and $\mathbf{D2}$ satisfy both properties $\mathbf{P1}$ and $\mathbf{P2}$, it follows by Theorem 5 that these definitions do not satisfy $\mathbf{P5}$. We have seen in Lemma 6.1 that definition $\mathbf{D5}$ does satisfy $\mathbf{P5}$, i.e. that every formula is equivalent to some conjunction of clauses with respect to $\mathbf{D5}$ and some disjunction of terms with respect to $\mathbf{D5}$. As every clause (resp. term) of $\mathbf{D5}$ is also a clause (resp. term) with respect to definitions $\mathbf{D3a}$, $\mathbf{D3b}$, and $\mathbf{D4}$ (by Lemma 6.2), it follows that every formula is equivalent to some conjunction of clauses and some disjunction of terms with respect to these definitions, which means they all satisfy $\mathbf{P5}$.

It is easy to see that property $\mathbf{P6}$ is satisfied by all of the definitions since all of our definitions are context-free grammars, and it is well-known that deciding membership for context-free grammars can be accomplished in polynomial time (cf. Younger, 1967).

From Lemma 6.3, we know that deciding entailment between clauses or terms with respect to either $\mathbf{D1}$ or $\mathbf{D2}$ is NP-complete (and hence not in P, unless P=NP). Entailment between clauses/terms is PSPACE-complete for $\mathbf{D5}$ (Lemma 6.4). As every clause (resp. term) of $\mathbf{D5}$ is also a clause (resp. term) with respect to definitions $\mathbf{D3a}$, $\mathbf{D3b}$, and $\mathbf{D4}$ (from Lemma 6.2), it follows that entailment between clauses or terms is PSPACE-hard for these definitions. Membership in PSPACE is immediate since entailment between arbitrary $\mathcal{K}$ formulae is in PSPACE. $\qquad\blacksquare$

We prove Theorem 9 in several steps:

**Lemma 9.1** *The notions of prime implicates and prime implicants induced by $\boldsymbol{D4}$ satisfy* ***Implicant-Implicate Duality***.

*Proof.* Suppose for a contradiction that we have a prime implicant $\kappa$ of some formula $\varphi$ which is not equivalent to the negation of a prime implicate of $\neg\varphi$. Let $\lambda$ be a clause which is equivalent to $\neg\kappa$ (there must exist such a clause because of property $\mathbf{P4}$, cf. Theorem 6). The clause $\lambda$ is an implicate of $\neg\varphi$ since $\kappa \models \varphi$ and $\lambda \equiv \neg\kappa$. As we have assumed that $\lambda$ is not a prime implicate, there must be some implicate $\lambda'$ of $\neg\varphi$ such that $\lambda' \models \lambda$ and $\lambda \not\models \lambda'$. But then let $\kappa'$ be a term equivalent to $\neg\lambda'$ (here again we use $\mathbf{P4}$). Now $\kappa'$ must be an implicant of $\varphi$ since $\neg\varphi \models \neg\kappa'$. Moreover, $\kappa'$ is strictly weaker than $\kappa$ since $\lambda' \models \lambda$ and $\lambda \not\models \lambda'$ and $\kappa \equiv \neg\lambda$ and $\kappa' \equiv \neg\lambda'$. But this means that $\kappa$ cannot be a prime implicant, contradicting our earlier assumption. Hence, we can conclude that every prime implicant of a formula $\varphi$ is equivalent to the negation of some prime implicate of $\neg\varphi$. The proof that every prime implicate of a formula $\varphi$ is equivalent to the negation of a prime implicant of $\neg\varphi$ proceeds analogously. $\qquad\blacksquare$

**Lemma 9.2** *If clauses and terms are defined according to definition $\boldsymbol{D4}$, then every implicate $\lambda$ of a formula $\varphi$ is entailed by some implicate $\lambda'$ of $\varphi$ with $var(\lambda') \subseteq var(\varphi)$ and with depth at most $\delta(\varphi) + 1$, and every implicant $\kappa$ of $\varphi$ entails an implicant $\kappa'$ of $\varphi$ with $var(\kappa') \subseteq var(\varphi)$ and depth at most $\delta(\varphi) + 1$.*

*Proof.* We intend to show that the following statement holds: for any formula $\varphi$ and any implicate $\lambda$ of $\varphi$, there exists a clause $\lambda'$ such that $\varphi \models \lambda' \models \lambda$ and $var(\lambda') \subseteq var(\varphi)$ and $\delta(\lambda) \leq \delta(\varphi) + 1$. So let $\varphi$ be an arbitrary formula, and let $\lambda$ be some implicate of $\varphi$. If $\varphi$ is a tautology, then we can set $\lambda' = a \vee \neg a$ (where $a \in var(\varphi)$). If $\lambda \equiv \bot$, then we can set $\lambda' = \Diamond(a \wedge \neg a)$ (where $a \in var(\varphi)$), as this clause verifies all of the necessary conditions. Now we consider the case where neither $\varphi$ nor $\lambda$ is a tautology or a falsehood, and we show how to construct the clause $\lambda'$. The first thing we do is use **Dnf-4** to rewrite $\varphi$ as a disjunction of satisfiable terms $T_i$ with respect to **D4** such that the $T_i$ contain only the variables appearing in $\varphi$ and have depth at most $\delta(\varphi)$:

$$\varphi = T_1 \vee ... \vee T_z$$

As $\varphi \models \lambda$, it must be the case that $T_i \models \lambda$ for every $T_i$. Our aim is to find a clause $\lambda_i$ for each of the terms $T_i$ such that $T_i \models \lambda_i \models \lambda$ and $var(\lambda_i) \subseteq var(T_i)$ and $\delta(\lambda_i) \leq \delta(T_i)$. So consider some $T_i$. Since $T_i$ is a term, it has the form $\gamma_1 \wedge ... \wedge \gamma_k \wedge \Diamond\psi_1 \wedge ... \wedge \Diamond\psi_m \wedge \Box\chi_1 \wedge ... \wedge \Box\chi_n$, where $\gamma_1, ..., \gamma_k$ are propositional literals. As $\lambda$ is a clause, it must be of the form $\rho_1 \vee ... \vee \rho_p \vee \Diamond\epsilon_1 \vee ... \vee \Diamond\epsilon_q \vee \Box\zeta_1 \vee ... \vee \Box\zeta_r$, where $\rho_1, ..., \rho_p$ are propositional literals. As $T_i \models \lambda$, it must be the case that the formula

$$\gamma_1 \wedge ... \wedge \gamma_k \wedge \Diamond\psi_1 \wedge ... \wedge \Diamond\psi_m \wedge \Box\chi_1 \wedge ... \wedge \Box\chi_n \wedge$$
$$\neg\rho_1 \wedge ... \wedge \neg\rho_p \wedge \Box\neg\epsilon_1 \wedge ... \wedge \Box\neg\epsilon_q \wedge \Diamond\neg\zeta_1 \wedge ... \wedge \Diamond\neg\zeta_r$$

is unsatisfiable. By Theorem 1, one of the following must hold:

(a) there exists $\gamma_u$ and $\rho_v$ such that $\gamma_u \equiv \rho_v$

(b) there exists $\psi_u$ such that $\psi_u \wedge \chi_1 \wedge ... \wedge \chi_n \wedge \neg\epsilon_1 \wedge ... \wedge \neg\epsilon_q \models \bot$

(c) there exists $\zeta_u$ such that $\neg\zeta_u \wedge \chi_1 \wedge ... \wedge \chi_n \wedge \neg\epsilon_1 \wedge ... \wedge \neg\epsilon_q \models \bot$

Now if (a) holds, we can set $\lambda_i = \gamma_u$ since $T_i \models \gamma_u \models \lambda$, $\delta(\gamma_u) = 0 \leq \delta(T_i)$, and $var(\gamma_u) \subseteq var(T_i)$. If it is (b) that holds, then it must be the case that

$$\psi_u \wedge \chi_1 \wedge ... \wedge \chi_n \models \epsilon_1 \vee ... \vee \epsilon_q$$

and hence that

$$\Diamond(\psi_u \wedge \chi_1 \wedge ... \wedge \chi_n) \models \Diamond\epsilon_1 \vee ... \vee \Diamond\epsilon_q \models \lambda$$

We can set $\lambda_i = \Diamond(\psi_u \wedge \chi_1 \wedge ... \wedge \chi_n)$, since $T_i \models \Diamond(\psi_u \wedge \chi_1 \wedge ... \wedge \chi_n) \models \lambda$, $\delta(\Diamond(\psi_u \wedge \chi_1 \wedge ... \wedge \chi_n)) \leq \delta(T_i)$, and $var(\Diamond(\psi_u \wedge \chi_1 \wedge ... \wedge \chi_n)) \subseteq var(T_i)$. Finally, if (c) holds, then it must be the case that

$$\chi_1 \wedge ... \wedge \chi_n \models \epsilon_1 \vee ... \vee \epsilon_q \vee \zeta_u$$

and hence that

$$\Box(\chi_1 \wedge ... \wedge \chi_n) \models \Diamond\epsilon_1 \vee ... \vee \Diamond\epsilon_q \vee \Box\zeta_u \models \lambda$$

So we can set $\lambda_i = \Box(\chi_1 \wedge ... \wedge \chi_n)$, as $T_i \models \Box(\chi_1 \wedge ... \wedge \chi_n) \models \lambda$, $\delta(\Box(\chi_1 \wedge ... \wedge \chi_n)) \leq \delta(T_i)$, and $var(\Box(\chi_1 \wedge ... \wedge \chi_n)) \subseteq var(T_i)$. Thus, we have shown that for every $T_i$, there is some $\lambda_i$ such that $T_i \models \lambda_i \models \lambda$ and $var(\lambda_i) \subseteq var(T_i)$ and $\delta(\lambda_i) \leq \delta(T_i)$. But then $\lambda_1 \vee ... \vee \lambda_z$ is a clause implied by every $T_i$, and hence by $\varphi$, and such that $var(\lambda_i) \subseteq \cup_i var(T_i) \subseteq var(\varphi)$ and $\delta(\lambda_i) \leq max_i \delta(T_i) \leq \delta(\varphi)$.

Now let $\kappa$ be an implicant of $\varphi$, and let $\lambda$ be the formula **Nnf**$(\neg\kappa)$. We know that the NNF transformation is equivalence-preserving, hence $\lambda \equiv \neg\kappa$, and it is straightforward to show that $\lambda$ must be a clause with respect to **D4**. But then $\lambda$ is an implicate of $\neg\varphi$, so there must be some clause $\lambda'$ with $var(\lambda') \subseteq var(\neg\varphi) = var(\varphi)$ and depth at most $\delta(\neg\varphi) + 1 = \delta(\varphi) + 1$ such that $\neg\varphi \models \lambda' \models \lambda$. Let $\kappa'$ be **Nnf**$(\neg\lambda')$. It can be easily verified that $\kappa'$ is a term. Moreover, by properties of the NNF transformation, we have $\kappa' \equiv \neg\lambda'$, $var(\kappa') = var(\neg\lambda') = var(\lambda')$, and $\delta(\kappa') = \delta(\neg\lambda') = \delta(\lambda')$. But then $\kappa'$ is a term such that $var(\kappa') \subseteq var(\varphi)$, $\delta(\kappa') \leq \delta(\varphi) + 1$, and $\kappa \models \kappa' \models \varphi$. $\qquad\square$

**Lemma 9.3** *The notions of prime implicates and prime implicants induced by* ***D4*** *satisfy* ***Finiteness***.

*Proof.* Consider an arbitrary formula $\varphi$. From Lemma 9.2, we know that for each prime implicate $\lambda$ of $\varphi$, there must be an implicate $\lambda'$ of $\varphi$ containing only those propositional atoms appearing in $\varphi$ and such that $\delta(\lambda') \leq \delta(\varphi) + 1$ and $\lambda' \models \lambda$. But since $\lambda$ is a prime implicate, we must also have $\lambda \models \lambda'$ and hence $\lambda \equiv \lambda'$. Thus, every prime implicate of $\varphi$ is equivalent to some clause built from the finite set of propositional symbols in $\varphi$ and having depth at most $\delta(\varphi) + 1$. As there are only finitely many non-equivalent formulae on a finite alphabet and with fixed depth, it follows that there can be only finitely many distinct prime implicates. By Lemma 9.1, every prime implicant of $\varphi$ is equivalent to the negation of some prime implicate of $\neg\varphi$. It follows then that every formula can only have finitely many distinct prime implicants. $\qquad\square$

**Lemma 9.4** *The notions of prime implicates and prime implicants induced by* ***D4*** *satisfy* ***Covering***.

*Proof.* Let $\varphi$ be an arbitrary formula. From Lemma 9.2, we know that every implicate of $\varphi$ is entailed by some implicate of $\varphi$ whose propositional variables are contained in $var(\varphi)$ and whose depth is at most $\delta(\varphi) + 1$. Now consider the following set

$$\Sigma = \{\sigma \mid \varphi \models \sigma, \sigma \text{ is a clause}, var(\sigma) \subseteq var(\varphi), \delta(\sigma) \leq \delta(\varphi) + 1\}$$

and define another set $\Pi$ from $\Sigma$ as follows:

$$\Pi = \{\sigma \in \Sigma \mid \not\exists \sigma' \in \Sigma. \ \sigma' \models \sigma \text{ and } \sigma \not\models \sigma'\}$$

In other words, $\Pi$ is the set of all of the logically strongest implicates of $\varphi$ having depth at most $\delta(\varphi) + 1$ and built from the propositional letters in $\varphi$. We claim the following:

(1) every $\pi \in \Pi$ is a prime implicate of $\varphi$

(2) for every implicate $\lambda$ of $\varphi$, there is some $\pi \in \Pi$ such that $\pi \models \lambda$

We begin by proving (1). Suppose that (1) does not hold, that is, that there is some $\pi \in \Pi$ which is not a prime implicate of $\varphi$. Since $\pi$ is by definition an implicate of $\varphi$, it follows that there must be some implicate $\lambda$ of $\varphi$ such that $\lambda \models \pi$ and $\pi \not\models \lambda$. But by Lemma 9.2, there is some implicate $\lambda'$ of $\varphi$ such that $\delta(\lambda') \leq \delta(\varphi) + 1$, $var(\lambda') \subseteq var(\varphi)$, and $\lambda' \models \lambda$. But that means that $\lambda'$ is an element of $\Sigma$ which implies but is not implied by $\pi$, contradicting the assumption that $\pi$ is in $\Pi$. We can thus conclude that every element of $\Pi$ must be a prime implicate of $\varphi$.

For (2): let $\lambda$ be some implicate of $\varphi$. Then by Lemma 9.2, there exists some clause $\lambda' \in \Sigma$ such that $\lambda' \models \lambda$. If $\lambda' \in \Pi$, we are done. Otherwise, there must exist some $\sigma \in \Sigma$ such that $\sigma \models \lambda'$ and $\lambda' \not\models \sigma$. If $\sigma \in \Pi$, we are done, otherwise, we find another stronger member of $\Sigma$. But as $\Sigma$ has finitely many elements modulo equivalence, after a finite number of steps, we will find some element which is in $\Pi$ and which implies $\lambda$. Since we have just seen that all members of $\Pi$ are prime implicates of $\varphi$, it follows that every implicate of $\varphi$ is implied by some prime implicate of $\varphi$.

For the second part of **Covering**, let $\kappa$ be an implicant of $\varphi$, and let $\lambda$ be a clause equivalent to $\neg\kappa$ (there must be one because **D4** satisfies **P4**). Now since $\kappa \models \varphi$, we must also have $\neg\varphi \models \lambda$. According to what we have just shown, there must be some prime implicate $\pi$ of $\neg\varphi$ such that $\neg\varphi \models \pi \models \lambda$. By Lemma 9.1, $\pi$ must be equivalent to the negation of some prime implicant $\rho$ of $\varphi$. But since $\rho \equiv \neg\pi$ and $\pi \models \lambda$ and $\lambda \equiv \neg\kappa$, it follows that $\kappa \models \rho$, completing the proof. $\qquad\square$

**Lemma 9.5** *The notions of prime implicates and prime implicants induced by **D4** satisfy* ***Equivalence***.

*Proof.* Let $\varphi$ be some formula in $\mathcal{K}$, and suppose that $\mathfrak{M}$ is a model of every prime implicate of $\varphi$. As **D4** is known to satisfy property **P5** (by Theorem 6), we can find a conjunction of clauses which is equivalent to $\varphi$. By **Covering** (Lemma 9.3), each of these clauses is implied by some prime implicate of $\varphi$, so $\mathfrak{M}$ must be a model of each of these clauses. It follows that $\mathfrak{M}$ is a model of $\varphi$. For the other direction, we simply note that by the definition of prime implicates if $\mathfrak{M}$ is a model of $\varphi$, then it must also be a model of every prime implicate of $\varphi$. We have thus shown that $\mathfrak{M}$ is a model of $\varphi$ if and only if it is a model of every prime implicate of $\varphi$. Using a similar argument, we can show that $\mathfrak{M}$ is a model of $\varphi$ if and only if it is a model of some prime implicant of $\varphi$. $\qquad\square$

**Lemma 9.6** *The notions of prime implicates and prime implicants induced by **D4** satisfy* ***Distribution***.

*Proof.* Let $\lambda$ be a prime implicate of $\varphi_1 \vee ... \vee \varphi_n$. Now for each $\varphi_i$, we must have $\varphi_i \models \lambda$. From **Covering**, we know that there must exist some prime implicate $\lambda_i$ for each $\varphi_i$ such that $\lambda_i \models \lambda$. This means that the formula $\lambda_1 \vee ... \vee \lambda_n$ (which is a clause because it is a disjunction of clauses) entails $\lambda$. But since $\lambda$ is a prime implicate, it must also be the case that $\lambda \models \lambda_1 \vee ... \vee \lambda_n$, and hence $\lambda \equiv \lambda_1 \vee ... \vee \lambda_n$. The proof for prime implicants is entirely similar. $\qquad\square$

**Theorem 9** *The notions of prime implicates and prime implicants induced by definition **D4** satisfy **Finiteness**, **Covering**, **Equivalence**, **Implicant-Implicate Duality**, and **Distribution***.

*Proof.* Follows directly from Lemmas 9.1-9.6. $\qquad\square$

**Theorem 10** *The notions of prime implicates and prime implicants induced by definitions **D1** and **D2** do not satisfy **Equivalence***.

*Proof.* The proof is the same for both definitions. Suppose that **Equivalence** holds. Then for every formula $\varphi$, the set $\Pi$ of prime implicates of $\varphi$ is equivalent to $\varphi$. But this means that the set $\Pi \cup \{\neg \varphi\}$ is inconsistent, and hence by compactness of $\mathcal{K}$ (cf. Blackburn et al., 2001, p. 86) that there is some finite subset $S \subseteq \Pi \cup \{\neg \varphi\}$ which is inconsistent. If $\varphi \not\equiv \bot$, then we know that the set $S$ must contain $\neg \varphi$ because the set of prime implicates of $\varphi$ cannot be inconsistent. But then the conjunction of elements in $S \setminus \{\neg \varphi\}$ is a conjunction of clauses which is equivalent to $\varphi$. It follows that every formula $\varphi$ is equivalent to some conjunction of clauses. As we have shown earlier in the proof of Theorem 5 that there are formulae which are not equivalent to a conjunction of clauses with respect to **D1** or **D2**, it follows that **Equivalence** cannot hold for these definitions. $\square$

**Theorem 11** *The notions of prime implicates and prime implicants induced by definitions **D3a**, **D3b**, and **D5** do not satisfy **Finiteness**.*

*Proof.* Suppose that clauses are defined with respect to definition **D3a**, **D3b**, or **D5** (the proof is the same for all three definitions). Consider the formula $\varphi = \Box(a \wedge b)$. It follows from Theorem 3 that $\varphi$ implies $\lambda_k = \Box(\Diamond^k a) \vee \Diamond(a \wedge b \wedge \Box^k \neg a)$ for every $k \geq 1$. As the formulae $\lambda_k$ are clauses (with respect to **D3a**, **D3b**, and **D5**), the $\lambda_k$ are all implicates of $\varphi$. To complete the proof, we show that every $\lambda_k$ is a prime implicate of $\varphi$. Since the $\lambda_k$ are mutually non-equivalent (because $\Box^p \neg a \not\models \Box^q \neg a$ whenever $p \neq q$), it follows that $\varphi$ has infinitely many prime implicates modulo equivalence.

Consider some $\lambda_k$ and some implicate $\mu = \Diamond \psi_1 \vee ... \vee \Diamond \psi_m \vee \Box \chi_1 \vee ... \vee \Box \chi_n$ of $\varphi$ that implies it (by Theorem 2 there cannot be any propositional literals in $\mu$). Using Theorem 3 and the fact that $\varphi \models \mu \models \lambda_k$, we get the following:

(a) $a \wedge b \models \chi_i \vee \psi_i \vee ... \vee \psi_m$ for some $\chi_i$

(b) $\chi_i \models (\Diamond^k a) \vee (a \wedge b \wedge \Box^k \neg a)$ for every $\chi_i$

(c) $\psi_1 \vee ... \vee \psi_m \models a \wedge b \wedge \Box^k \neg a$

Let $\chi_i$ be such that $a \wedge b \models \chi_i \vee \psi_i \vee ... \vee \psi_m$. We remark that $\chi_i$ must be satisfiable since otherwise we can combine (a) and (c) to get $a \wedge b \models a \wedge b \wedge \Box^k \neg a$. Now by (b), we know that $\chi_i \models (\Diamond^k a) \vee (a \wedge b \wedge \Box^k \neg a)$ and hence that $\chi_i \wedge (\Box^k \neg a) \wedge (\neg a \vee \neg b \vee \Diamond^k a)$ is inconsistent. It follows that both $\chi_i \wedge (\Box^k \neg a) \wedge \neg a$ and $\chi_i \wedge (\Box^k \neg a) \wedge \neg b$ are inconsistent. Using Theorem 1, we find that either $\chi_i \models \Diamond^k a$ or $\chi_i \models a \wedge b$. As $\chi_i$ is a satisfiable clause with respect to definitions **D3a**, **D3b**, and **D5**, it cannot imply $a \wedge b$, so we must have $\chi_i \models \Diamond^k a$. By putting (a) and (c) together, we find that

$$a \wedge b \wedge \neg \chi_i \models \psi_1 \vee ... \vee \psi_m \models a \wedge b \wedge \Box^k \neg a$$

It follows that $\neg \chi_i \models \Box^k \neg a$, i.e. $\Diamond^k a \models \chi_i$. We thus have $\chi_i \equiv \Diamond^k a$ and $\psi_1 \vee ... \vee \psi_m \equiv a \wedge b \wedge \Box^k \neg a$. As $\Diamond^k a \models \chi_i$ and $a \wedge b \wedge \Box^k \neg a \models \psi_1 \vee ... \vee \psi_m$, by Theorem 3 we get $\Box(\Diamond^k a) \vee \Diamond(a \wedge b \wedge \Box^k \neg a) \models \Box \chi_i \vee \Diamond \psi_i \vee ... \vee \Diamond \psi_m \models \mu$ and hence $\lambda_k \equiv \mu$. We have thus shown that any implicate of $\varphi$ which implies $\lambda_k$ must be equivalent to $\lambda_k$. This means that each $\lambda_k$ is a prime implicate of $\varphi$, completing the proof. $\square$

Lemmas 12, 13, and 14 follow easily from known properties of the disjunctive normal form transformation in propositional logic (cf. Bienvenu, 2009, ch. 2).

In the proof of Theorem 16, we will make use of the following lemmas:

**Lemma 16.1** *The algorithm* **GenPI** *always terminates.*

*Proof.* We know from Lemma 12 that the algorithm **Dnf-4** always terminates and returns a finite set of formulae. This means that there are only finitely many terms $T$ to consider. For each $T$, the set $\Delta(T)$ contains only finitely many elements (this is immediate given the definition of $\Delta(T)$), which means that the set CANDIDATES also has finite cardinality. In the final step, we compare at most once each pair of elements in CANDIDATES. As the comparison always terminates, and there are only finitely many pairs to check, it follows that the algorithm **GenPI** terminates. □

**Lemma 16.2** *The algorithm* **GenPI** *outputs exactly the set of prime implicates of the input formula.*

*Proof.* We first prove that every prime implicate of a satisfiable term $T$ is equivalent to some element in $\Delta(T)$. Let $T = \gamma_1 \wedge ... \wedge \gamma_k \wedge \Diamond \psi_1 \wedge ... \wedge \Diamond \psi_m \wedge \Box \chi_1 \wedge ... \wedge \Box \chi_n$ be some satisfiable term, and let $\lambda = \rho_1 \vee ... \vee \rho_p \vee \Diamond \epsilon_1 \vee ... \vee \Diamond \epsilon_q \vee \Box \zeta_1 \vee ... \vee \Box \zeta_r$ be one of its prime implicates. We restrict our attention to the interesting case in which both $T$ and $\lambda$ are non-tautologous. As $T \models \lambda$, it must be the case that

$$\gamma_1 \wedge ... \wedge \gamma_k \wedge \Diamond \psi_1 \wedge ... \wedge \Diamond \psi_m \wedge \Box \chi_1 \wedge ... \wedge \Box \chi_n \wedge$$
$$\neg \rho_1 \wedge ... \wedge \neg \rho_p \wedge \Box \neg \epsilon_1 \wedge ... \wedge \Box \neg \epsilon_q \wedge \Diamond \neg \zeta_1 \wedge ... \wedge \Diamond \neg \zeta_r$$

is unsatisfiable. By Theorem 1, one of the following must hold:

(a) there exists $\gamma_u$ and $\rho_v$ such that $\gamma_u \equiv \rho_v$

(b) there exists $\psi_u$ such that $\psi_u \wedge \chi_1 \wedge ... \wedge \chi_n \models \epsilon_1 \vee ... \vee \epsilon_q$

(c) there exists $\zeta_u$ such that $\chi_1 \wedge ... \wedge \chi_n \models \zeta_u \vee \epsilon_1 \vee ... \vee \epsilon_q$

If (a) holds, then $\gamma_u \models \lambda$, so $\lambda$ must be equivalent to $\gamma_u$ or else we would have found a stronger implicate, contradicting our assumption that $\lambda$ is a prime implicate of $T$. But then the result holds since $\gamma_u$ is in $\Delta(T)$. If (b) holds, then the formula $\Diamond(\psi_u \wedge \chi_1 \wedge ... \wedge \chi_n)$ is an implicate of $T$ which implies $\lambda$, so $\lambda \equiv \Diamond(\psi_u \wedge \chi_1 \wedge ... \wedge \chi_n)$. We are done since $\Diamond(\psi_u \wedge \chi_1 \wedge ... \wedge \chi_r)$ is a member of $\Delta(T)$. Finally we consider the case where (c) holds. In this case, $\Box(\chi_1 \wedge ... \wedge \chi_n)$ is an implicate of $T$ which implies $\lambda$, and so is equivalent to $\lambda$ (as $\lambda$ is a prime implicate). But then we have the desired result since $\Box(\chi_1 \wedge ... \wedge \chi_n)$ is one of the elements in $\Delta(T)$. Thus we can conclude that every prime implicate of a term $T$ is equivalent to some element in $\Delta(T)$. By Lemma 13, the elements in **Dnf-4**$(\varphi)$ are terms, and their disjunction is equivalent to $\varphi$. As **D4** satisfies **Distribution**, it follows that every prime implicate of the input $\varphi$ is equivalent to some element in CANDIDATES. This means that if an element $\lambda_i$ in CANDIDATES is not a prime implicate of $\varphi$, then there is some prime implicate $\pi$ of $\varphi$ that implies but is not implied by $\lambda_i$, and hence some $\lambda_j \in$ CANDIDATES such that $\lambda_j \models \lambda_i$ and $\lambda_i \not\models \lambda_j$. Thus, during the comparison phase, this clause will be removed from CANDIDATES. Now suppose that the clause $\lambda$ is a prime implicate of $\varphi$. Then we know that there must be some $\lambda_i \in$ CANDIDATES such that $\lambda_i \equiv \lambda$, and moreover, we can choose $\lambda_i$ so that there is no $\lambda_j$ with $j < i$ such that $\lambda_j \models \lambda_i$. When in the final step we compare $\lambda_i$ with all the clauses $\lambda_j$ with $j \neq i$, we will never find that $\lambda_j \models \lambda_i$ for $j < i$, nor that $\lambda_j \models \lambda_i \not\models \lambda_j$ for some $j > i$, otherwise $\lambda$ would not be a prime implicate. It follows then that $\lambda_i$ remains in the set CANDIDATES which is returned by the algorithm. We have thus shown that the set of formulae output by **GenPI** on input $\varphi$ is precisely the set of prime implicates of $\varphi$. □

**Theorem 16** *The algorithm* **GenPI** *always terminates and outputs exactly the set of prime implicates of the input formula.*

*Proof.* Follows directly from Lemmas 16.1 and 16.2. ∎

**Theorem 17** *The length of the smallest clausal representation of a prime implicate of a formula is at most single exponential in the length of the formula.*

*Proof.* Prime implicates generated by **GenPI** can have at most $2^{|\varphi|}$ disjuncts as there are at most $2^{|\varphi|}$ terms in **Dnf-4**$(\varphi)$ by Lemma 14. Moreover, each disjunct has length at most $2|\varphi|$ (also by Lemma 14). This gives us a total of $2|\varphi| * 2^{|\varphi|}$ symbols, to which we must add the at most $2^{|\varphi|} - 1$ disjunction symbols connecting the disjuncts. We thus find that the length of the smallest representation of a prime implicate of a formula $\varphi$ is at most $2|\varphi| * 2^{|\varphi|} + (2^{|\varphi|} - 1)$. ∎

**Theorem 18** *The length of the smallest clausal representation of a prime implicate of a formula can be exponential in the length of the formula.*

*Proof.* Consider the formula

$$\varphi = \bigwedge_{i=1}^{n} (\Box a_{i,1} \vee \Box a_{i,2})$$

and the clause

$$\lambda = \bigvee_{(j_1,\ldots,j_n) \in \{1,2\}^n} \Box(a_{1,j_1} \wedge a_{2,j_2} \wedge \ldots \wedge a_{n,j_n})$$

where $a_{k,l} \neq a_{m,p}$ whenever $k \neq m$ or $l \neq p$. It is not difficult to see that $\varphi$ and $\lambda$ are equivalent, which means that $\lambda$ must be a prime implicate of $\varphi$. All that remains to be shown is that any clause equivalent to $\lambda$ must have length at least $|\lambda|$. This yields the result since $\lambda$ clearly has size exponential in $n$, whereas the length of $\varphi$ is only linear in $n$.

Let $\lambda'$ be a shortest clause which is equivalent to $\lambda$. As $\lambda'$ is equivalent to $\lambda$, it follows from Theorem 2 that $\lambda'$ is a disjunction of $\Box$-literals and of inconsistent $\Diamond$-literals. But since $\lambda'$ is assumed to be a shortest representation of $\lambda$, it cannot contain any inconsistent $\Diamond$-literals or any redundant $\Box$-literals, since we could remove them to find an equivalent shorter clause. So $\lambda'$ must be of the form $\Box\chi_1 \vee \ldots \vee \Box\chi_m$, where $\chi_l \not\equiv \chi_k$ whenever $l \neq k$. Now since $\lambda' \models \lambda$, every disjunct $\Box\chi_p$ must also imply $\lambda$. As $\lambda$ is a disjunction of $\Box$-literals, it follows from Theorem 3 that every disjunct $\Box\chi_p$ of $\lambda'$ implies some disjunct $\Box\delta_q$ of $\lambda$. But that means that every $\Box\chi_p$ must have length at least $2n + 1$, since each $\chi_p$ is a satisfiable formula which implies a conjunction of $n$ distinct propositional variables. We also know that every disjunct $\Box\delta_q$ of $\lambda$ implies some disjunct $\Box\chi_p$ of $\lambda'$ since $\lambda \models \lambda'$. We now wish to show that no two disjuncts of $\lambda$ imply the same disjunct of $\lambda'$. Suppose that this is not the case, that is, that there are distinct disjuncts $\Box\delta_1$ and $\Box\delta_2$ of $\lambda$ and some disjunct $\Box\chi_p$ of $\lambda'$ such that $\Box\delta_1 \models \Box\chi_p$ and $\Box\delta_2 \models \Box\chi_p$. Now since $\Box\delta_1$ and $\Box\delta_2$ are distinct disjuncts, there must be some $i$ such that $\Box\delta_1 \models a_{i,1}$ and $\Box\delta_2 \models a_{i,2}$ or $\Box\delta_1 \models a_{i,2}$ and $\Box\delta_2 \models a_{i,1}$. We know that $\Box\chi_p \models \Box\delta_q$ for some $\delta_q$, and that every $\delta_q$ implies either $a_{i1}$ or $a_{i2}$, so either $\Box\chi_p \models \Box a_{i1}$ or $\Box\chi_p \models \Box a_{i2}$. But we know that the $\Box\delta_q$ each imply either $\Box a_{i,1}$ or $\Box a_{i,2}$ but not both, so one of $\Box\delta_1$ and $\Box\delta_2$ must not imply $\Box\chi_p$. This contradicts our earlier assumption that $\Box\delta_1 \models \Box\chi_p$ and $\Box\delta_2 \models \Box\chi_p$, so each disjunct of $\lambda$ must imply a distinct disjunct of $\lambda'$. We have thus demonstrated that $\lambda'$ contains just as many disjuncts as $\lambda$. As we have already shown that the disjuncts of $\lambda'$ are no shorter than the disjuncts of $\lambda$, it follows that $|\lambda'| \geq |\lambda|$, and hence $|\lambda'| = |\lambda|$. We conclude that every clause equivalent to $\lambda$ has length at least $|\lambda|$, completing the proof. $\qquad\blacksquare$

For **Theorem 19**, we will prove that the following clause

$$\lambda = \bigvee_{(q_1,\ldots,q_n)\in\{\Diamond,\Box\}^n} \Box q_1 \ldots q_n\, c$$

is a prime implicate (with respect to both **D1** and **D2**) of the formula

$$\varphi = \left(\Box\Diamond(b_0 \wedge b_1) \vee \Box\Box(b_0 \wedge b_1)\right) \wedge \bigwedge_{i=2}^{n}\left(\Box^i\Diamond b_i \vee \Box^i\Box b_i\right)$$
$$\wedge \bigwedge_{i=1}^{n-1}\Box^{i+1}\left((b_{i-1} \wedge b_i) \to \Box b_i\right) \wedge \Box^{n+1}\left((b_{n-1} \wedge b_n) \to c\right)$$

and moreover that there is no shorter way to represent $\lambda$.

The proof of **Theorem 19** makes use of the following lemmas.

**Lemma 19.1** *Let $l_1 \vee \ldots \vee l_m$ be a **D1**-clause which implies $q_1 \ldots q_n a$, where $q_i \in \{\Box, \Diamond\}$ and $a$ is a propositional variable. Then $l_1 \vee \ldots \vee l_m \equiv q_1 \ldots q_n a$.*

*Proof.* In the proof, we will make use of the fact that every **D1**-clause is satisfiable. This is very straightforwardly shown by structural induction. The base case is propositional literals, which are clearly satisfiable. For the induction step, we consider a **D1**-clause $\lambda$ such that all its proper sub-clauses are satisfiable. There are three possibilities: either $\lambda$ is of the form $\Box\psi$ or $\Diamond\psi$ where $\psi$ is a satisfiable **D1**-clause, or a disjunction $\psi_1 \vee \psi_2$ of satisfiable **D1**-clauses $\psi_1$ and $\psi_2$. In all three cases, we find that $\lambda$ must also be satisfiable.

The proof of the lemma is by induction on $n$. When $n = 0$, we have just $l_1 \vee \ldots \vee l_m \models a$. According to Theorem 2, every disjunct of $l_1 \vee \ldots \vee l_m$ must be either $a$ or some unsatisfiable formula. But we have shown in the previous paragraph that every **D1**-clause is satisfiable, so $l_1 \vee \ldots \vee l_m \equiv a$.

Now suppose the result holds whenever $n \leq k$, and suppose that we have $l_1 \vee \ldots \vee l_m \models q_1 \ldots q_{k+1} a$. For every $l_i$, we must have $l_i \models q_1 \ldots q_{k+1} a$, and hence $\models \neg l_i \vee q_1 \ldots q_{k+1} a$. Using Theorem 1, we arrive at the following four possibilities:

(a) $\models q_1...q_{k+1}a$

(b) $l_i \equiv \bot$

(c) $q_1 = \Diamond$ and $l_i \equiv \Diamond l_i'$ and $l_i' \models q_2...q_{k+1}a$

(d) $q_1 = \Box$ and $l_i \equiv \Box l_i'$ and $l_i' \models q_2...q_{k+1}a$

We can eliminate case (a) since $\not\models q_1...q_{k+1}a$ for every string of modalities $q_1...q_{k+1}$. We can also eliminate (b) since all of the $l_i$ must be satisfiable as they are **D1**-clauses. We remark that if (c) holds, then according to the induction hypothesis, $l_i \equiv \Diamond q_2...q_{k+1}a$. Similarly, if (d) holds, then $l_i \equiv \Box q_2...q_{k+1}a$. It follows then that each $l_i$ is equivalent to $q_1...q_{k+1}a$, and so $l_1 \vee ... \vee l_m \equiv q_1...q_{k+1}a$. $\qquad\square$

**Lemma 19.2** *Fix* $(q_1, ..., q_n) \in \{\Box, \Diamond\}^n$, *and let* $T = \Box q_1(b_0 \wedge b_1) \wedge (\bigwedge_{k=2}^n \Box^k q_k b_k) \wedge \bigwedge_{k=1}^{n-1} \Box^{k+1}((b_{k-1} \wedge b_k) \to \Box b_k) \wedge \Box^{n+1}((b_{n-1} \wedge b_n) \to c)$. *Then* $T \models \Box r_1...r_n c$ *if and only if* $r_k = q_k$ *for all* $1 \le k \le n$.

*Proof.* We begin by showing that for all $1 \le i \le n-1$ the formula

$$b_{i-1} \wedge b_i \wedge (\bigwedge_{k=i+1}^n \Box^{k-i-1} q_k b_k) \wedge (\bigwedge_{k=i}^{n-1} \Box^{k-i}((b_{k-1} \wedge b_k) \to \Box b_k)) \wedge \Box^{n-i}((b_{n-1} \wedge b_n) \to c)$$

entails the formula $r_{i+1}...r_n c$ just in the case that $q_{i+1}...q_n = r_{i+1}...r_n$.

The proof is by induction on $i$. The base case is $i = n-1$. We have

$$b_{n-2} \wedge b_{n-1} \wedge q_n b_n \wedge ((b_{n-2} \wedge b_{n-1}) \to \Box b_{n-1}) \wedge \Box((b_{n-1} \wedge b_n) \to c) \models r_n c \qquad (1)$$

if and only if

$$b_{n-2} \wedge b_{n-1} \wedge q_n b_n \wedge \Box b_{n-1} \wedge \Box((b_{n-1} \wedge b_n) \to c) \models r_n c$$

if and only if (Theorem 1) either

$$q_n = \Diamond \text{ and } r_n = \Box \text{ and } b_{n-1} \wedge ((b_{n-1} \wedge b_n) \to c) \models c$$

or

$$q_n = r_n \text{ and } b_{n-1} \wedge b_n \wedge ((b_{n-1} \wedge b_n) \to c) \models c$$

As $b_{n-1} \wedge ((b_{n-1} \wedge b_n) \to c) \not\models c$, we cannot have the first alternative. It follows then that if Equation (1) holds, then the second alternative must hold, in which case we get $q_n = r_n$, as desired. For the other direction, we simply note that $b_{n-1} \wedge b_n \wedge ((b_{n-1} \wedge b_n) \to c) \models c$ is a valid entailment, which means $q_n = r_n$ implies Equation (1).

Next let us suppose that the above statement holds for all $1 < j \le i \le n-1$, and let us prove the statement holds when $i = j - 1$. Then

$$b_{j-2} \wedge b_{j-1} \wedge (\bigwedge_{k=j}^n \Box^{k-j} q_k b_k) \wedge (\bigwedge_{k=j-1}^{n-1} \Box^{k-j+1}(b_{k-1} \wedge b_k \to \Box b_k))$$
$$\wedge \Box^{n-j+1}((b_{n-1} \wedge b_n) \to c) \qquad\qquad \models r_j...r_n c \qquad (2)$$

if and only if one of the following holds:

(a) $q_j = \diamond$ and $r_j = \square$ and

$$b_{j-1} \wedge \left( \bigwedge_{k=j+1}^{n} \square^{k-j-1} q_k \, b_k \right) \wedge \left( \bigwedge_{k=j}^{n-1} \square^{k-j}((b_{k-1} \wedge b_k) \to \square b_k) \right) \wedge \square^{n-j}((b_{n-1} \wedge b_n) \to c)$$

$$\models r_{j+1}...r_n c$$

(b) $q_j = r_j$ and

$$b_{j-1} \wedge b_j \wedge \left( \bigwedge_{k=j+1}^{n} \square^{k-j-1} q_k b_k \right) \wedge \left( \bigwedge_{k=j}^{n-1} \square^{k-j}((b_{k-1} \wedge b_k) \to \square b_k) \right) \wedge \square^{n-j}((b_{n-1} \wedge b_n) \to c)$$

$$\models r_{j+1}...r_n c$$

We will first show that the entailment in (a) does not hold. Consider the model $\mathfrak{M} = \langle \mathcal{W}, \mathcal{R}, v \rangle$ defined as follows:

- $\mathcal{W} = \{w_j, ..., w_n\}$

- $\mathcal{R} = \{(w_j, w_{j+1}), ..., (w_{n-1}, w_n)\}$

- $v(c, w) = false$ for all $w \in \mathcal{W}$

- for $w \neq w_j$: $v(b_k, w) = true$ if and only if $w = w_k$

- $v(b_k, w_j) = true$ if and only if $k = j - 1$

Notice that since each world (excepting $w_n$) has exactly one successor, the $\square$- and $\diamond$-quantifiers have the same behaviour (except at $w_n$). It can easily be verified that $\mathfrak{M}, w_j$ satisfies the left-hand side of the above entailment for any tuple $q_{j+1}...q_n$: we have $\mathfrak{M}, w_j \models b_{j-1}$ by definition, we have $\mathfrak{M}, w_j \models \bigwedge_{k=j+1}^{n} \square^{k-j-1} q_k b_k$ because $\mathfrak{M}, w_k \models b_k$ for $k \neq j$, we have $\mathfrak{M}, w_j \models \bigwedge_{k=j}^{n-1} \square^{k-j}((b_{k-1} \wedge b_k) \to \square b_k))$ since $\mathfrak{M}, w_j \not\models b_j$ and $\mathfrak{M}, w_k \not\models b_{k-1}$ for $k \neq j$, and finally we have $\mathfrak{M}, w_j \models \square^{n-j}((b_{n-1} \wedge b_n) \to c)$ since $w_n \not\models b_{n-1}$. However, the right-hand side $r_{j+1}...r_n c$ is not satisfied at $w_j$: the only world accessible from $w_j$ in $n - j$ steps is $w_n$ which does not satisfy $c$.

We have just shown that case (a) cannot hold, which means that Equation (2) holds if and only if (b) does. But if we apply the induction hypothesis to the entailment in (b), we find that it holds just in the case that $q_{j+1}...q_n = r_{j+1}...r_n$. It follows then that Equation (2) holds if and only if $q_j...q_n = r_j...r_n$, as desired. This completes our proof of the above statement.

We now proceed to the proof of the lemma. By Theorem 1,

$$\square q_1(b_0 \wedge b_1) \wedge \left( \bigwedge_{k=2}^{n} \square^k q_k b_k \right) \wedge \left( \bigwedge_{k=1}^{n-1} \square^{k+1}((b_{k-1} \wedge b_k) \to \square b_k) \right) \wedge \square^{n+1}((b_{n-1} \wedge b_n) \to c)$$

$$\models \square r_1...r_n c$$

holds just in the case that

$$q_1(b_0 \wedge b_1) \wedge \left( \bigwedge_{k=2}^{n} \square^{k-1} q_k b_k \right) \wedge \bigwedge_{k=1}^{n-1} \square^k((b_{k-1} \wedge b_k) \to \square b_k) \wedge \square^n((b_{n-1} \wedge b_n) \to c)$$

$$\models r_1...r_n c$$

which in turn holds if and only if one of the following statements holds:

(i) $q_1 = \Diamond$ and $r_1 = \Box$ and

$$(\bigwedge_{k=2}^{n} \Box^{k-2} q_k b_k) \wedge (\bigwedge_{k=1}^{n-1} \Box^{k-1}((b_{k-1} \wedge b_k) \rightarrow \Box b_k)) \wedge \Box^{n-1}((b_{n-1} \wedge b_n) \rightarrow c) \models r_2...r_n c$$

(ii) $q_1 = r_1$ and

$$b_0 \wedge b_1 \wedge (\bigwedge_{k=2}^{n} \Box^{k-2} q_k b_k) \wedge (\bigwedge_{k=1}^{n-1} \Box^{k-1}((b_{k-1} \wedge b_k) \rightarrow \Box b_k)) \wedge \Box^{n-1}((b_{n-1} \wedge b_n) \rightarrow c)$$
$$\models r_2...r_n c$$

We remark that if we set $j = 1$ in (a) above, then the left-hand side of the entailment in (i) is logically weaker than that in (a), and the right-hand side matches that in (a). As we have already shown that the entailment in (a) does not hold, it follows that the entailment in (i) cannot hold either. Thus, we find that the desired entailment relation in the statement of the lemma holds if and only if (ii) does. This completes the proof since we have already shown in the induction above that the entailment in (ii) holds if and only if $q_2...q_n = r_2...r_n$, i.e. (ii) is true just in the case that $q_1...q_n = r_1 = r_n$. $\qquad\square$

**Lemma 19.3** *There is no **D1**-clause equivalent to $\lambda$ and with strictly smaller size than $\lambda$.*

*Proof.* Let $\lambda'$ be a **D1**-clause which is equivalent to $\lambda$. Suppose furthermore that $\lambda'$ is a shortest such clause. As $\lambda$ is non-tautologous and contains only $\Box$-literals as disjuncts, it follows that every disjunct of $\lambda'$ must be either unsatisfiable or a $\Box$-literal (cf. Theorem 2). But **D1**-clauses are always satisfiable (cf. proof of Lemma 19.1), so $\lambda'$ must contain only $\Box$-literals.

Since $\lambda' \models \lambda$, every disjunct $\Box l$ of $\lambda'$ must imply some disjunct $\Box q_1...q_n c$ of $\lambda$. Also, every disjunct $\Box l$ of $\lambda'$ must be implied by some disjunct $\Box q_1...q_n c$ of $\lambda$, since otherwise we could remove $\Box l$ from $\lambda'$ while preserving the equivalence between $\lambda$ and $\lambda'$.

It follows then that each disjunct of $\lambda'$ is implied by some disjunct of $\lambda$ and implies some disjunct of $\lambda$. But since the disjuncts of $\lambda$ do not imply each other (because of Lemma 19.1), it follows that each disjunct of $\lambda'$ is equivalent to some disjunct of $\lambda$, and moreover that every disjunct of $\lambda$ is equivalent to some disjunct of $\lambda'$.

This completes the proof since it is clear that the disjuncts $\Box q_1...q_n c$ of $\lambda$ cannot be more compactly represented.

Our proof works equally well for **D2**, since every **D2**-clause is also a **D1**-clause. $\qquad\square$

**Theorem 19** *If prime implicates are defined using either **D1** or **D2**, then the length of the smallest clausal representation of a prime implicate of a formula can be exponential in the length of the formula.*

*Proof.* We begin with definition **D1**. Let $\lambda$ and $\varphi$ be as defined on page 112. We begin by distributing $\vee$ over $\wedge$ in order to transform $\varphi$ into an equivalent disjunction of **D4**-terms:

$$\varphi \equiv \bigvee_{(q_1,...,q_n) \in \{\Box, \Diamond\}^n} T_{q_1,...,q_n}$$

where $T_{q_1,\ldots,q_n}$ is equal to

$$\Box q_1(b_0 \wedge b_1) \wedge (\bigwedge_{i=2}^{n} \Box^i q_i b_i) \wedge \bigwedge_{i=1}^{n-1} \Box^{i+1} ((b_{i-1} \wedge b_i) \to \Box b_i) \wedge \Box^{n+1} ((b_{n-1} \wedge b_n) \to c)$$

By Lemma 19.2, $T_{q_1,\ldots,q_n} \models \Box q_1 \ldots q_n c$, and hence $T_{q_1,\ldots,q_n} \models \lambda$. We thus have $\varphi \models \lambda$.

We now show that there is no stronger clause with respect to **D1** which is implied by $\varphi$. Let $\lambda'$ be a **D1**-clause such that $\varphi \models \lambda' \models \lambda$. As $\lambda$ is a non-tautologous disjunction of $\Box$-literals, we know from Lemma 2 that every disjunct of $\lambda'$ must be of the form $\Box l$ where $l$ is a **D1**-clause such that $l \models r_1 \ldots r_n c$ for some quantifier string $r_1 \ldots r_n$. But according to Lemma 19.1, if $l \models r_1 \ldots r_n c$, then $l$ is equivalent to $r_1 \ldots r_n c$. It follows that $\lambda'$ is equivalent to a clause having only disjuncts of the forms $\Box r_1 \ldots r_n c$.

As $\varphi \models \lambda'$, it must be the case that each of the terms $T_{q_1,\ldots,q_n}$ implies $\lambda'$, or equivalently $T_{q_1,\ldots,q_n} \wedge \neg \lambda' \models \bot$. As we have shown above that the disjuncts of $\lambda'$ are all $\Box$-literals, it follows from Theorem 1 that each term implies some disjunct of $\lambda'$. Moreover, we know from the preceding paragraph that each of the disjuncts of $\lambda'$ is equivalent to some formula of the form $\Box r_1 \ldots r_n c$. By Lemma 19.2, the only formula of this type which is implied by $T_{q_1,\ldots,q_n}$ is the formula $\Box q_1 \ldots q_n c$. This means that for every tuple of quantifiers $(q_1, \ldots, q_n)$, there is a disjunct of $\lambda'$ which is equivalent to $\Box q_1 \ldots q_n c$. It follows that every disjunct of $\lambda$ is equivalent to some disjunct of $\lambda'$, giving us $\lambda \models \lambda'$. We can thus conclude that $\lambda$ is a prime implicate of $\varphi$.

This completes the proof, since we have already shown in Lemma 19.3 that there is no shorter **D1**-clause which is equivalent to $\lambda$ than $\lambda$ itself.

The above proof also works for definition **D2** since every **D2**-clause is also a **D1**-clause. In particular this means that any **D2**-clause which is a prime implicate with respect to **D1** is also a prime implicate with respect to **D2**, and that any **D2**-clause which is shortest among all equivalent **D1**-clauses is also shortest among **D2**-clauses. $\square$

**Theorem 20** *The number of non-equivalent prime implicates of a formula is at most double exponential in the length of the formula.*

*Proof.* We know from Theorem 16 that every prime implicate of $\varphi$ is equivalent to some clause returned by **GenPI**. Every such clause is of the form $\bigvee_{T \in \mathbf{Dnf\text{-}4}(\varphi)} \theta_T$ where $\theta_T \in \Delta(T)$. As there can be at most $2^{|\varphi|}$ terms in $\mathbf{Dnf\text{-}4}(\varphi)$ by Lemma 14, these clauses can have no more than $2^{|\varphi|}$ disjuncts. Moreover, there are at most $2|\varphi|$ choices for each disjunct $\theta_T$ since the cardinality of $\Delta(T)$ is bounded above by the size of $T$, which we know from Lemma 1.3 to be no more than $2|\varphi|$. It follows then that there are at most $(2|\varphi|)^{2^{|\varphi|}}$ clauses returned by **GenPI**, hence at most $(2|\varphi|)^{2^{|\varphi|}}$ non-equivalent prime implicates of $\varphi$. $\square$

**Theorem 21** *The number of non-equivalent prime implicates of a formula may be double exponential in the length of the formula.*

*Proof.* Let $n$ be some natural number, and let $a_{11}$, $a_{12}$, ..., $a_{n1}$, $a_{n2}$, $b_{11}$, $b_{12}$, $b_{12}$, ..., $b_{n1}$, $b_{n2}$ be $4n$ distinct propositional variables. Consider the formula $\varphi$ defined as

$$\bigwedge_{i=1}^{n} ((\Diamond a_{i1} \wedge \Box b_{i1}) \vee (\Diamond a_{i2} \wedge \Box b_{i2}))$$

It is not hard to see that there will be $2^n$ terms in **Dnf-4**$(\varphi)$, corresponding to the $2^n$ ways of deciding for each $i \in \{1, ..., n\}$ whether to take the first or second disjunct. Each term $T \in$ **Dnf-4**$(\varphi)$ will be of the form

$$\bigwedge_{i=1}^{n} (\diamond a_{i\,f(i,T)} \wedge \Box b_{i\,f(i,T)})$$

where $f(i,T) \in \{1, 2\}$ for all $i$. For each $T$, denote by $\mathcal{D}(T)$ the set of formulae $\{\diamond(a_{f(i,T)} \wedge b_{1\,f(1,T)} \wedge ... \wedge b_{n\,f(n,T)})) \,|\, 1 \leq i \leq n\}$. Now consider the set of clauses $\mathcal{C}$ defined as

$$\{ \bigvee_{T \in \mathbf{Dnf\text{-}4}(\varphi)} d_T \,|\, d_T \in \mathcal{D}(T) \}$$

Notice that there are $n^{2^n}$ clauses in $\mathcal{C}$ since each clause corresponds to a choice of one of the $n$ elements in $\mathcal{D}(T)$ for each of the $2^n$ terms $T$ in **Dnf-4**$(\varphi)$. This number is double exponential in $|\varphi|$ since the length of $\varphi$ is linear in $n$. In order to complete the proof, we show that (i) all of the clauses in $\mathcal{C}$ are prime implicates of $\varphi$ and (ii) that the clauses in $\mathcal{C}$ are mutually non-equivalent.

We begin by showing that $\lambda_1 \not\models \lambda_2$ for every pair of distinct elements $\lambda_1$ and $\lambda_2$ in $\mathcal{C}$. This immediately gives us (ii) and will prove useful in the proof of (i). Let $\lambda_1$ and $\lambda_2$ be distinct clauses in $\mathcal{C}$. As $\lambda_1$ and $\lambda_2$ are distinct, there must be some term $T \in$ **Dnf-4**$(\varphi)$ for which $\lambda_1$ and $\lambda_2$ choose different elements from $\mathcal{D}(T)$. Let $d_1$ be the element from $\mathcal{D}(T)$ appearing as a disjunct in $\lambda_1$, let $d_2$ be the element in $\mathcal{D}(T)$ which is a disjunct in $\lambda_2$, and let $a_{j,k}$ be the $a$-literal which appears in $d_2$ (and hence not in $d_1$). Consider the formula $\rho = \Box(\neg a_{j,k} \wedge \neg b_{1,k_1} \wedge ... \wedge \neg b_{n,k_n})$, where the tuple $(k_1, ..., k_n)$ is just like the tuple associated with $T$ except that the 1's and 2's are inversed. Clearly $d_1 \wedge \rho$ is consistent, since the variables in $\rho$ do not appear in $d_1$. But $\rho$ is inconsistent with every disjunct in $\lambda_2$, since by construction every disjunct in $\lambda_2$ contains a literal whose negation appears in $\rho$. It follows that $\lambda_2 \models \neg \rho$ but $\lambda_1 \not\models \neg \rho$, and hence $\lambda_1 \not\models \lambda_2$.

We now prove (i). Let $\lambda$ be a clause in $\mathcal{C}$, and let $\pi$ be a prime implicate of $\varphi$ which implies $\lambda$. By Theorem 16, we know that $\pi$ must be equivalent to one of the clauses output by **GenPI**, and more specifically to a clause output by **GenPI** which is a disjunction of $\diamond$-literals (because of Theorem 2). We remark that the set $\mathcal{C}$ is composed of exactly those candidate clauses which are disjunctions of $\diamond$-literals, so $\pi$ must be equivalent to some clause in $\mathcal{C}$. But we have just shown that the only element in $\mathcal{C}$ which implies $\lambda$ is $\lambda$ itself. It follows that $\pi \equiv \lambda$, which means that $\lambda$ is a prime implicate of $\varphi$. $\qquad\Box$

**Theorem 22** *If prime implicates are defined using either **D1** or **D2**, then the number of non-equivalent prime implicates of a formula may be double exponential in the length of the formula.*

*Proof.* Let $\lambda$ and $\varphi$ be as defined on page 112. Set $\varphi'$ equal to the formula obtained from $\varphi$ by replacing $c$ in the last conjunct of $\varphi$ by $c \wedge d$. Set $\Sigma$ equal to the set of clauses that can be obtained from $\lambda$ by replacing zero or more occurrences of $c$ by $d$. For example, if $n = 1$, then $\Sigma = \{\Box\diamond c \vee \Box\Box c, \Box\diamond d \vee \Box\Box c, \Box\diamond c \vee \Box\Box d, \Box\diamond d \vee \Box\Box d\}$. There are $2^{2^n}$ elements in $\Sigma$ since we choose for each of the $2^n$ disjuncts of $\lambda$ whether to change $c$ into $d$. We intend to show that the clauses in $\Sigma$ are all pairwise non-equivalent prime implicates of $\varphi'$. The proof that every element in $\Sigma$ is indeed a prime implicate of $\varphi'$ (with respect to both **D1** and **D2**) proceeds quite similarly to the proof that $\lambda$ is a prime implicate of $\varphi$ (see proof of Theorem 19), so we will not repeat it here. Instead we will show that all of the elements in $\Sigma$ are pairwise non-equivalent. To do so, we consider any two distinct elements $\alpha$ and $\beta$ of $\Sigma$. Since $\alpha$ and $\beta$ are distinct, there must be some string of quantifiers $q_1...q_n$ such that $\alpha$ has a disjunct $\Box q_1...q_n\gamma$ ($\gamma \in \{c, d\}$) which is not a disjunct of $\beta$. Now if $\alpha \models \beta$, then we would have $\Box q_1...q_n\gamma \models \beta$, and hence $\Box q_1...q_n\gamma \models \Box r_1...r_n\zeta$ for some disjunct $r_1...r_n\zeta$ of $\beta$. But by using Lemma 19.1, we see that this can only happen if $r_1...r_n = q_1...q_n$ and $\gamma = \zeta$, i.e. if $\Box q_1...q_n$ is a disjunct of $\beta$. This is a contradiction, so we must have $\alpha \not\models \beta$. It follows that the elements of $\Sigma$ are pairwise non-equivalent, and hence that $\varphi'$ possesses a double exponential number of prime implicates. $\qquad\square$

**Theorem 23** *There exists an algorithm which runs in single-exponential space in the size of the input and incrementally outputs, without duplicates, the set of prime implicates of the input formula.*

*Proof.* Let the sets $\mathcal{T}$ and CANDIDATES and the function $\Delta$ be defined as in Figure 3. We assume that $\mathcal{T}$ is ordered: $\mathcal{T} = \{T_1, ..., T_n\}$. For each $T_i \in \mathcal{T}$, we let $max_i$ denote the number of elements in $\Delta(T_i)$, and we assume an ordering on the elements of $\Delta(T_i)$: $\Delta(T_i) = \{\tau_{i,1}, ..., \tau_{i,max_i}\}$. Notice that the tuples in $\{1, .., max_1\} \times ... \times \{1, ..., max_n\}$ can be ordered using the standard lexicographic ordering $<_{lex}$: $(a_1, ..., a_n) <_{lex} (b_1, ..., b_n)$ if and only if there is some $1 \leq j \leq n$ such that $a_j < b_j$ and $a_k \leq b_k$ for all $1 \leq k \leq j - 1$. Now set $maxindex = \Pi_{i=1}^n max_i$, and let $f : \{1, ..., max_1\} \times ... \times \{1, ..., max_n\} \to \{1, ..., maxindex\}$ be the bijection defined as follows: $f(a_1, ..., a_n) = m$ if and only if $(a_1, ..., a_n)$ is the $m$-th tuple in the lexicographic ordering of $\{1, .., max_1\} \times ... \times \{1, ..., max_n\}$. We will denote by $\lambda_m$ the unique clause of form $\tau_{1,a_1} \vee ... \vee \tau_{n,a_n}$ such that $f(a_1, ..., a_n) = m$. We remark that given an index $m \in \{1, ..., maxindex\}$ and the sets $\Delta(T_1), ..., \Delta(T_n)$, it is possible to generate in polynomial space (in the size of the sets $\Delta(T_1), ..., \Delta(T_n)$) the clause $\lambda_m$. We make use of this fact in our modified version of algorithm GENPI, which is defined as follows:

**Function IterGenPI($\varphi$)**
(1) Same as in **GenPI**.
(2) Same as in **GenPI**.
(3) For $i = 1$ to $maxindex$: if $\lambda_j \not\models \lambda_i$ for all $j < i$ *and either* $\lambda_j \not\models \lambda_i$ or $\lambda_i \models \lambda_j$
   for every $i < j \leq maxindex$, then output $\lambda_i$.

The proofs of termination, correctness, and completeness of **IterGenPI** are very similar to corresponding results for **GenPI** (Theorem 16), so we will omit the details. We will instead focus on the spatial complexity of **IterGenPI**. The first step of **IterGenPI** clearly runs in single-exponential space in $|\varphi|$, since deciding the satisfiability of $\varphi$ takes only polynomial space in $|\varphi|$, and generating the elements in **Dnf-4($\varphi$)** takes at most single-exponential space in $|\varphi|$ (refer to Lemma 14). Step 2 also uses no more than single-exponential space in $|\varphi|$, since each of the sets $\Delta(T)$ associated with a term $T_i \in \mathcal{T}$ has polynomial size in $T_i$. Finally, for Step 3, we use the above observation that the generation of a given $\lambda_i$ from its index $i$ can be done in polynomial space in the size of the sets $\Delta(T_1)$, ..., $\Delta(T_n)$, and hence in single-exponential space in $|\varphi|$. This is sufficient since for the comparisons in Step 3, we only need to keep two candidate clauses in memory at any one time, and deciding whether one candidate clause entails another can be accomplished in single-exponential space (since both clauses have single-exponential size in $|\varphi|$). □

**Theorem 24** *Prime implicate recognition is* Pspace-*hard.*

*Proof.* The reduction is simple: a formula $\varphi$ is unsatisfiable if and only if $\Diamond(a \wedge \neg a)$ is a prime implicate of $\varphi$. This suffices as the problem of checking the unsatisfiability of formulae in $\mathcal{K}$ is known to be Pspace-complete. □

We will need the following two lemmas for Theorem 25:

**Lemma 25.1** *Let $\varphi$ be a formula from $\mathcal{K}$, and let $\lambda = \gamma_1 \vee ... \vee \gamma_k \vee \Diamond\psi_1 \vee ... \vee \Diamond\psi_m \vee \Box\chi_1 \vee ... \vee \Box\chi_n$ ($\gamma_j$ propositional literals) be a non-tautologous clause. Suppose furthermore that there is no literal $l$ in $\lambda$ such that $\lambda \equiv \lambda \setminus \{l\}$. If $\lambda \in \Pi(\varphi)$, then $\gamma_1 \vee ... \vee \gamma_k \in \Pi(\varphi \wedge \neg(\lambda \setminus \{\gamma_1, ..., \gamma_k\}))$ and $\Diamond(\psi_1 \vee ... \vee \psi_n) \in \Pi(\varphi \wedge \neg(\lambda \setminus \{\Diamond\psi_1, ..., \Diamond\psi_m\}))$ and for every $i$, $\Box(\chi_i \wedge \neg\psi_1 \wedge ... \wedge \neg\psi_m) \in \Pi(\varphi \wedge \neg(\lambda \setminus \{\Box\chi_i\}))$.*

*Proof.* We will prove the contrapositive: if $\gamma_1 \vee ... \vee \gamma_k \notin \Pi(\varphi \wedge \neg(\lambda \setminus \{\gamma_1, ..., \gamma_k\}))$ or $\Diamond(\psi_1 \vee ... \vee \psi_n) \notin \Pi(\varphi \wedge \neg(\lambda \setminus \{\Diamond\psi_1, ..., \Diamond\psi_m\}))$ or there is some $i$ for which $\Box(\chi_i \wedge \neg\psi_1 \wedge ... \wedge \neg\psi_m) \notin \Pi(\varphi \wedge \neg(\lambda \setminus \{\Box\chi_i\}))$, then $\lambda \notin \Pi(\varphi)$. We will only consider the case where $\varphi \models \lambda$ because if $\varphi \not\models \lambda$ then we immediately get $\lambda \notin \Pi(\varphi)$.

Let us first suppose that $\gamma_1 \vee ... \vee \gamma_k \notin \Pi(\varphi \wedge \neg(\lambda \setminus \{\gamma_1, ..., \gamma_k\}))$. Since $\varphi \models \lambda$, we must also have $\varphi \wedge \neg(\lambda \setminus \{\gamma_1, ..., \gamma_k\}) \models \gamma_1 \vee ... \vee \gamma_k$, so $\gamma_1 \vee ... \vee \gamma_k$ is an implicate of $\varphi \wedge \neg(\lambda \setminus \{\gamma_1, ..., \gamma_k\})$. As $\gamma_1 \vee ... \vee \gamma_k$ is known not to be a prime implicate of $\varphi \wedge \neg(\lambda \setminus \{\gamma_1, ..., \gamma_k\})$, it follows that there must be some clause $\lambda'$ such that $\varphi \wedge \neg(\lambda \setminus \{\gamma_1, ..., \gamma_k\}) \models \lambda' \models \gamma_1 \vee ... \vee \gamma_k \not\models \lambda'$. Now consider the clause $\lambda'' = \lambda' \vee \Diamond\psi_1 \vee ... \vee \Diamond\psi_m \vee \Box\chi_1 \vee ... \vee \Box\chi_n$. We know that $\varphi \models \lambda''$ since $\varphi \wedge \neg(\lambda \setminus \{\gamma_1, ..., \gamma_k\}) \models \lambda'$, and that $\lambda'' \models \lambda$ because $\lambda' \models \gamma_1 \vee ... \vee \gamma_k$. We also have $\lambda \not\models \lambda''$ since $\lambda'$ must be equivalent to a propositional clause (by Theorem 2) and the propositional part of $\lambda$ (namely $\gamma_1 \vee ... \vee \gamma_k$) does not imply $\lambda'$. It follows then that $\varphi \models \lambda'' \models \lambda \not\models \lambda''$, so $\lambda \notin \Pi(\varphi)$.

Next suppose that $\Diamond(\psi_1 \vee ... \vee \psi_n) \notin \Pi(\varphi \wedge \neg(\lambda \setminus \{\Diamond\psi_1, ..., \Diamond\psi_m\}))$. Now $\Diamond(\psi_1 \vee ... \vee \psi_n)$ must be an implicate of $\varphi \wedge \neg(\lambda \setminus \{\Diamond\psi_1, ..., \Diamond\psi_m\})$ since we have assumed that $\varphi \models \lambda$. As $\Diamond(\psi_1 \vee ... \vee \psi_n)$ is not a prime implicate of $\varphi \wedge \neg(\lambda \setminus \{\Diamond\psi_1, ..., \Diamond\psi_m\})$, it follows that there is some $\lambda'$ such that $\varphi \wedge \neg(\lambda \setminus \{\Diamond\psi_1, ..., \Diamond\psi_m\}) \models \lambda' \models \Diamond(\psi_1 \vee ... \vee \psi_n) \not\models \lambda'$. Let $\lambda'' = \gamma_1 \vee ... \vee \gamma_k \vee \lambda' \vee \Box\chi_1 \vee ... \vee \Box\chi_n$. Because of Theorem 2, we know that $\lambda'$ is a disjunction of $\Diamond$-literals, so according to Theorem 3 we must have $\lambda \not\models \lambda''$ since $\Diamond(\psi_1 \vee ... \vee \psi_n) \not\models \lambda'$. We also know that $\varphi \models \lambda''$ since $\varphi \wedge \neg(\lambda \setminus \{\Diamond\psi_1, ..., \Diamond\psi_m\}) \models \lambda'$ and that $\lambda'' \models \lambda$ since $\lambda' \models \Diamond(\psi_1 \vee ... \vee \psi_n)$. That means that $\varphi \models \lambda'' \models \lambda \not\models \lambda''$, so $\lambda \notin \Pi(\varphi)$.

Finally consider the case where there is some $i$ for which $\Box(\chi_i \wedge \neg\psi_1 \wedge ... \wedge \neg\psi_m) \notin \Pi(\varphi \wedge \neg(\lambda \setminus \{\Box\chi_i\}))$. We know that $\varphi \models \lambda$ and hence that $\varphi \wedge \neg(\lambda \setminus \{\Box\chi_i\}) \models \Box\chi_i$. Moreover, since $\neg(\lambda \setminus \{\Box\chi_i\}) \models \neg\Diamond\psi_j$ for all $j$, we have $\varphi \wedge \neg(\lambda \setminus \{\Box\chi_i\}) \models \Box(\chi_i \wedge \neg\psi_1 \wedge ... \wedge \neg\psi_m)$. Thus, if $\Box(\chi_i \wedge \neg\psi_1 \wedge ... \wedge \neg\psi_m) \notin \Pi(\varphi \wedge \neg(\lambda \setminus \{\Box\chi_i\}))$, it must mean that there is some $\lambda'$ such that $\varphi \wedge \neg(\lambda \setminus \{\Box\chi_i\}) \models \lambda' \models \Box(\chi_i \wedge \neg\psi_1 \wedge ... \wedge \neg\psi_m) \not\models \lambda'$. By assumption, $\lambda$ is not a tautology, so $\Box(\chi_i \wedge \neg\psi_1 \wedge ... \wedge \neg\psi_m)$ cannot be a tautology either. As $\lambda' \models \Box(\chi_i \wedge \neg\psi_1 \wedge ... \wedge \neg\psi_m)$ and $\Box(\chi_i \wedge \neg\psi_1 \wedge ... \wedge \neg\psi_m)$ is not a tautology, it follows from Theorem 2 that $\lambda'$ is equivalent to some formula $\Box\zeta_1 \vee ... \vee \Box\zeta_p$. Let $\lambda'' = \gamma_1 \vee ... \vee \gamma_k \vee \Diamond\psi_1 \vee ... \vee \Diamond\psi_m \vee \Box\chi_1 \vee ... \vee \Box\chi_{i-1} \vee (\Box\zeta_1 \vee ... \vee \Box\zeta_p) \vee \Box\chi_{i+1} \vee ... \vee \Box\chi_n$. As $\varphi \wedge \neg(\lambda \setminus \{\Box\chi_i\}) \models \Box\zeta_1 \vee ... \vee \Box\zeta_p$, it must be the case that $\varphi \models \lambda''$. Also, we know that there can be no $j$ such that $\chi_i \models \zeta_j \vee \psi_1 \vee ... \vee \psi_m$ because otherwise we would have $\chi_i \wedge \neg\psi_1 \wedge ... \wedge \neg\psi_m \models \zeta_j$ and hence $\Box(\chi_i \wedge \neg\psi_1 \wedge ... \wedge \neg\psi_m) \models \Box\zeta_1 \vee ... \vee \Box\zeta_p$. Similarly, there can be no $k \neq i$ such that $\Box\chi_i \models \Box(\chi_k \vee \psi_1 \vee ... \vee \psi_m)$ because this would mean that $\lambda \equiv \lambda \setminus \{\Box\chi_i\}$, contradicting our assumption that there are no superfluous disjuncts in $\lambda$. It follows then by Theorem 3 that $\lambda \not\models \lambda''$. Thus, $\varphi \models \lambda'' \models \lambda \not\models \lambda''$, which means $\lambda \notin \Pi(\varphi)$. □

**Lemma 25.2** *Let $\varphi$ be a formula of $\mathcal{K}$, and let $\lambda = \gamma_1 \vee ... \vee \gamma_k \vee \Diamond\psi_1 \vee ... \vee \Diamond\psi_m \vee \Box\chi_1 \vee ... \vee \Box\chi_n$ ($\gamma_j$ propositional literals) be a non-tautologous clause. Suppose furthermore that there is no literal $l$ in $\lambda$ such that $\lambda \equiv \lambda \setminus \{l\}$. Then if $\lambda \not\equiv \Pi(\varphi)$, either $\gamma_1 \vee ... \vee \gamma_k \not\in \Pi(\varphi \wedge \neg(\lambda \setminus \{\gamma_1, ..., \gamma_k\}))$ or $\Diamond(\psi_1 \vee ... \vee \psi_m) \not\in \Pi(\varphi \wedge \neg(\gamma_1 \vee ... \vee \gamma_k \vee \Box(\psi_1 \vee ... \vee \psi_m) \vee ... \vee \Box(\chi_n \vee \psi_1 \vee ... \vee \psi_m))$ or $\Box(\chi_i \wedge \neg\psi_1 \wedge ... \wedge \neg\psi_m) \not\in \Pi(\varphi \wedge \neg(\lambda \setminus \{\Box\chi_i\}))$ for some $i$.*

*Proof.* We will only consider the case where $\varphi \models \lambda$ because if $\varphi \not\models \lambda$ then we immediately get the result. Suppose then that $\lambda \notin \Pi(\varphi)$ and $\varphi \models \lambda$. By Definition 7, there must be some $\lambda' = \gamma_1' \vee ... \vee \gamma_o' \vee \Diamond\psi_1' \vee ... \vee \Diamond\psi_p' \vee \Box\chi_1' \vee ... \vee \Box\chi_q'$ such that $\varphi \models \lambda' \models \lambda \not\models \lambda'$. Since $\lambda \models \lambda'$, by Proposition 3 we know that either $\gamma_1 \vee ... \vee \gamma_k \models \gamma_1' \vee ... \vee \gamma_o'$ or $\psi_1 \vee ... \vee \psi_m \not\models \psi_1' \vee ... \vee \psi_p'$ or there is some $i$ for which $\chi_i \not\models \chi_j' \vee \psi_1' \vee ... \vee \psi_p'$ for all $j$.

We begin with the case where $\gamma_1 \vee ... \vee \gamma_k \not\models \gamma_1' \vee ... \vee \gamma_o'$. As $\lambda' \models \lambda$, by Theorem 3, $\psi_1' \vee ... \vee \psi_p' \models \psi_1 \vee ... \vee \psi_m$ and for every $i$ there is some $j$ such that $\chi_i' \models \psi_1 \vee ... \vee \psi_m \vee \chi_j$. It follows then (also by Theorem 3) that $\varphi \models \lambda' \models \gamma_1' \vee ... \vee \gamma_o' \vee \Diamond\psi_1 \vee ... \vee \Diamond\psi_m \vee \Box\chi_1 \vee ... \vee \Box\chi_n$, and hence that $\varphi \wedge \neg(\lambda \setminus \{\gamma_1, ..., \gamma_k\}) \models \gamma_1' \vee ... \vee \gamma_o'$. As $\gamma_1' \vee ... \vee \gamma_o' \models \gamma_1 \vee ... \vee \gamma_k \not\models \gamma_1' \vee ... \vee \gamma_o'$, we have found an implicate of $\varphi \wedge \neg(\lambda \setminus \{\gamma_1, ..., \gamma_k\})$ which is stronger than $\gamma_1 \vee ... \vee \gamma_k$, so $\gamma_1 \vee ... \vee \gamma_k \notin \Pi(\varphi \wedge \neg(\lambda \setminus \{\gamma_1, ..., \gamma_k\}))$.

Next suppose that $\psi_1 \vee ... \vee \psi_m \not\models \psi_1' \vee ... \vee \psi_p'$. As $\lambda' \models \lambda$, it follows from Theorem 3 that $\gamma_1' \vee ... \vee \gamma_o' \models \gamma_1 \vee ... \vee \gamma_k$ and that for every $i$ there is some $j$ such that $\chi_i' \models \psi_1 \vee ... \vee \psi_m \vee \chi_j$. We thereby obtain $\varphi \models \lambda' \models \gamma_1 \vee ... \vee \gamma_k \vee \Diamond\psi_1' \vee ... \vee \Diamond\psi_p' \vee \Box(\chi_1 \vee \psi_1 \vee ... \vee \psi_m) \vee ... \vee \Box(\chi_n \vee \psi_1 \vee ... \vee \psi_m)$. From this, we can infer that $\varphi \wedge \neg(\gamma_1 \vee ... \vee \gamma_k \vee \Box(\chi_1 \vee \psi_1 \vee ... \vee \psi_m) \vee ... \vee \Box(\chi_n \vee \psi_1 \vee ... \vee \psi_m)) \models \Diamond\psi_1' \vee ... \vee \Diamond\psi_p' \models \Diamond\psi_1 \vee ... \vee \Diamond\psi_m \not\models \Diamond\psi_1' \vee ... \vee \Diamond\psi_p'$. As $\Diamond\psi_1 \vee ... \vee \Diamond\psi_m \equiv \Diamond(\psi_1 \vee ... \vee \psi_m)$, it follows that $\Diamond(\psi_1 \vee ... \vee \psi_m) \notin \Pi(\varphi \wedge \neg(\gamma_1 \vee ... \vee \gamma_k \vee \Box(\chi_1 \vee \psi_1 \vee ... \vee \psi_m) \vee ... \vee \Box(\chi_n \vee \psi_1 \vee ... \vee \psi_m)))$.

Finally suppose that $\chi_i \not\models \chi_j' \vee \psi_1' \vee ... \vee \psi_p'$ for all $j$ and furthermore that $\psi_1 \vee ... \vee \psi_m \models \psi_1' \vee ... \vee \psi_p'$ (we have already shown the result holds when $\psi_1 \vee ... \vee \psi_m \not\models \psi_1' \vee ... \vee \psi_p'$). Now $\Box(\chi_i \wedge \neg\psi_1 \wedge ... \wedge \neg\psi_m)$ is an implicate of $\varphi \wedge \neg(\lambda \setminus \{\Box\chi_i\})$) so to show that $\Box(\chi_i \wedge \neg\psi_1 \wedge ... \wedge \neg\psi_m)$ is not a prime implicate of $\varphi \wedge \neg(\lambda \setminus \{\Box\chi_i\})$), we must find some stronger implicate. Consider the set $S = \{s \in \{1, ..., q\} : \chi_s' \models \chi_i \vee \psi_1 \vee ... \vee \psi_m$ and $\chi_s' \not\models \chi_k \vee \psi_1 \vee ... \vee \psi_m$ for $k \neq i\}$. We note that there must be at least one element in $S$ as we have assumed $\varphi \not\models \lambda \setminus \{\Box\chi_i\}$. Now since $\gamma_1' \vee ... \vee \gamma_o' \models \gamma_1 \vee ... \vee \gamma_k$, $\psi_1' \vee ... \vee \psi_p' \models \psi_1 \vee ... \vee \psi_m$, for every $s \not\in S$ there is some $r \neq i$ such that $\chi_s' \models \chi_r \vee \psi_1 \vee ... \vee \psi_m$, and $\chi_s' \models \chi_s'$ for $s \in S$, we get $\varphi \models \lambda' \models \gamma_1 \vee ... \vee \gamma_k \vee \Diamond\psi_1 \vee ... \vee \Diamond\psi_m \vee (\bigvee_{j \neq i} \Box\chi_j) \vee (\bigvee_{s \in S} \Box\chi_s')$. It follows that $\varphi \wedge \neg(\lambda \setminus \{\Box\chi_i\}) \models \bigvee_{s \in S} \Box(\chi_s' \wedge \neg\psi_1 \wedge ... \wedge \neg\psi_m)$, which means $\bigvee_{s \in S} \Box(\chi_s' \wedge \neg\psi_1 \wedge ... \wedge \neg\psi_m)$ is an implicate of $\varphi \wedge \neg(\lambda \setminus \{\Box \chi_i\})$. Moreover, $\bigvee_{s \in S} \Box(\chi'_s \wedge \neg \psi_1 \wedge ... \wedge \neg \psi_m) \models \Box(\chi_i \wedge \neg \psi_1 \wedge ... \wedge \neg \psi_m)$ since by construction $\chi'_s \models \chi_i \vee \psi_1 \vee ... \vee \psi_m$ for every $s \in S$.

It remains to be shown that $\Box(\chi_i \wedge \neg \psi_1 \wedge ... \wedge \neg \psi_m) \not\models \bigvee_{s \in S} \Box(\chi'_s \wedge \neg \psi_1 \wedge ... \wedge \neg \psi_m)$. Suppose for a contradiction that the contrary holds. Then $\Box(\chi_i \wedge \neg \psi_1 \wedge ... \wedge \neg \psi_m) \models \bigvee_{s \in S} \Box(\chi'_s \wedge \neg \psi_1 \wedge ... \wedge \neg \psi_m)$, so by Theorem 1, there must be some $s \in S$ for which $\chi_i \wedge \neg \psi_1 \wedge ... \wedge \neg \psi_m \models \chi'_s \wedge \neg \psi_1 \wedge ... \wedge \neg \psi_m$. But then $\chi_i \models \chi'_s \vee \psi_1 \vee ... \vee \psi_m$, and thus $\chi_i \models \chi'_s \vee \psi'_1 \vee ... \vee \psi'_p$ since we have assumed $\psi_1 \vee ... \vee \psi_m \models \psi'_1 \vee ... \vee \psi'_p$. This contradicts our earlier assumption that $\chi_i \not\models \chi'_j \vee \psi'_1 \vee ... \vee \psi'_p$ for all $j$. Thus, we have shown that $\Box(\chi_i \wedge \neg \psi_1 \wedge ... \wedge \neg \psi_m) \not\models \bigvee_{s \in S} \Box(\chi'_s \wedge \neg \psi_1 \wedge ... \wedge \neg \psi_m)$, so $\Box(\chi_i \wedge \neg \psi_1 \wedge ... \wedge \neg \psi_m) \notin \Pi(\varphi \wedge \neg(\lambda \setminus \{\Box \chi_i\}))$. $\qquad \blacksquare$

**Theorem 25** *Let $\varphi$ be a formula of $\mathcal{K}$, and let $\lambda = \gamma_1 \vee ... \vee \gamma_k \vee \Diamond \psi_1 \vee ... \vee \Diamond \psi_n \vee \Box \chi_1 \vee ... \vee \Box \chi_m$ ($\gamma_j$ propositional literals) be a non-tautologous clause such that (a) $\chi_i \equiv \chi_i \vee \psi_1 \vee ... \vee \psi_n$ for all $i$, and (b) there is no literal $l$ in $\lambda$ such that $\lambda \equiv \lambda \setminus \{l\}$. Then $\lambda \in \Pi(\varphi)$ if and only if the following conditions hold:*

1. *$\gamma_1 \vee ... \vee \gamma_k \in \Pi(\varphi \wedge \neg(\lambda \setminus \{\gamma_1, ..., \gamma_k\}))$*

2. *$\Box(\chi_i \wedge \neg \psi_1 \wedge ... \wedge \neg \psi_n) \in \Pi(\varphi \wedge \neg(\lambda \setminus \{\Box \chi_i\}))$ for every $i$*

3. *$\Diamond(\psi_1 \vee ... \vee \psi_n) \in \Pi(\varphi \wedge \neg(\lambda \setminus \{\Diamond \psi_1, ..., \Diamond \psi_n\}))$*

*Proof.* The forward direction was shown in Lemma 25.1. The other direction follows from Lemma 25.2 together with the hypothesis that $\chi_i \equiv \chi_i \vee \psi_1 \vee ... \vee \psi_n$ for all $i$ (which ensures that $\varphi \wedge \neg(\gamma_1 \vee ... \vee \gamma_k \vee \Box(\chi_1 \vee \psi_1 \vee ... \vee \psi_m) \vee ... \vee \Box(\chi_n \vee \psi_1 \vee ... \vee \psi_m)) \equiv \varphi \wedge \neg(\lambda \setminus \{\Diamond \psi_1, ..., \Diamond \psi_n\}))$. $\qquad \blacksquare$

**Theorem 26** *Let $\varphi$ be a formula of $\mathcal{K}$, and let $\gamma$ be a non-tautologous propositional clause such that $\varphi \models \gamma$ and such that there is no literal $l$ in $\gamma$ such that $\gamma \equiv \gamma \setminus \{l\}$. Then $\gamma \in \Pi(\varphi)$ if and only if $\varphi \not\models \gamma \setminus \{l\}$ for all $l$ in $\gamma$.*

*Proof.* Consider a formula $\varphi$ and a non-tautologous propositional clause $\lambda$ such that $\varphi \models \lambda$ and such that there is no literal $l$ in $\lambda$ such that $\lambda \equiv \lambda \setminus \{l\}$. Suppose that $\varphi \models \lambda \setminus \{l\}$ for some literal $l$ in $\lambda$. As we know that $\lambda \not\equiv \lambda \setminus \{l\}$, it follows that $\lambda \setminus \{l\}$ is an implicate of $\varphi$ which is strictly stronger than $\lambda$, so $\lambda$ is not a prime implicate of $\varphi$. For the other direction, suppose that $\lambda \notin \Pi(\varphi)$. Then it must be the case that there is some clause $\rho$ such that $\varphi \models \rho \models \lambda \not\models \rho$. Since $\rho \models \lambda$, it follows from Theorem 2 that each literal in $\rho$ is a propositional literal of $\lambda$ or is inconsistent. If all of the literals in $\rho$ are inconsistent, then both $\rho$ and $\varphi$ must be inconsistent, so clearly $\varphi \models \gamma \setminus \{l\}$ for every $l$ in $\gamma$. Otherwise, $\rho$ is equivalent to a propositional clause, and more specifically to a propositional clause containing only those literals appearing in $\lambda$ (since $\rho \models \lambda$). As $\rho$ is strictly stronger than $\lambda$, there must be some literal $l$ in $\lambda$ which does not appear in $\rho$. But that means $\rho \models \lambda \setminus \{l\}$ and so $\varphi \models \lambda \setminus \{l\}$, completing the proof. $\qquad \blacksquare$

**Theorem 27** *Let $\varphi$ be a formula of $\mathcal{K}$, and let $\lambda = \Box \chi$ be a non-tautologous clause such that $\varphi \models \lambda$. Then $\lambda \in \Pi(\varphi)$ if and only if there exists some term $T \in \mathbf{Dnf\text{-}4}(\varphi)$ such that $\chi \models \beta_T$, where $\beta_T$ is the conjunction of formulae $\psi$ such that $\Box \psi$ is in $T$.*

*Proof.* Let $\varphi$ be some formula, and let $\lambda = \square\chi$ be a non-tautologous clause such that $\varphi \models \lambda$. For the first direction, suppose that there is no term $T \in \mathbf{Dnf\text{-}4}(\varphi)$ such that $\chi \models \beta_T$, where $\beta_T$ is the conjunction of formulae $\psi$ such that $\square\psi$ is in $T$. There are two cases: either there are no terms in $\mathbf{Dnf\text{-}4}(\varphi)$ because $\varphi$ is unsatisfiable, or there are terms but none satisfy the condition. In the first case, $\square\chi$ is not a prime implicate of $\varphi$, since any contradictory clause (e.g. $\Diamond(a \wedge \neg a)$) is stronger. In the second case, consider the clause $\lambda' = \bigvee_T \square\beta_T$, where $\beta_T$ is the conjunction of formulae $\psi$ such that $\square\psi$ is in $T$. Now for every $T$ we must have $\square\beta_T \models \square\chi$, otherwise we would have $T \not\models \square\chi$, and hence $\varphi \not\models \square\chi$. Moreover, $\varphi \models \bigvee_T \square\beta_T$ since $T \models \square\beta_T$ for every $T$. But by Theorem 1, $\square\chi \not\models \bigvee_T \square\beta_T$ since $\chi \not\models \beta_T$ for all $T$. So we have $\varphi \models \lambda' \models \lambda \not\models \lambda'$, which means that $\lambda$ is not a prime implicate of $\varphi$.

For the other direction, suppose that $\square\chi$ is not a prime implicate of $\varphi$ and that $\varphi \not\models \bot$. Then $\mathbf{Dnf\text{-}4}(\varphi)$ is non-empty. As $\varphi \models \square\chi$, we must have $T \models \square\chi$ for all $T \in \mathbf{Dnf\text{-}4}(\varphi)$, so $\bigvee_T \square\beta_T$ also implies $\square\chi$. We now show that $\bigvee_T \square\beta_T$ is a prime implicate of $\varphi$. We let $\kappa$ be some implicate of $\varphi$ which implies $\bigvee_T \square\beta_T$. Now since $\kappa \models \bigvee_T \square\beta_T$ and $\bigvee_T \square\beta_T$ is non-tautologous, it follows from Theorem 2 that $\kappa \equiv \square\zeta_1 \vee ... \vee \square\zeta_n$ for some formulae $\zeta_i$. As $\varphi \models \kappa$, we must have $T \models \square\zeta_1 \vee ... \vee \square\zeta_n$ for all $T \in \mathbf{Dnf\text{-}4}(\varphi)$. But that can only be the case if $\square\beta_T \models \square\zeta_1 \vee ... \vee \square\zeta_n$ for all $T$, which means $\bigvee_T \square\beta_T \models \square\zeta_1 \vee ... \vee \square\zeta_n$. As $\bigvee_T \square\beta_T$ implies every implicate of $\varphi$ that implies it, $\bigvee_T \square\beta_T$ must be a prime implicate of $\varphi$. But this means that $\square\chi \not\models \bigvee_T \square\beta_T$, since we have assumed that $\square\chi$ is not a prime implicate of $\varphi$. It follows from Theorem 1 that $\chi \not\models \beta_T$ for all $T \in \mathbf{Dnf\text{-}4}(\varphi)$. $\qquad\square$

In order to show Theorem 28 we will need the following lemmas:

**Lemma 28.1** *If $\Diamond\psi$ is an implicate of $\varphi$ which is not a prime implicate, the algorithm* **Test$\Diamond$PI** *returns* ***no*** *on input $(\Diamond\psi, \varphi)$.*

*Proof.* Suppose that $\Diamond\psi$ is not a prime implicate of $\varphi$. If $\varphi$ is unsatisfiable, then $\psi$ must be satisfiable, so we will return **no** in the first step. If $\varphi$ is satisfiable, then since we have assumed that $\Diamond\psi$ is an implicate of $\varphi$, there must be some clause $\lambda$ such that $\varphi \models \lambda \models \Diamond\psi$ but $\Diamond\psi \not\models \lambda$. As $\lambda \models \Diamond\psi$, it follows from Theorem 2 that $\lambda$ is equivalent to a disjunction of $\Diamond$-formulae, and hence to some clause $\Diamond\psi'$.

We know from Lemma 13 that $\varphi$ is equivalent to the disjunction of terms in $\mathbf{Dnf\text{-}4}(\varphi)$. It must thus be the case that $T_i \models \Diamond\psi'$ for all $T_i \in \mathbf{Dnf\text{-}4}(\varphi)$. Since each $T_i$ is a satisfiable conjunction of propositional literals and $\square$- and $\Diamond$-formulae, it follows that there exists a set $\{\Diamond\eta_i, \square\mu_{i,1}, ..., \square\mu_{i,k(i)}\}$ of conjuncts of $T_i$ such that $\Diamond(\eta_i \wedge \mu_{i,1} \wedge ... \wedge \mu_{i,k(i)}) \models \Diamond\psi'$, otherwise $T_i$ would fail to imply $\Diamond\psi'$. Moreover, all of the elements of $\{\Diamond\eta_i, \square\mu_{i,1}, ..., \square\mu_{i,k(i)}\}$ must appear in the NNF of $\varphi$ outside modal operators, so the formulae $\eta_i, \mu_{i,1}, ..., \mu_{i,k(i)}$ must all be elements of the set $\mathcal{X}$. It is immediate that both

$$\Diamond\bigvee_i (\eta_i \wedge \mu_{i,1} \wedge ... \wedge \mu_{i,k(i)}) \models \Diamond\psi' \models \Diamond\psi \tag{3}$$

and

$$\Diamond\psi \not\models \Diamond\bigvee_i (\eta_i \wedge \mu_{i,1} \wedge ... \wedge \mu_{i,k(i)})$$

The latter implies that the formula $\Diamond \psi \wedge \neg(\Diamond \bigvee_i(\eta_i \wedge \mu_{i,1} \wedge ... \wedge \mu_{i,k(i)}))$ must be consistent, which means that

$$\psi \wedge \neg(\bigvee_i(\eta_i \wedge \mu_{i,1} \wedge ... \wedge \mu_{i,k(i)})) \equiv \psi \wedge \bigwedge_i(\neg\eta_i \vee \neg\mu_{i,1} \vee ... \vee \neg\mu_{i,k(i)})$$

must be consistent as well. But then it must be the case that we can select for each $i$ some $\sigma_i \in \{\eta_i, \mu_{i,1}, ..., \mu_{i,k(i)}\}$ such that $\psi \wedge \bigwedge_i \neg\sigma_i$ is consistent. Let $S$ be the set of $\sigma_i$. The set $S$ satisfies the condition of the algorithm since:

- $S \subseteq \mathcal{X}$

- $\psi \not\models \bigvee_{\sigma \in S} \sigma$ (because we know $\psi \wedge \bigwedge_i \neg\sigma_i$ to be consistent)

- for each $T_i \in \mathbf{Dnf\text{-}4}(\varphi)$, the conjuncts $\Diamond\eta_i, \Box\mu_{i,1}, ..., \Box\mu_{i,k(i)}$ of $T_i$ are such that:

    - $\{\eta_i, \mu_{i,1}, ..., \mu_{i,k(i)}\} \cap S \neq \emptyset$ (since $S$ contains $\sigma_i \in \{\eta_i, \mu_{i,1}, ..., \mu_{i,k(i)}\}$)
    - $\Diamond(\eta_i \wedge \mu_{i,1} \wedge ... \wedge \mu_{i,k(i)}) \models \Diamond\psi$ (follows from (3) above)

Since there exists a set $S \subseteq \mathcal{X}$ satisfying these conditions, the algorithm returns **no**. $\qquad\square$

**Lemma 28.2** *If the algorithm* **Test$\Diamond$PI** *returns* **no** *on input* ($\Diamond\psi$, $\varphi$)*, then* $\Diamond\psi$ *is not a prime implicate of* $\varphi$.

*Proof.* Suppose **Test$\Diamond$PI** returns **no** on input ($\Diamond\psi$, $\varphi$). If this happens during the first step, it must be the case that $\varphi$ is unsatisfiable and $\Diamond\psi$ is unsatisfiable, in which case $\Diamond\psi$ is not a prime implicate of $\varphi$. The other possibility is that the algorithm returns **no** in Step 3, which means there must be some $S \subseteq \mathcal{X}$ satisfying:

(a) $\psi \not\models \bigvee_{\lambda \in S} \lambda$
(b) for each $T_i \in \mathbf{Dnf\text{-}4}(\varphi)$, there exist conjuncts $\Diamond\eta_i, \Box\mu_{i,1}, ..., \Box\mu_{i,k(i)}$ of $T_i$
    such that:
    (i) $\{\eta_i, \mu_{i,1}, ..., \mu_{i,k(i)}\} \cap S \neq \emptyset$
    (ii) $\Diamond(\eta_i \wedge \mu_{i,1} \wedge ... \wedge \mu_{i,k(i)}) \models \Diamond\psi$

Let $\alpha$ be the clause $\bigvee_i \Diamond(\eta_i \wedge \mu_{i,1} \wedge ... \wedge \mu_{i,k(i)})$. We remark that for each $T_i$, we have $T_i \models \Diamond(\eta_i \wedge \mu_{i,1} \wedge ... \wedge \mu_{i,k(i)})$, and hence $\bigvee_i T_i \models \bigvee_i \Diamond(\eta_i \wedge \mu_{i,1} \wedge ... \wedge \mu_{i,k(i)})$. From the definition of $\mathbf{Dnf\text{-}4}(\varphi)$, we also have $\varphi \equiv \bigvee_i T_i$. It immediately follows that $\varphi \models \bigvee_i \Diamond(\eta_i \wedge \mu_{i,1} \wedge ... \wedge \mu_{i,k(i)})$ and hence $\varphi \models \alpha$. From 2 (b) (ii), we have that $\Diamond(\eta_i \wedge \mu_{i,1} \wedge ... \wedge \mu_{i,k(i)}) \models \Diamond\psi$ for every $i$, and hence $\bigvee_i \Diamond(\eta_i \wedge \mu_{i,1} \wedge ... \wedge \mu_{i,k(i)}) \models \Diamond\psi$ which yields $\alpha \models \Diamond\psi$. From 2 (b) (i), we have that $\{\eta_i, \mu_{i,1}, ..., \mu_{i,k(i)}\} \cap S \neq \emptyset$ and hence that for every $i$ there is some $\lambda \in S$ such that $\eta_i \wedge \mu_{i,1} \wedge ... \wedge \mu_{i,k(i)} \models \lambda$. From this we can infer that $\bigvee_i \Diamond(\eta_i \wedge \mu_{i,1} \wedge ... \wedge \mu_{i,k(i)}) \models \bigvee_{\lambda \in S} \Diamond\lambda$, and hence $\alpha \models \bigvee_{\lambda \in S} \Diamond\lambda$. But we know from 2 (a) and Theorem 1 that $\Diamond\psi \not\models \bigvee_{\lambda \in S} \lambda$. It follows then that $\Diamond\psi \not\models \alpha$. Putting all this together, we see that there exists a clause $\alpha$ such that $\varphi \models \alpha \models \Diamond\psi$ but $\Diamond\psi \not\models \alpha$, and hence that $\Diamond\psi$ is not a prime implicate of $\varphi$. $\qquad\square$

**Theorem 28** *Let* $\varphi$ *be a formula, and let* $\Diamond\psi$ *be an implicate of* $\varphi$. *Then the algorithm* **Test$\Diamond$PI** *returns* **yes** *on input* ($\Diamond\psi$, $\varphi$) *if and only if* $\Diamond\psi$ *is a prime implicate of* $\varphi$.

*Proof.* It is clear that **Test$\diamond$PI** terminates since unsatisfiability testing and the NNF transformation always terminate, and there are only finitely many $S$ and $T_i$. Lemmas 28.1 and 28.2 show us that the algorithm always gives the correct response. $\qquad\blacksquare$

**Theorem 29** *The algorithm* **Test$\diamond$PI** *runs in polynomial space.*

*Proof.* We remark that the sum of the lengths of the elements in $\mathcal{X}$ is bounded by the length of the formula $\text{NNF}(\varphi)$, and hence by Lemma 14 the sum of the lengths of the elements of a particular $S \subseteq \mathcal{X}$ cannot exceed $2|\varphi|$. Testing whether $\psi \not\models \bigvee_{\lambda \in S} \lambda$ can thus be accomplished in polynomial space in the length of $\varphi$ and $\psi$ as it involves testing the satisfiability of the formula $\psi \wedge \bigwedge_{\lambda \in S} \neg\lambda$ whose length is clearly polynomial in $\varphi$ and $\psi$.

Now let us turn to Step 3 (b). We notice that it is not necessary to keep all of the $T_i$ in memory at once, since we can generate the terms $T_i$ one at a time using only polynomial space by Lemma 12. By Lemma 14, the length of any $T_i$ in **Dnf-4**$(\varphi)$ can be at most $2|\varphi|$. It follows that checking whether $\{\eta_i, \mu_{i,1}, ..., \mu_{i,k(i)}\} \cap S \neq \emptyset$, or whether $\diamond(\eta_i \wedge \mu_{i,1} \wedge ... \wedge \mu_{i,k(i)}) \models \diamond\psi$ can both be accomplished in polynomial space in the length of $\varphi$ and $\psi$. We conclude that the algorithm **Test$\diamond$PI** runs in polynomial space. $\qquad\blacksquare$

In order to show Theorem 32, we use the following lemmas:

**Lemma 32.1** *If $\lambda$ is a clause that is not a prime implicate of $\varphi$, then* **TestPI** *outputs* **no** *on this input.*

*Proof.* Let us begin by considering a formula $\lambda$ which is a clause but that is not a prime implicate of $\varphi$. There are two possible reasons for this: either $\lambda$ is not an implicate of $\varphi$, or it is an implicate but there exists some stronger implicate. In the first case, **TestPI** returns **no** in Step 1, as desired. We will now focus on the case where $\lambda$ is an implicate but not a prime implicate. We begin by treating the limit cases where one or both of $\varphi$ and $\lambda$ is a tautology or contradiction. Given that we know $\lambda$ to be a non-prime implicate of $\varphi$, there are only two possible scenarios: either $\not\models \varphi$ and $\models \lambda$, or $\varphi \models \bot$ and $\lambda \not\models \bot$. In both cases, the algorithm returns **no** in Step 2.

If $\lambda$ is an implicate of $\varphi$, and neither $\varphi$ nor $\lambda$ is a tautology or contradiction, then the algorithm will continue on to Step 3. In this step, any redundant literals will be deleted from $\lambda$, and if $\lambda$ contains $\diamond$-literals, we add an extra disjunct to the $\square$-literals so that $\lambda$ satisfies the syntactic requirements of Theorem 25. Let $\gamma_1 \vee ...\gamma_k \vee \diamond\psi_1 \vee ... \vee \diamond\psi_m \vee \square\chi_1 \vee ... \vee \square\chi_n$ be the clause $\lambda$ at the end of Step 3 once all modifications have been made. As the transformations in Step 3 are equivalence-preserving (Theorem 1), the modified $\lambda$ is equivalent to the original, so $\lambda$ is still a non-tautologous non-prime implicate of $\varphi$. This means $\varphi$ and $\lambda$ now satisfy all of the conditions of Theorem 25. It follows then that one of the following holds:

(a) $\gamma_1 \vee ... \vee \gamma_k \notin \Pi(\varphi \wedge \neg(\lambda \setminus \{\gamma_1, ..., \gamma_k\})$

(b) $\square(\chi_i \wedge \neg\psi_1 \wedge ... \wedge \neg\psi_n) \notin \Pi(\varphi \wedge \neg(\lambda \setminus \{\square\chi_i\}))$ for some $i$

(c) $\diamond(\psi_1 \vee ... \vee \psi_n) \notin \Pi(\varphi \wedge \neg(\lambda \setminus \{\diamond\psi_1, ..., \diamond\psi_n\}))$

Suppose that (a) holds. Now $\gamma_1 \vee ... \vee \gamma_k$ is a non-tautologous propositional clause implied by $\varphi \wedge \neg(\lambda \setminus \{\gamma_1, ..., \gamma_k\})$ which contains no redundant literals. This means that $\varphi \wedge \neg(\lambda \setminus \{\gamma_1, ..., \gamma_k\})$ and $\gamma_1 \vee ... \vee \gamma_k$ satisfy the conditions of Theorem 26. According to this theorem, as $\gamma_1 \vee ... \vee \gamma_k \notin \Pi(\varphi \wedge \neg(\lambda \setminus \{\gamma_1, ..., \gamma_k\})$, then there must be some $\gamma_j$ such that $\varphi \wedge \neg(\lambda \setminus \{\gamma_1, ..., \gamma_k\}) \models \gamma_1 \vee ... \vee \gamma_{j-1} \vee \gamma_{j+1} \vee ... \vee \gamma_k$. This means that $\varphi \models \lambda \setminus \{\gamma_j\}$, so the algorithm returns **no** in Step 4.

Suppose next that (b) holds, and let $i$ be such that $\Box(\chi_i \wedge \neg\psi_1 \wedge ... \wedge \neg\psi_n) \notin \Pi(\varphi \wedge \neg(\lambda \setminus \{\Box\chi_i\}))$. By Theorem 27, this means that there is no $T \in \mathbf{Dnf\text{-}4}(\varphi)$ such that $\Box(\chi_i \wedge \neg\psi_1 \wedge ... \wedge \neg\psi_n)$ entails the conjunction of $\Box$-formulae conjuncts of $T$. It follows that the algorithm returns **no** in Step 5.

Finally consider the case where neither (a) nor (b) holds but (c) does. Then in Step 6, we will call $\mathbf{Test}\Diamond\mathbf{PI}(\Diamond(\bigvee_{i=1}^m \psi_i), \varphi \wedge \neg(\lambda \setminus \{\Diamond\psi_1, ..., \Diamond\psi_m\}))$. As $\Diamond(\bigvee_{i=1}^m \psi_i)$ is not a prime implicate of $\varphi \wedge \neg(\lambda \setminus \{\Diamond\psi_1, ..., \Diamond\psi_m\}))$ and we have shown $\mathbf{Test}\Diamond\mathbf{PI}$ to be correct (Theorem 28), $\mathbf{Test}\Diamond\mathbf{PI}$ will return **no**, so $\mathbf{TestPI}$ will return **no** as well. As we have covered each of the possible cases, we can conclude that if $\lambda$ is a clause that is not a prime implicate of $\varphi$, then $\mathbf{TestPI}$ outputs **no**. $\qquad\blacksquare$

**Lemma 32.2** *If* $\mathbf{TestPI}$ *outputs* **no** *with input* $(\lambda, \varphi)$ *and* $\lambda$ *is a clause, then* $\lambda$ *is not a prime implicate of* $\varphi$.

*Proof.* There are 5 different ways for $\mathbf{TestPI}$ to return **no** (these occur in Steps 1, 2, 4, 5, and 6). Let us consider each of these in turn. The first way that the algorithm can return **no** is in Step 1 if we find that $\varphi \not\models \lambda$. This is correct since $\lambda$ cannot be a prime implicate if it is not a consequence of $\varphi$. In Step 2, we return **no** if $\varphi$ is unsatisfiable but $\lambda$ is not, or if $\lambda$ is a tautology but $\varphi$ is not. This is also correct since in both cases $\lambda$ cannot be a prime implicate since there exist stronger implicates (any contradictory clause if $\varphi \equiv \bot$, and any non-tautologous implicate of $\varphi$ if $\lambda \equiv \top$). In Step 3, we may modify $\lambda$, but the resulting formula is equivalent to the original, and so it is a prime implicate just in the case that the original clause was. Let $\gamma_1 \vee ... \gamma_k \vee \Diamond\psi_1 \vee ... \vee \Diamond\psi_m \vee \Box\chi_1 \vee ... \vee \Box\chi_n$ be the clause at the end of Step 3. Now in Step 4, we return **no** if we find some propositional literal $l$ in $\lambda$ for which $\varphi \models \lambda \setminus \{l\}$. Now since in Step 3, we have removed redundant literals from $\lambda$, we can be sure that $\lambda \setminus \{l\}$ is strictly stronger than $\lambda$. So we have $\varphi \models \lambda \setminus \{l\} \models \lambda$ and $\lambda \not\models \lambda \setminus \{l\}$, which means that $\lambda$ is not a prime implicate of $\varphi$. We now consider Step 5 of $\mathbf{TestPI}$. In this step, we return **no** if for some disjunct $\Box\chi_i$ there is no term $T$ in $\mathbf{Dnf\text{-}4}(\varphi \wedge \neg(\lambda \setminus \{\Box\chi_i\}))$ for which $\Box(\chi_i \wedge \neg\psi_1 \wedge ... \wedge \neg\psi_m)$ entails the conjunction of $\Box$-literals in $T$. According to Theorem 27, this means that $\Box(\chi_i \wedge \neg\psi_1 \wedge ... \wedge \neg\psi_m)$ is not a prime implicate of $\varphi \wedge \neg(\lambda \setminus \{\Box\chi_i\})$, which means that $\lambda$ is not a prime implicate of $\varphi$ by Theorem 25. . In this step, we return **no** if $\mathbf{Test}\Diamond\mathbf{PI}$ returns **no** on input $(\Diamond(\bigvee_{i=1}^k \psi_i), \varphi \wedge \neg(\lambda \setminus \{\Diamond\psi_1, ..., \Diamond\psi_m\}))$. By Theorem 28, we know that this happens just in the case that $\Diamond(\bigvee_{i=1}^k \psi_i)$ is not a prime implicate of $\varphi \wedge \neg(\lambda \setminus \{\Diamond\psi_1, ..., \Diamond\psi_m\})$. It follows from Theorem 25 that $\lambda$ is not a prime implicate of $\varphi$. $\qquad\blacksquare$

**Theorem 32** *The algorithm* $\mathbf{TestPI}$ *always terminates, and it returns* **yes** *on input* $(\lambda, \varphi)$ *if and only if* $\lambda$ *is a prime implicate of* $\varphi$.

*Proof.* The algorithm **TestPI** clearly terminates because Steps 1 to 5 involve a finite number of syntactic operations on $\lambda$ and a finite number of entailment checks. Moreover, the call to **Test◇PI** in Step 6 is known to terminate (Theorem 28). Correctness and completeness have already been shown in Lemmas 32.1 and 32.2. □

We make use of the following lemma in the proof of Theorem 34:

**Lemma 34.1** *The algorithm* **TestPI** *provided in Figure 5 runs in polynomial space in the length of the input.*

*Proof.* It is clear that steps 1 through 5 can be carried out in polynomial space in the length of the input, since they simply involve testing the satisfiability of formulae whose lengths are polynomial in $|\lambda| + |\varphi|$. Step 6 can also be carried out in polynomial space since by Theorem 29 deciding whether the formula $\Diamond(\bigvee_{i=1}^{m} \psi_i)$ is a prime implicate of $\varphi \wedge \neg(\lambda \setminus \{\psi_1, ..., \psi_m\})$ takes only polynomial space in $|\Diamond(\bigvee_{i=1}^{m} \psi_i)| + |\varphi \wedge \neg(\lambda \setminus \{\Diamond\psi_1, ..., \Diamond\psi_m\}))|$, and hence in $|\lambda| + |\varphi|$. We can thus conclude that the algorithm **TestPI** runs in polynomial space in the length of the input. □

**Theorem 34** *Prime implicate recognition is in* PSPACE.

*Proof.* We have show in Theorem 32 that **TestPI** always terminates and returns **yes** whenever the clause is a prime implicate and **no** otherwise. This means that **TestPI** is a decision procedure for prime implicate recognition. Since the algorithm has been shown to run in polynomial space (Lemma 34.1), we can conclude that prime implicate recognition is in PSPACE. □

**Corollary 35** *Prime implicate recognition is* PSPACE-*complete.*

*Proof.* Follows directly from Theorems 24 and 34. □